\documentclass[twocolumn,aps,prc,superscriptaddress,showpacs,floatfix]{revtex4}

\usepackage{epsfig}
\usepackage{amsmath}
\usepackage{multirow}
\usepackage{graphicx}
\usepackage{amsmath}
\usepackage{feynmf}
\usepackage[normalem]{ulem}  
\usepackage[dvips]{color} 

\renewcommand\sout{\bgroup \color{red} \ULdepth=-.5ex \ULset}

\newcommand{\Ex}[2]{\ifmmode{#1\times10^{#2}}\else{$#1\times10^{#2}$}\fi}

\begin{document}
\title{Dibaryons with two strange quarks and one heavy flavor in a constituent quark model}

\author{Aaron Park}
\email{aaron.park@yonsei.ac.kr}\affiliation{Department of Physics and Institute of Physics and Applied Physics, Yonsei University, Seoul 120-749, Korea}
\author{Woosung Park}\email{diracdelta@yonsei.ac.kr}\affiliation{Department of Physics and Institute of Physics and Applied Physics, Yonsei University, Seoul 120-749, Korea}
\author{Su Houng~Lee}\email{suhoung@yonsei.ac.kr}\affiliation{Department of Physics and Institute of Physics and Applied Physics, Yonsei University, Seoul 120-749, Korea}
\date{\today}
\begin{abstract}
We investigate the symmetry property and the stability of dibaryons containing two strange quarks and one heavy flavor with $I=\frac{1}{2}$. We construct the wave function of the dibaryon in two ways. First, we directly construct the color and spin state of the dibaryon starting from the four possible SU(3) flavor state. Second, we consider the states composed of five light quarks, and then construct the wave function of the dibaryon by adding one heavy quark. The stability of the dibaryon against the strong decay into two baryons is discussed by using variational method in a constituent quark model with confining and hyperfine potential. We find that for all configurations with S=0,1,2, the ground states of the dibaryons are the sum of two baryons,  and there are no compact bound state that is stable against the strong decay.
\end{abstract}


\maketitle

\section{Introduction}
Investigating the stability of multiquark hadrons have been pursued in various models after Jaffe suggested the possible existence of such particles in QCD~\cite{Jaffe:1976ig, Jaffe:1976ih, Jaffe:1976yi}. The observation of many charmonium-like states, such as $X(3872)$, $Z_c(3900)$ and $Z^+(4430)$, and of heavy pentaquark states\cite{Aaij:2015tga} revived great interest in the studies of multiquark hadrons and/or of molecular bound states containing heavy quark hadrons.   Additionally, a new particle, called $X(5568)$ was recently observed by the D0 collaboration in the $B^0_s \pi^\pm$ invariant mass spectrum with 5.1$\sigma$ significance~\cite{D0:2016mwd}. $X(5568)$ may be the first observed tetraquark which has four different flavors: up, down, strange and bottom. If all the flavors are different, for any typical two body interaction, one can always find the most attractive combination so that one has advantage to form a bound multiquark state.

The stability of the dibaryons with heavy quarks were studied already in several models; those based on chromomagnetic models~\cite{Leandri:1993zg,Leandri:1995zm} and chiral constituent quark model~\cite{Pepin:1998ih,Carames:2015sya}. Furthermore, Huang et al. studied $H$-like dibaryon states containing heavy quarks instead of strange quarks within the framework of the quark delocalization color screening model~\cite{Huang:2013rla}. Dibaryons within the diquark models with heavy quarks are also considered~\cite{Lee:2009rt}.

Most of the models studying the possible existence of dibaryons are looking at the most attractive color spin interaction channel\cite{Jaffe:1976ig, Jaffe:1976ih, Jaffe:1976yi}.  For example, for the $H$ dibaryon, the attraction in the color spin interaction is larger than those coming from two $\Lambda$'s, which is the most attractive two baryon channel that the dibaryon can decay.  However, it should be noted that whether such attraction really leads to a stable compact dibaryon states is determined by whether the attraction is strong enough to overcome the extra repulsion coming from bringing all the quarks together into a compact configuration.  As we will discuss later, the magnitude of each effect  depends on the masses of the quarks involved that can only be systematically studied within a complete model that consistently treats the kinetic terms and the interaction terms within one framework.

In this work, to investigate the subtle interplay between the two competing effects, in a simple but consistent model, we will
study the stability of $uudssQ$ dibaryon using variational method in a constituent quark model. This is a generalization of H-dibaryon to include one heavy quark so that it contains the most attractive color-spin interaction channel but at the same time, reduced kinetic energy from combining six quarks in a compact configuration. In particular, we focus our attention on $I=\frac{1}{2}$ because the states with lowest isospin are most attractive bound for a given quark system~\cite{Leandri:1993zg}.

Moreover, we will demonstrate how to consistently construct the color spin flavor wave functions that contains the $uudssQ$ quarks.
There are two ways of constructing the wave function of dibaryon with one heavy quark. First, we can directly construct the color and spin wave function of dibaryon starting from the four possible SU(3) flavor state. Or we can consider the color and spin state of $q^5$ and then construct the wave function of dibaryon by adding one heavy quark. We show the two approaches lead to identical wave functions, showing the consistency of our approach.   Technically, the second approach is more convenient to obtain the wave function compared to using the  first approach, because the former utilises Clebsch-Gordan coefficients of $S_6$ while the second approach uses that of  $S_5$.

This paper is organized as follows. We first present the Hamiltonian and calculate the masses of baryons to determine the fitting parameters of the model in Sec.\ref{section2}. In Sec.\ref{stability}, we explain why we choose the dibaryon with one heavy flavor in terms of the relation between hyperfine potential and the stability condition. In Sec.\ref{section3}, we construct the spatial wave function of the dibaryon. In Sec.\ref{section4}, we classify $q^5$ with SU(3) flavor symmetry and construct the color and spin wave function of dibaryon using the first method. In Sec.\ref{section5}, we construct the color and spin wave function of dibaryon using the second method. In Sec.\ref{section6}, we calculate the wave function of dibaryon and show that their results are the same in both methods. In Sec.\ref{section7}, we represent the numerical results obtained from the variational method, and finally summarize the results in Sec.\ref{section8}. The appendices include some details of the calculations.

\section{Hamiltonian}

We take a nonrelativistic Hamiltonian with the confinement and hyperfine potential given by
\begin{eqnarray}
H=\sum_{i=1}^{6}(m_{i}+\frac{\textbf{p}^2_i}{2m_i})-\frac{3}{16}\sum_{i<j}^{6}
\lambda^c_i\lambda^c_j(V^{C}_{ij}+V^{SS}_{ij}), \label{Hamiltonian}
\end{eqnarray}
where $m_i$'s are the quark masses, and $\lambda^c_i/2$ are the color operator of the $i$'th quark for the color SU(3), and $V^{C}_{ij}$ and $V^{SS}_{ij}$ are the confinement and hyperfine potential, respectively.
\begin{eqnarray}
V^{C}_{ij}=-\frac{\kappa}{r_{ij}}+\frac{r_{ij}}{a_0}-D. \label{vc_ij-01}
\end{eqnarray}
The hyperfine term which effectively splits the multiplets of baryon with respect to spin is expressed as
\begin{eqnarray}
V^{SS}_{ij}=\frac{\hbar^2c^2{\kappa}^{\prime}}{m_im_jc^4}\frac{1}{(r_{0ij})^2r_{ij}}e^{-(r_{ij})^2/(r_{0ij})^2}
{\sigma}_i\cdot{\sigma}_j.
\end{eqnarray}
Here, $r_{ij}$ is the distance between interquarks, $\mid\textbf{r}_i-\textbf{r}_j\mid$, and $r_{0ij}$ are chosen to depend on
the masses of interquarks given by
\begin{eqnarray}
r_{0ij}=1/(\alpha+\beta \frac{m_im_j}{m_i+m_j}).
\end{eqnarray}
We choose to keep the isospin symmetry by requiring that $m_u$=$m_d$. In the Hamiltonian, the parameters have been chosen so that the fitted mass of baryons are comparable with those of experiments.
\begin{table}
\caption{ Parameters fitted to the experimental  baryon masses using the variational method with a single Gaussian. The respective units are given in the third row.}
\begin{center}
\begin{tabular}{|c|c|c|c|c|c|c|c|c|c|c}
\hline \hline
  $\kappa$  & $\kappa '$ & $a_0$  & $D$ & $\alpha$ &$\beta$ & $m_q$  & $m_s$ & $m_c$  & $m_b$   \\
\hline
  0.59  & 0.5 & 5.386  & 0.960  &  2.6    & 0.552 & 0.343  & 0.632 & 1.93 & 5.3 \\
  & & $ \mbox{GeV}^{-2}$ & $\mbox{GeV}$   & $(\mbox{fm})^{-1}$  &     & $\mbox{GeV}$   & $\mbox{GeV}$ & $\mbox{GeV}$ & $\mbox{GeV}$  \\
\hline \hline
\end{tabular}
\end{center}
\label{parameter-01}
\end{table}

When we calculate the expectation value of the potential terms for baryon with certain symmetry,
it is convenient to introduce the following three Jacobian coordinates. Then it reduces our problem to the two body-system in the center
of mass frame :
\begin{itemize}
\item Coordinate I :
\begin{align}
\pmb{x_1}=&\frac{1}{\sqrt{2}}(\textbf{r}_1-\textbf{r}_2),\nonumber\\
\pmb{x_2}=&\frac{m_1+m_2}{\sqrt{2m_1^2+2m_2^2+2m_1 m_2}}(\frac{m_1}{m_1+m_2}\textbf{r}_1 \nonumber\\
&+\frac{m_2}{m_1+m_2}\textbf{r}_2-\textbf{r}_3).
\end{align}
\item Coordinate II :
\begin{align}
\pmb{y_1}=&\frac{1}{\sqrt{2}}(\textbf{r}_2-\textbf{r}_3),\nonumber\\
\pmb{y_2}=&\frac{m_2+m_3}{\sqrt{2m_2^2+2m_3^2+2m_2 m_3}}(\frac{m_2}{m_2+m_3}\textbf{r}_2 \nonumber\\
&+\frac{m_3}{m_2+m_3}\textbf{r}_3-\textbf{r}_1).
\end{align}
\item Coordinate III :
\begin{align}
\pmb{z_1}=&\frac{1}{\sqrt{2}}(\textbf{r}_3-\textbf{r}_1),\nonumber\\
\pmb{z_2}=&\frac{m_1+m_3}{\sqrt{2m_1^2+2m_3^2+2m_1 m_3}}(\frac{m_1}{m_1+m_3}\textbf{r}_1 \nonumber\\
&+\frac{m_3}{m_1+m_3}\textbf{r}_3-\textbf{r}_2).
\end{align}
\end{itemize}
By using simple Gaussian function, we calculate the baryon masses containing charm or bottom quark. The method to calculate the color, flavor and spin basis was explained in Ref.~\cite{Park:2015nha}.
\begin{table}
\caption{This table shows the masses of baryons obtained from the variational method. The third row indicates the experimental data. ( unit : $\rm{GeV}$ ) }
\begin{center}
\begin{tabular}{c|c|c|c|c|c|c|c}
\hline \hline
(I,S)      & ($\frac{1}{2}$,$\frac{1}{2}$) &($\frac{1}{2}$,$\frac{3}{2}$)& (0,$\frac{1}{2}$) & (1,$\frac{1}{2}$) &(1,$\frac{3}{2}$) &($\frac{1}{2}$,$\frac{1}{2}$) &($\frac{1}{2}$,$\frac{3}{2}$)    \\
           & N, P     & $\Delta$    & $\Lambda$   & $\Sigma$     & $\Sigma^*$   & $\Xi$    &  $\Xi^*$     \\
\hline
Mass  & 0.977 & 1.23 & 1.12 & 1.2 & 1.38 & 1.324 & 1.52      \\
\hline
Exp       & 0.938   & 1.232  & 1.115  & 1.189 & 1.382 & 1.315 & 1.532  \\
\hline \hline
\end{tabular}
\end{center}
\label{baryon_mass-01}
\begin{center}
\begin{tabular}{c|c|c|c|c|c|c|c}
\hline \hline
(I,S)      &  (0,$\frac{1}{2}$) & (1,$\frac{1}{2}$) & (1,$\frac{3}{2}$) & ($\frac{1}{2}$,$\frac{1}{2}$) & ($\frac{1}{2}$,$\frac{3}{2}$) & (0,$\frac{1}{2}$) & (0,$\frac{3}{2}$)   \\
           & $\Lambda_c$   & $\Sigma_c$   & $\Sigma_c^*$ & $\Xi_c$ & $\Xi_c^*$ & $\Omega_c$ & $\Omega_c^*$     \\
\hline
Mass   & 2.285 & 2.45 & 2.526 & 2.476 & 2.649 & 2.687 & 2.763   \\
\hline
Exp       & 2.286 & 2.453 & 2.518 & 2.468 & 2.646  & 2.695 & 2.766  \\
\hline \hline
\end{tabular}
\end{center}
\label{baryon_mass-02}
\begin{center}
\begin{tabular}{c|c|c|c|c|c|c|c}
\hline \hline
(I,S)      &  (0,$\frac{1}{2}$) & (1,$\frac{1}{2}$) & (1,$\frac{3}{2}$) & ($\frac{1}{2}$,$\frac{1}{2}$) & ($\frac{1}{2}$,$\frac{3}{2}$) & (0,$\frac{1}{2}$) & (0,$\frac{3}{2}$)  \\
           & $\Lambda_b$ & $\Sigma_b$ & $\Sigma_b^*$ & $\Xi_b$ & $\Xi_b^*$ & $\Omega_b$ & $\Omega_b^*$   \\
\hline
Mass  & 5.608 & 5.809 & 5.839 & 5.787 & 5.95 & 6.019 & 6.053  \\
\hline
Exp    & 5.619 & 5.811 & 5.832 & 5.792 & 5.949 & 6.048 &  \\
\hline \hline
\end{tabular}
\end{center}
\label{baryon_mass-03}
\end{table}
\label{section2}

\section{Color-Spin interaction and the Stability condition}
\label{stability}
It is well known that color-spin interaction is an important factor in  investigating the stability of multiquark system. In SU(3) flavor symmetry, there is a simple formula~\cite{Aerts:1977rw} from which one can easily calculate the expectation value of hyperfine potential.
\begin{align}
  &H_{SS}=-\sum_{i<j}^N \lambda_i^c \lambda_j^c \sigma_i \cdot \sigma_j \nonumber\\
  &= N(N-10)+\frac{4}{3}S(S+1)+2C_C+4C_F
\label{eq:hyperfine-formula}
\end{align}
where $C_F=\frac{1}{4}\lambda^F \lambda^F$. For the flavor singlet H-dibaryon, $H_{SS}=-24$, and for $\Lambda$, $H_{SS}$= -8. Hence,  H-dibaryon is more attractive than $\Lambda \Lambda$ system in terms of color-spin interaction. This is the basis for a possible stable H-dibaryon.

At the same time, it is interesting to point out that we can split the dibaryon into five quarks and one quark system. By using Eq.~(\ref{eq:hyperfine-formula}) we can calculate the expectation value of hyperfine potential of five quarks in H-dibaryon. We represent the expectation values of hyperfine potential for H-dibaryon with flavor singlet and $\Lambda \Lambda$ in Table~\ref{table:hyperfine1}. In that case, the flavor and color state of five quarks are anti-triplet and $C_C=C_F=\frac{4}{3}$ for anti-triplet state so it leads to $H_{SS}=-16$ which is the same as $H_{SS}^{\Lambda \Lambda}$. So it shows that the interaction between sixth quark and the other quarks give more attractive effect than $\Lambda \Lambda$ system and agrees with our recent work~\cite{Park:2016cmg}. Unfortunately, H-dibaryon is not stable in our model when we consider the Hamiltonian as the repulsion coming from kinetic energy and confinement potential dominates over the attraction coming from hyperfine potential as we bring six quarks to compact configuration.

If we replace the sixth quark with heavy quark, then the situation becomes more subtle. In infinite heavy quark mass limit, the contribution from sixth quark becomes zero because hyperfine potential has $1/m_Q$ factor. And in that case, the expectation value of hyperfine potential is the same as that of $\Lambda$ and diquark system. However, heavy quark mass is not infinite, so $1/m_Q$ factor weakens the attractive effect, but it will also reduce the kinetic energy if it doesn't change the interquark distances. So we can consider the dibaryon with heavy flavor and it may lead more chance to form the stable state than H-dibaryon. It should be noted that anti-triplet flavor state is the most attractive color-spin interaction when we consider five quarks only.
\begin{table}
\caption{The expectation value of $-\sum_{i<j}\langle \lambda^c_i \lambda^c_j \sigma_i \cdot \sigma_j \rangle $ for H-dibaryon with flavor singlet and $\Lambda \Lambda$.}
\begin{center}
\begin{tabular}{c|c|c}
\hline
\hline
$-\sum_{i<j}\langle \lambda^c_i \lambda^c_j \sigma_i \cdot \sigma_j \rangle $ & $i<j=1-5$ & $i=1-5,j=6$ \\
\hline
H-dibaryon, $F^1$ & -16 & -8 \\
\hline
\hline
\end{tabular}
\end{center}
\begin{center}
\begin{tabular}{c|c|c|c}
\hline
\hline
$-\sum_{i<j}\langle \lambda^c_i \lambda^c_j \sigma_i \cdot \sigma_j \rangle $ & $i<j=1-3$ & $i,j=4-5$ & i=4-5,j=6 \\
\hline
$\Lambda \Lambda$ & -8 & -8 & 0 \\
\hline
\hline
\end{tabular}
\end{center}
\label{table:hyperfine1}
\end{table}


\section{Spatial function}
In order to construct an antisymmetric wave function of the dibaryon, we choose the spatial function to be symmetric such that the  rest of the wave function represented by color $\otimes$ flavor $\otimes$ spin should be antisymmetric. Here, we calculate in the flavor SU(3) breaking case and fix the position of each quarks on $u(1)u(2)d(3)s(4)s(5)Q(6)$. So our wave function should have the specific symmetry property which is antisymmetric among 1,2 and 3, and at the same time antisymmetric between 4 and 5. And among various Jacobi coordinates, we choose the baryon-baryon configuration because it is convenient to investigate the strong decay mode.
\begin{align}
&\sum_{i=1}^6 \frac{1}{2} m_i \dot{\textbf{r}}_i^2-\frac{1}{2}M \dot{\textbf{r}}_{CM}^2=\sum_{i=1}^5 \frac{1}{2}M_i \dot{\textbf{x}}_i^2 \nonumber\\
&\mathrm{where}~M_1=M_2=m, ~M_3=m_s, ~M_4=\frac{3m_s m_Q}{2m_s+m_Q}, \nonumber\\
&M_5=\frac{2m(5m_s^2+2m_s m_Q+2m_Q^2)}{(3m+2m_s+m_Q)(2m_s+m_Q)}
\label{mass}
\end{align}
The Jacobian coordinates are given by
\begin{eqnarray}
\pmb{x_1}&=&\frac{1}{\sqrt{2}}(\textbf{r}_1-\textbf{r}_2), \nonumber\\
\pmb{x_2}&=&\sqrt{\frac{2}{3}}(\frac{1}{2}\textbf{r}_1+\frac{1}{2}\textbf{r}_2-\textbf{r}_3),\nonumber\\
\pmb{x_3}&=&\frac{1}{\sqrt{2}}(\textbf{r}_4-\textbf{r}_5), \nonumber\\
\pmb{x_4}&=&\sqrt{\frac{2}{3}}(\frac{1}{2}\textbf{r}_4+\frac{1}{2}\textbf{r}_5-\textbf{r}_6),\nonumber\\
\pmb{x_5}&=&\frac{\sqrt{3}(2m_s+m_Q)}{\sqrt{10m_s^2+4m_s m_Q+4m_Q^2}}(\frac{1}{3}\textbf{r}_1+\frac{1}{3}\textbf{r}_2+\frac{1}{3}\textbf{r}_3 \nonumber\\
&&-\frac{m_s}{2m_s+m_Q}\textbf{r}_4 -\frac{m_s}{2m_s+m_Q}\textbf{r}_5-\frac{m_Q}{2m_s+m_Q}\textbf{r}_6).\nonumber\\
\label{eq-jac2}
\end{eqnarray}

Then, we can construct the spatial wave function of the dibaryon in a single Gaussian form that can accommodate the required symmetry property:
\begin{equation}
R=\mathrm{exp}[-a(x_1^2+x_2^2)-b x_3^2-c x_4^2-d x_5^2],
\label{spatial function}
\end{equation}
where $a$, $b$, $c$ and $d$ are the variational parameters. The spatial function in Eq.~(\ref{spatial function}) is symmetric among 1,2 and 3, and at the same time symmetric between 4 and 5. We will denote this symmetry property of the spatial function by [123][45]6.
Considering the dibaryon to be formed by bringing together a baryon composed of particle [123] and baryon composed of [45]6, one notes that the additional kinetic term will involve coordinate $\pmb{x_5}$ with mass $M_5$.  Hence, for fixed $\pmb{x_5}$, the additional kinetic term becomes smaller when $m_Q$ increases but only becomes zero when more than one quark becomes heavy as can be seen in  Eq.~(\ref{mass}).  This additional kinetic term has to be smaller than the additional attraction coming from the color spin interaction for the dibaryon to form a stable compact state.
\label{section3}

\section{Classification of $q^5Q$ with SU(3) flavor symmetry}
In this section, we directly construct the color and spin wave function of the dibaryon from the four possible SU(3) flavor state.
\label{section4}
\subsection{Flavor state of $q^5$}
Here, we classify the flavor states in terms of $SU(3)_F$ symmetry and will break the flavor symmetry later. Since spatial wave function is symmetric, we have to construct the flavor, color and spin wave function to be antisymmetric. Under the general group $SU(18)_{CFS}$ totally antisymmetric multiplet of $[1^5]_{FCS}$ can be decomposed as
\begin{align}
[1^5]_{FCS}=&([\bar{3}]_{F}, [420]_{CS}) \oplus ([6]_{F}, [336]_{CS}) \oplus \nonumber\\
&([\bar{15}]_{F}, [210]_{CS}) \oplus ([24]_{F}, [84]_{CS}) \oplus \nonumber\\
&([21]_{F}, [\bar{6}]_{CS}).
\end{align}
In this article, we consider only $I=\frac{1}{2}$, so that we exclude $[21]_F$ flavor state.  Hence,  there are four possible flavor states as follows.
\begin{eqnarray}
&&[\bar{3}]_F=
\begin{tabular}{|c|c|}
\hline
\quad \quad & \quad \quad  \\
\cline{1-2}
\quad \quad & \quad \quad  \\
\cline{1-2}
\multicolumn{1}{|c|}{\quad}  \\
\cline{1-1}
\end{tabular}
,~
[6]_F=
\begin{tabular}{|c|c|c|}
\hline
\quad \quad & \quad \quad & \quad \quad \\
\hline
\multicolumn{1}{|c|}{\quad}  \\
\cline{1-1}
\multicolumn{1}{|c|}{\quad}  \\
\cline{1-1}
\end{tabular}
,~
[\bar{15}]_F=
\begin{tabular}{|c|c|c|}
\hline
\quad \quad & \quad \quad & \quad \quad \\
\hline
\multicolumn{1}{|c|}{\quad} & \multicolumn{1}{|c|}{\quad}  \\
\cline{1-2}
\end{tabular}, \nonumber\\
&&[24]_F=
\begin{tabular}{|c|c|c|c|}
\hline
\quad \quad & \quad \quad & \quad \quad & \quad \quad \\
\hline
\multicolumn{1}{|c|}{\quad}  \\
\cline{1-1}
\end{tabular}. \nonumber
\end{eqnarray}

For each flavor state, we can determine the possible Young tableau of color and spin state of $q^5Q$. According to group theory, for a given Young tableau, the fully antisymmetric state can be constructed by multiplying the
Young tableau by its conjugate of the Young tableau, where the conjugate representation of a given Young tableau can be obtained by exchanging the row and column in the Young tableau.
Additionally, the Young tableau of color and spin state that can contribute to the final state depends on the spin of the  dibaryon.
For a given spin sate, the possible color and spin state can be obtained by taking the direct product of the color singlet dibaryon configuration to spin state and taking the conjugate, with the addition of the 6'th quark, of the fixed flavor state.
 We represent the possible flavor, color and spin state of $q^5Q$ with $I=\frac{1}{2}$ for each spin states as follows.
\begin{widetext}
\begin{itemize}
\item S=0 : 4 states
\begin{eqnarray}
\begin{tabular}{|c|c|}
\hline
\quad \quad & \quad \quad  \\
\cline{1-2}
\quad \quad & \quad \quad  \\
\cline{1-2}
\multicolumn{1}{|c|}{\quad}  \\
\cline{1-1}
\end{tabular}_F
\otimes
\begin{tabular}{|c|c|c|}
\hline
\quad \quad & \quad \quad & \quad \quad   \\
\hline
\quad \quad & \quad \quad & \quad \quad   \\
\hline
\end{tabular}_{~CS},~
\begin{tabular}{|c|c|c|}
\hline
\quad \quad & \quad \quad & \quad \quad \\
\hline
\multicolumn{1}{|c|}{\quad}  \\
\cline{1-1}
\multicolumn{1}{|c|}{\quad}  \\
\cline{1-1}
\end{tabular}_F
\otimes
\begin{tabular}{|c|c|c|c|}
\hline
\quad \quad & \quad \quad & \quad \quad & \quad \quad  \\
\hline
\multicolumn{1}{|c|}{\quad}  \\
\cline{1-1}
\multicolumn{1}{|c|}{\quad}  \\
\cline{1-1}
\end{tabular}_{~CS},~
\begin{tabular}{|c|c|c|}
\hline
\quad \quad & \quad \quad & \quad \quad \\
\hline
\multicolumn{1}{|c|}{\quad} & \multicolumn{1}{|c|}{\quad} \\
\cline{1-2}
\end{tabular}_F
\otimes
\begin{tabular}{|c|c|}
\hline
\quad \quad & \quad \quad  \\
\hline
\quad \quad & \quad \quad  \\
\hline
\multicolumn{1}{|c|}{\quad}  \\
\cline{1-1}
\multicolumn{1}{|c|}{\quad}  \\
\cline{1-1}
\end{tabular}_{~CS},~
\begin{tabular}{|c|c|c|c|}
\hline
\quad \quad & \quad \quad & \quad \quad & \quad \quad \\
\hline
\multicolumn{1}{|c|}{\quad} \\
\cline{1-1}
\end{tabular}_F
\otimes
\begin{tabular}{|c|c|}
\hline
\quad \quad & \quad \quad  \\
\hline
\quad \quad & \quad \quad  \\
\hline
\multicolumn{1}{|c|}{\quad}  \\
\cline{1-1}
\multicolumn{1}{|c|}{\quad}  \\
\cline{1-1}
\end{tabular}_{~CS}. \nonumber
\end{eqnarray}
\item S=1 : 8 states
\begin{eqnarray}
&&\begin{tabular}{|c|c|}
\hline
\quad \quad & \quad \quad  \\
\cline{1-2}
\quad \quad & \quad \quad  \\
\cline{1-2}
\multicolumn{1}{|c|}{\quad}  \\
\cline{1-1}
\end{tabular}_F
\otimes
\begin{tabular}{|c|c|c|}
\hline
\quad \quad & \quad \quad & \quad \quad   \\
\hline
\multicolumn{1}{|c|}{\quad} & \multicolumn{1}{|c|}{\quad}   \\
\cline{1-2}
\multicolumn{1}{|c|}{\quad} \\
\cline{1-1}
\end{tabular}_{~CS},~
\begin{tabular}{|c|c|c|}
\hline
\quad \quad & \quad \quad & \quad \quad \\
\hline
\multicolumn{1}{|c|}{\quad}  \\
\cline{1-1}
\multicolumn{1}{|c|}{\quad}  \\
\cline{1-1}
\end{tabular}_F
\otimes
\begin{tabular}{|c|c|c|}
\hline
\quad \quad & \quad \quad & \quad \quad \\
\hline
\multicolumn{1}{|c|}{\quad}  \\
\cline{1-1}
\multicolumn{1}{|c|}{\quad}  \\
\cline{1-1}
\multicolumn{1}{|c|}{\quad}  \\
\cline{1-1}
\end{tabular}_{~CS},~
\begin{tabular}{|c|c|c|}
\hline
\quad \quad & \quad \quad & \quad \quad \\
\hline
\multicolumn{1}{|c|}{\quad} & \multicolumn{1}{|c|}{\quad} \\
\cline{1-2}
\end{tabular}_F
\otimes
\begin{tabular}{|c|c|}
\hline
\quad \quad & \quad \quad  \\
\hline
\quad \quad & \quad \quad  \\
\hline
\quad \quad & \quad \quad  \\
\hline
\end{tabular}_{~CS},~
\begin{tabular}{|c|c|c|c|}
\hline
\quad \quad & \quad \quad & \quad \quad & \quad \quad \\
\hline
\multicolumn{1}{|c|}{\quad} \\
\cline{1-1}
\end{tabular}_F
\otimes
\begin{tabular}{|c|c|}
\hline
\quad \quad & \quad \quad  \\
\hline
\multicolumn{1}{|c|}{\quad}  \\
\cline{1-1}
\multicolumn{1}{|c|}{\quad}  \\
\cline{1-1}
\multicolumn{1}{|c|}{\quad}  \\
\cline{1-1}
\multicolumn{1}{|c|}{\quad}  \\
\cline{1-1}
\end{tabular}_{~CS}, \nonumber\\
&&\begin{tabular}{|c|c|}
\hline
\quad \quad & \quad \quad  \\
\cline{1-2}
\quad \quad & \quad \quad  \\
\cline{1-2}
\multicolumn{1}{|c|}{\quad}  \\
\cline{1-1}
\end{tabular}_F
\otimes
\begin{tabular}{|c|c|c|c|}
\hline
\quad \quad & \quad \quad & \quad \quad & \quad \quad \\
\hline
\multicolumn{1}{|c|}{\quad} & \multicolumn{1}{|c|}{\quad}   \\
\cline{1-2}
\end{tabular}_{~CS},~
\begin{tabular}{|c|c|c|}
\hline
\quad \quad & \quad \quad & \quad \quad \\
\hline
\multicolumn{1}{|c|}{\quad}  \\
\cline{1-1}
\multicolumn{1}{|c|}{\quad}  \\
\cline{1-1}
\end{tabular}_F
\otimes
\begin{tabular}{|c|c|c|}
\hline
\quad \quad & \quad \quad & \quad \quad \\
\hline
\multicolumn{1}{|c|}{\quad} & \multicolumn{1}{|c|}{\quad} \\
\cline{1-2}
\multicolumn{1}{|c|}{\quad}  \\
\cline{1-1}
\end{tabular}_{~CS},~
\begin{tabular}{|c|c|c|}
\hline
\quad \quad & \quad \quad & \quad \quad \\
\hline
\multicolumn{1}{|c|}{\quad} & \multicolumn{1}{|c|}{\quad} \\
\cline{1-2}
\end{tabular}_F
\otimes
\begin{tabular}{|c|c|c|}
\hline
\quad \quad & \quad \quad & \quad \quad \\
\hline
\multicolumn{1}{|c|}{\quad} & \multicolumn{1}{|c|}{\quad} \\
\cline{1-2}
\multicolumn{1}{|c|}{\quad}  \\
\cline{1-1}
\end{tabular}_{~CS},~
\begin{tabular}{|c|c|c|c|}
\hline
\quad \quad & \quad \quad & \quad \quad & \quad \quad \\
\hline
\multicolumn{1}{|c|}{\quad} \\
\cline{1-1}
\end{tabular}_F
\otimes
\begin{tabular}{|c|c|c|}
\hline
\quad \quad & \quad \quad & \quad \quad \\
\hline
\multicolumn{1}{|c|}{\quad}  \\
\cline{1-1}
\multicolumn{1}{|c|}{\quad}  \\
\cline{1-1}
\multicolumn{1}{|c|}{\quad}  \\
\cline{1-1}
\end{tabular}_{~CS}. \nonumber
\end{eqnarray}
\item S=2 : 5 states
\begin{eqnarray}
&&\begin{tabular}{|c|c|}
\hline
\quad \quad & \quad \quad  \\
\cline{1-2}
\quad \quad & \quad \quad  \\
\cline{1-2}
\multicolumn{1}{|c|}{\quad}  \\
\cline{1-1}
\end{tabular}_F
\otimes
\begin{tabular}{|c|c|c|}
\hline
\quad \quad & \quad \quad & \quad \quad   \\
\hline
\multicolumn{1}{|c|}{\quad} & \multicolumn{1}{|c|}{\quad}   \\
\cline{1-2}
\multicolumn{1}{|c|}{\quad} \\
\cline{1-1}
\end{tabular}_{~CS},~
\begin{tabular}{|c|c|c|}
\hline
\quad \quad & \quad \quad & \quad \quad \\
\hline
\multicolumn{1}{|c|}{\quad}  \\
\cline{1-1}
\multicolumn{1}{|c|}{\quad}  \\
\cline{1-1}
\end{tabular}_F
\otimes
\begin{tabular}{|c|c|c|}
\hline
\quad \quad & \quad \quad & \quad \quad   \\
\hline
\multicolumn{1}{|c|}{\quad} & \multicolumn{1}{|c|}{\quad}   \\
\cline{1-2}
\multicolumn{1}{|c|}{\quad} \\
\cline{1-1}
\end{tabular}_{~CS},~
\begin{tabular}{|c|c|c|}
\hline
\quad \quad & \quad \quad & \quad \quad \\
\hline
\multicolumn{1}{|c|}{\quad} & \multicolumn{1}{|c|}{\quad} \\
\cline{1-2}
\end{tabular}_F
\otimes
\begin{tabular}{|c|c|}
\hline
\quad \quad & \quad \quad  \\
\hline
\quad \quad & \quad \quad  \\
\hline
\multicolumn{1}{|c|}{\quad} \\
\cline{1-1}
\multicolumn{1}{|c|}{\quad} \\
\cline{1-1}
\end{tabular}_{~CS},~
\begin{tabular}{|c|c|c|c|}
\hline
\quad \quad & \quad \quad & \quad \quad & \quad \quad \\
\hline
\multicolumn{1}{|c|}{\quad} \\
\cline{1-1}
\end{tabular}_F
\otimes
\begin{tabular}{|c|c|}
\hline
\quad \quad & \quad \quad  \\
\hline
\quad \quad & \quad \quad  \\
\hline
\multicolumn{1}{|c|}{\quad} \\
\cline{1-1}
\multicolumn{1}{|c|}{\quad} \\
\cline{1-1}
\end{tabular}_{~CS}, \nonumber\\
&&\begin{tabular}{|c|c|c|}
\hline
\quad \quad & \quad \quad & \quad \quad \\
\hline
\multicolumn{1}{|c|}{\quad} & \multicolumn{1}{|c|}{\quad} \\
\cline{1-2}
\end{tabular}_F
\otimes
\begin{tabular}{|c|c|c|}
\hline
\quad \quad & \quad \quad & \quad \quad   \\
\hline
\multicolumn{1}{|c|}{\quad} & \multicolumn{1}{|c|}{\quad}   \\
\cline{1-2}
\multicolumn{1}{|c|}{\quad} \\
\cline{1-1}
\end{tabular}_{~CS}. \nonumber
\end{eqnarray}
\item S=3 : 1 state
\begin{eqnarray}
\begin{tabular}{|c|c|c|}
\hline
\quad \quad & \quad \quad & \quad \quad \\
\hline
\multicolumn{1}{|c|}{\quad} & \multicolumn{1}{|c|}{\quad} \\
\cline{1-2}
\end{tabular}_F
\otimes
\begin{tabular}{|c|c|}
\hline
\quad \quad & \quad \quad  \\
\hline
\quad \quad & \quad \quad  \\
\hline
\quad \quad & \quad \quad  \\
\hline
\end{tabular}_{~CS}. \nonumber
\end{eqnarray}
\end{itemize}
\end{widetext}
After $SU(3)_F$ breaking, there are only two flavor bases.
\begin{eqnarray}
|F_1 \rangle=(
\begin{tabular}{|c|c|}
\hline
1 & 2 \\
\hline
3 \\
\cline{1-1}
\end{tabular}
,~
\begin{tabular}{|c|c|}
\hline
4 & 5 \\
\hline
\end{tabular}
,~
\begin{tabular}{|c|}
\hline
6 \\
\hline
\end{tabular}
),~
|F_2 \rangle=(
\begin{tabular}{|c|c|}
\hline
1 & 3 \\
\hline
2 \\
\cline{1-1}
\end{tabular}
,~
\begin{tabular}{|c|c|}
\hline
4 & 5 \\
\hline
\end{tabular}
,~
\begin{tabular}{|c|}
\hline
6 \\
\hline
\end{tabular}
).
\end{eqnarray}
\subsection{Flavor, color and spin state of $q^5Q$}
Here, we fix $u$ and $d$ quarks to be 1,2, and 3, and two $s$ quarks to be 4 and 5. Since there is a flavor symmetry between strange quarks, the color and spin wave function should be antisymmetric between 4 and 5. The details to construct color and spin wave function which has specific symmetry property was explained in Ref.~\cite{Park:2016cmg}.

For example, in the case of $[\bar{3}]_F$ and S=0 state, the Young tableau of color and spin state should be [3,3].   Furthermore, the color and spin state have to be antisymmetric between 4 and 5 because of their flavor symmetry. Hence, by using Young-Yamanouchi representation and permutation property, we can construct the flavor, color and spin wave function which has the required symmetry.

\begin{itemize}
\item S=0
\begin{eqnarray}
\phi^{S=0}_1&=&\frac{1}{\sqrt{2}}|F_1 \rangle \otimes (
\frac{1}{2}
\begin{tabular}{|c|c|c|}
\hline
1 & 3 & 5 \\
\hline
2 & 4 & 6 \\
\hline
\end{tabular}_{~CS}
-\frac{\sqrt{3}}{2}
\begin{tabular}{|c|c|c|}
\hline
1 & 3 & 4 \\
\hline
2 & 5 & 6 \\
\hline
\end{tabular}_{~CS}
) \nonumber\\
&&
-\frac{1}{\sqrt{2}}|F_2 \rangle \otimes (\frac{1}{2}
\begin{tabular}{|c|c|c|}
\hline
1 & 2 & 5 \\
\hline
3 & 4 & 6 \\
\hline
\end{tabular}_{~CS}
-\frac{\sqrt{3}}{2}
\begin{tabular}{|c|c|c|}
\hline
1 & 2 & 4 \\
\hline
3 & 5 & 6 \\
\hline
\end{tabular}_{~CS}
) \nonumber
\end{eqnarray}
\begin{eqnarray}
\phi^{S=0}_2&=&\frac{1}{\sqrt{2}}|F_1 \rangle \otimes (
\frac{\sqrt{3}}{\sqrt{8}}
\begin{tabular}{|c|c|c|c|}
\hline
1 & 3 & 5 & 6 \\
\hline
2 \\
\cline{1-1}
4 \\
\cline{1-1}
\end{tabular}_{CS}
-\frac{\sqrt{5}}{\sqrt{8}}
\begin{tabular}{|c|c|c|c|}
\hline
1 & 3 & 4 & 6 \\
\hline
2 \\
\cline{1-1}
5 \\
\cline{1-1}
\end{tabular}_{CS}
) \nonumber\\
&&
-\frac{1}{\sqrt{2}}|F_2 \rangle \otimes (\frac{\sqrt{3}}{\sqrt{8}}
\begin{tabular}{|c|c|c|c|}
\hline
1 & 2 & 5 & 6 \\
\hline
3 \\
\cline{1-1}
4 \\
\cline{1-1}
\end{tabular}_{CS}
-\frac{\sqrt{5}}{\sqrt{8}}
\begin{tabular}{|c|c|c|c|}
\hline
1 & 2 & 4 & 6 \\
\hline
3 \\
\cline{1-1}
5 \\
\cline{1-1}
\end{tabular}_{CS}
) \nonumber
\end{eqnarray}
\begin{eqnarray}
\phi^{S=0}_3&=&\frac{1}{\sqrt{2}}|F_1 \rangle \otimes (
\frac{1}{2}
\begin{tabular}{|c|c|}
\hline
1 & 3  \\
\hline
2 & 5 \\
\hline
4 \\
\cline{1-1}
6 \\
\cline{1-1}
\end{tabular}_{CS}
-\frac{\sqrt{3}}{2}
\begin{tabular}{|c|c|}
\hline
1 & 3  \\
\hline
2 & 4 \\
\hline
5 \\
\cline{1-1}
6 \\
\cline{1-1}
\end{tabular}_{CS}
) \nonumber\\
&&
-\frac{1}{\sqrt{2}}|F_2 \rangle \otimes (
\frac{1}{2}
\begin{tabular}{|c|c|}
\hline
1 & 3  \\
\hline
2 & 5 \\
\hline
4 \\
\cline{1-1}
6 \\
\cline{1-1}
\end{tabular}_{CS}
-\frac{\sqrt{3}}{2}
\begin{tabular}{|c|c|}
\hline
1 & 3  \\
\hline
2 & 4 \\
\hline
5 \\
\cline{1-1}
6 \\
\cline{1-1}
\end{tabular}_{CS}
) \nonumber
\end{eqnarray}
\begin{eqnarray}
\phi^{S=0}_4=\frac{1}{\sqrt{2}}|F_1 \rangle \otimes
\begin{tabular}{|c|c|}
\hline
1 & 3 \\
\hline
2 & 6 \\
\hline
4 \\
\cline{1-1}
5 \\
\cline{1-1}
\end{tabular}_{CS}
-\frac{1}{\sqrt{2}}|F_2 \rangle \otimes
\begin{tabular}{|c|c|}
\hline
1 & 2 \\
\hline
3 & 6 \\
\hline
4 \\
\cline{1-1}
5 \\
\cline{1-1}
\end{tabular}_{CS}
\end{eqnarray}
\item S=1
\begin{eqnarray}
\phi^{S=1}_1&=&\frac{1}{\sqrt{2}}|F_1 \rangle \otimes (
\frac{1}{2}
\begin{tabular}{|c|c|c|}
\hline
1 & 3 & 5 \\
\hline
2 & 4 \\
\cline{1-2}
6 \\
\cline{1-1}
\end{tabular}_{~CS}
-\frac{\sqrt{3}}{2}
\begin{tabular}{|c|c|c|}
\hline
1 & 3 & 4 \\
\hline
2 & 5 \\
\cline{1-2}
6 \\
\cline{1-1}
\end{tabular}_{~CS}
) \nonumber\\
&&
-\frac{1}{\sqrt{2}}|F_2 \rangle \otimes (
\frac{1}{2}
\begin{tabular}{|c|c|c|}
\hline
1 & 2 & 5 \\
\hline
3 & 4 \\
\cline{1-2}
6 \\
\cline{1-1}
\end{tabular}_{~CS}
-\frac{\sqrt{3}}{2}
\begin{tabular}{|c|c|c|}
\hline
1 & 2 & 4 \\
\hline
3 & 5 \\
\cline{1-2}
6 \\
\cline{1-1}
\end{tabular}_{~CS}
) \nonumber
\end{eqnarray}
\begin{eqnarray}
\phi^{S=1}_2&=&\frac{1}{\sqrt{2}}|F_1 \rangle \otimes (
\frac{\sqrt{3}}{\sqrt{8}}
\begin{tabular}{|c|c|c|}
\hline
1 & 3 & 5 \\
\hline
2 \\
\cline{1-1}
4 \\
\cline{1-1}
6 \\
\cline{1-1}
\end{tabular}_{CS}
-\frac{\sqrt{5}}{\sqrt{8}}
\begin{tabular}{|c|c|c|}
\hline
1 & 3 & 4 \\
\hline
2 \\
\cline{1-1}
5 \\
\cline{1-1}
6 \\
\cline{1-1}
\end{tabular}_{CS}
) \nonumber\\
&&
-\frac{1}{\sqrt{2}}F_2 \otimes (
\frac{\sqrt{3}}{\sqrt{8}}
\begin{tabular}{|c|c|c|}
\hline
1 & 2 & 5 \\
\hline
3 \\
\cline{1-1}
4 \\
\cline{1-1}
6 \\
\cline{1-1}
\end{tabular}_{CS}
-\frac{\sqrt{5}}{\sqrt{8}}
\begin{tabular}{|c|c|c|}
\hline
1 & 2 & 4 \\
\hline
3 \\
\cline{1-1}
5 \\
\cline{1-1}
6 \\
\cline{1-1}
\end{tabular}_{CS}
) \nonumber
\end{eqnarray}
\begin{eqnarray}
\phi^{S=1}_3&=&\frac{1}{\sqrt{2}}|F_1 \rangle \otimes (
\frac{1}{2}
\begin{tabular}{|c|c|}
\hline
1 & 3  \\
\hline
2 & 5 \\
\hline
4 & 6 \\
\hline
\end{tabular}_{~CS}
-\frac{\sqrt{3}}{2}
\begin{tabular}{|c|c|}
\hline
1 & 3  \\
\hline
2 & 4 \\
\hline
5 & 6 \\
\hline
\end{tabular}_{~CS}
) \nonumber\\
&&
-\frac{1}{\sqrt{2}}|F_2 \rangle \otimes (
\frac{1}{2}
\begin{tabular}{|c|c|}
\hline
1 & 2  \\
\hline
3 & 5 \\
\hline
4 & 6 \\
\hline
\end{tabular}_{~CS}
-\frac{\sqrt{3}}{2}
\begin{tabular}{|c|c|}
\hline
1 & 2  \\
\hline
3 & 4 \\
\hline
5 & 6 \\
\hline
\end{tabular}_{~CS}
) \nonumber
\end{eqnarray}
\begin{eqnarray}
\phi^{S=1}_4=\frac{1}{\sqrt{2}}|F_1 \rangle \otimes
\begin{tabular}{|c|c|}
\hline
1 & 3 \\
\hline
2  \\
\cline{1-1}
4 \\
\cline{1-1}
5 \\
\cline{1-1}
6 \\
\cline{1-1}
\end{tabular}_{CS}
-\frac{1}{\sqrt{2}}|F_2 \rangle \otimes
\begin{tabular}{|c|c|}
\hline
1 & 2 \\
\hline
3  \\
\cline{1-1}
4 \\
\cline{1-1}
5 \\
\cline{1-1}
6 \\
\cline{1-1}
\end{tabular}_{CS} \nonumber
\end{eqnarray}
\begin{eqnarray}
\phi^{S=1}_5&=&\frac{1}{\sqrt{2}}|F_1 \rangle \otimes (
\frac{1}{2}
\begin{tabular}{|c|c|c|c|}
\hline
1 & 3 & 5 & 6 \\
\hline
2 & 4 \\
\cline{1-2}
\end{tabular}_{CS}
-\frac{\sqrt{3}}{2}
\begin{tabular}{|c|c|c|c|}
\hline
1 & 3 & 4 & 6 \\
\hline
2 & 5 \\
\cline{1-2}
\end{tabular}_{CS}
) \nonumber\\
&&
-\frac{1}{\sqrt{2}}|F_2 \rangle \otimes (
\frac{1}{2}
\begin{tabular}{|c|c|c|c|}
\hline
1 & 2 & 5 & 6 \\
\hline
3 & 4 \\
\cline{1-2}
\end{tabular}_{CS}
-\frac{\sqrt{3}}{2}
\begin{tabular}{|c|c|c|c|}
\hline
1 & 2 & 4 & 6 \\
\hline
3 & 5 \\
\cline{1-2}
\end{tabular}_{CS}
) \nonumber
\end{eqnarray}
\begin{eqnarray}
\phi^{S=1}_6&=&\frac{1}{\sqrt{2}}|F_1 \rangle \otimes (
\frac{\sqrt{3}}{\sqrt{8}}
\begin{tabular}{|c|c|c|}
\hline
1 & 3 & 5 \\
\hline
2 & 6 \\
\cline{1-2}
4 \\
\cline{1-1}
\end{tabular}_{CS}
-\frac{\sqrt{5}}{\sqrt{8}}
\begin{tabular}{|c|c|c|}
\hline
1 & 3 & 4 \\
\hline
2 & 6 \\
\cline{1-2}
5 \\
\cline{1-1}
\end{tabular}_{CS}
) \nonumber\\
&&
-\frac{1}{\sqrt{2}}|F_2 \rangle \otimes (
\frac{\sqrt{3}}{\sqrt{8}}
\begin{tabular}{|c|c|c|}
\hline
1 & 2 & 5 \\
\hline
3 & 6 \\
\cline{1-2}
4 \\
\cline{1-1}
\end{tabular}_{CS}
-\frac{\sqrt{5}}{\sqrt{8}}
\begin{tabular}{|c|c|c|}
\hline
1 & 2 & 4 \\
\hline
3 & 6 \\
\cline{1-2}
5 \\
\cline{1-1}
\end{tabular}_{CS}
) \nonumber
\end{eqnarray}
\begin{eqnarray}
\phi^{S=1}_7&=&\frac{1}{\sqrt{2}}|F_1 \rangle \otimes (
\frac{1}{2}
\begin{tabular}{|c|c|c|}
\hline
1 & 3 & 6 \\
\hline
2 & 5 \\
\cline{1-2}
4 \\
\cline{1-1}
\end{tabular}_{CS}
-\frac{\sqrt{3}}{2}
\begin{tabular}{|c|c|c|}
\hline
1 & 3 & 6 \\
\hline
2 & 4 \\
\cline{1-2}
5 \\
\cline{1-1}
\end{tabular}_{CS}
) \nonumber\\
&&
-\frac{1}{\sqrt{2}}|F_2 \rangle \otimes (
\frac{1}{2}
\begin{tabular}{|c|c|c|}
\hline
1 & 2 & 6 \\
\hline
3 & 5 \\
\cline{1-2}
4 \\
\cline{1-1}
\end{tabular}_{CS}
-\frac{\sqrt{3}}{2}
\begin{tabular}{|c|c|c|}
\hline
1 & 2 & 6 \\
\hline
3 & 4 \\
\cline{1-2}
5 \\
\cline{1-1}
\end{tabular}_{CS}
) \nonumber
\end{eqnarray}
\begin{eqnarray}
\phi^{S=1}_8=\frac{1}{\sqrt{2}}|F_1 \rangle \otimes
\begin{tabular}{|c|c|c|}
\hline
1 & 3 & 6 \\
\hline
2  \\
\cline{1-1}
4 \\
\cline{1-1}
5 \\
\cline{1-1}
\end{tabular}_{CS}
-\frac{1}{\sqrt{2}}|F_2 \rangle \otimes
\begin{tabular}{|c|c|c|}
\hline
1 & 2 & 6 \\
\hline
3  \\
\cline{1-1}
4 \\
\cline{1-1}
5 \\
\cline{1-1}
\end{tabular}_{CS}
\end{eqnarray}
\item S=2
\begin{eqnarray}
\phi^{S=2}_1&=&\frac{1}{\sqrt{2}}|F_1 \rangle \otimes (
\frac{1}{2}
\begin{tabular}{|c|c|c|}
\hline
1 & 3 & 5 \\
\hline
2 & 4 \\
\cline{1-2}
6 \\
\cline{1-1}
\end{tabular}_{~CS}
-\frac{\sqrt{3}}{2}
\begin{tabular}{|c|c|c|}
\hline
1 & 3 & 4 \\
\hline
2 & 5 \\
\cline{1-2}
6 \\
\cline{1-1}
\end{tabular}_{~CS}
) \nonumber\\
&&
-\frac{1}{\sqrt{2}}|F_2 \rangle \otimes (
\frac{1}{2}
\begin{tabular}{|c|c|c|}
\hline
1 & 2 & 5 \\
\hline
3 & 4 \\
\cline{1-2}
6 \\
\cline{1-1}
\end{tabular}_{~CS}
-\frac{\sqrt{3}}{2}
\begin{tabular}{|c|c|c|}
\hline
1 & 2 & 4 \\
\hline
3 & 5 \\
\cline{1-2}
6 \\
\cline{1-1}
\end{tabular}_{~CS}
) \nonumber
\end{eqnarray}
\begin{eqnarray}
\phi^{S=2}_2&=&\frac{1}{\sqrt{2}}|F_1 \rangle \otimes (
\frac{\sqrt{3}}{\sqrt{8}}
\begin{tabular}{|c|c|c|}
\hline
1 & 3 & 5 \\
\hline
2 & 6 \\
\cline{1-2}
4 \\
\cline{1-1}
\end{tabular}_{CS}
-\frac{\sqrt{5}}{\sqrt{8}}
\begin{tabular}{|c|c|c|}
\hline
1 & 3 & 4 \\
\hline
2 & 6 \\
\cline{1-2}
5 \\
\cline{1-1}
\end{tabular}_{CS}
) \nonumber\\
&&
-\frac{1}{\sqrt{2}}|F_2 \rangle \otimes (
\frac{\sqrt{3}}{\sqrt{8}}
\begin{tabular}{|c|c|c|}
\hline
1 & 2 & 5 \\
\hline
3 & 6 \\
\cline{1-2}
4 \\
\cline{1-1}
\end{tabular}_{CS}
-\frac{\sqrt{5}}{\sqrt{8}}
\begin{tabular}{|c|c|c|}
\hline
1 & 2 & 4 \\
\hline
3 & 6 \\
\cline{1-2}
5 \\
\cline{1-1}
\end{tabular}_{CS}
) \nonumber
\end{eqnarray}
\begin{eqnarray}
\phi^{S=2}_3&=&\frac{1}{\sqrt{2}}|F_1 \rangle \otimes (
\frac{1}{2}
\begin{tabular}{|c|c|}
\hline
1 & 3  \\
\hline
2 & 5 \\
\hline
4 \\
\cline{1-1}
6 \\
\cline{1-1}
\end{tabular}_{CS}
-\frac{\sqrt{3}}{2}
\begin{tabular}{|c|c|}
\hline
1 & 3  \\
\hline
2 & 4 \\
\hline
5 \\
\cline{1-1}
6 \\
\cline{1-1}
\end{tabular}_{CS}
) \nonumber\\
&&
-\frac{1}{\sqrt{2}}|F_2 \rangle \otimes (
\frac{1}{2}
\begin{tabular}{|c|c|}
\hline
1 & 3  \\
\hline
2 & 5 \\
\hline
4 \\
\cline{1-1}
6 \\
\cline{1-1}
\end{tabular}_{CS}
-\frac{\sqrt{3}}{2}
\begin{tabular}{|c|c|}
\hline
1 & 3  \\
\hline
2 & 4 \\
\hline
5 \\
\cline{1-1}
6 \\
\cline{1-1}
\end{tabular}_{CS}
) \nonumber
\end{eqnarray}
\begin{eqnarray}
\phi^{S=2}_4=\frac{1}{\sqrt{2}}|F_1 \rangle \otimes
\begin{tabular}{|c|c|}
\hline
1 & 3 \\
\hline
2 & 6 \\
\hline
4 \\
\cline{1-1}
5 \\
\cline{1-1}
\end{tabular}_{CS}
-\frac{1}{\sqrt{2}}|F_2 \rangle \otimes
\begin{tabular}{|c|c|}
\hline
1 & 2 \\
\hline
3 & 6 \\
\hline
4 \\
\cline{1-1}
5 \\
\cline{1-1}
\end{tabular}_{CS}
\end{eqnarray}
\begin{eqnarray}
\phi^{S=2}_5&=&\frac{1}{\sqrt{2}}|F_1 \rangle \otimes (
\frac{\sqrt{3}}{\sqrt{8}}
\begin{tabular}{|c|c|c|}
\hline
1 & 3 & 6 \\
\hline
2 & 5 \\
\cline{1-2}
4 \\
\cline{1-1}
\end{tabular}_{CS}
-\frac{\sqrt{5}}{\sqrt{8}}
\begin{tabular}{|c|c|c|}
\hline
1 & 3 & 6 \\
\hline
2 & 4 \\
\cline{1-2}
5 \\
\cline{1-1}
\end{tabular}_{CS}
) \nonumber\\
&&
-\frac{1}{\sqrt{2}}|F_2 \rangle \otimes (
\frac{\sqrt{3}}{\sqrt{8}}
\begin{tabular}{|c|c|c|}
\hline
1 & 2 & 6 \\
\hline
3 & 5 \\
\cline{1-2}
4 \\
\cline{1-1}
\end{tabular}_{CS}
-\frac{\sqrt{5}}{\sqrt{8}}
\begin{tabular}{|c|c|c|}
\hline
1 & 2 & 6 \\
\hline
3 & 4 \\
\cline{1-2}
5 \\
\cline{1-1}
\end{tabular}_{CS}
) \nonumber
\end{eqnarray}
\item S=3
\begin{eqnarray}
\phi^{S=3}_1&=&\frac{1}{\sqrt{2}}|F_1 \rangle \otimes (
\frac{1}{2}
\begin{tabular}{|c|c|}
\hline
1 & 3  \\
\hline
2 & 5 \\
\hline
4 & 6 \\
\hline
\end{tabular}_{~CS}
-\frac{\sqrt{3}}{2}
\begin{tabular}{|c|c|}
\hline
1 & 3  \\
\hline
2 & 4 \\
\hline
5 & 6 \\
\hline
\end{tabular}_{~CS}
) \nonumber\\
&&
-\frac{1}{\sqrt{2}}|F_2 \rangle \otimes (
\frac{1}{2}
\begin{tabular}{|c|c|}
\hline
1 & 2  \\
\hline
3 & 5 \\
\hline
4 & 6 \\
\hline
\end{tabular}_{~CS}
-\frac{\sqrt{3}}{2}
\begin{tabular}{|c|c|}
\hline
1 & 2  \\
\hline
3 & 4 \\
\hline
5 & 6 \\
\hline
\end{tabular}_{~CS}
). \nonumber
\end{eqnarray}
\end{itemize}
Using the Clebsch-Gordon coefficients which are presented in appendix, we can obtain the flavor, color and spin state of the dibaryon.
\label{subsec:section4b}

\section{Classification of $q^5$ with SU(3) flavor symmetry}
\label{section5}
\subsection{Flavor and spin state of $q^5$}
We can construct the wave function of the dibaryon in another way. Here, we consider the state of five light quarks first and then, we will add a heavy quark later. The totally antisymmetric multiplet of $[1^5]_{CFS}$ can be
decomposed as
\begin{align}
[1^5]_{CFS}=&([\bar{3}]_{C}, [420]_{FS}) \oplus ([6]_{C}, [336]_{FS}) \oplus \nonumber\\
&([\bar{15}]_{C}, [210]_{FS}) \oplus ([24]_{C}, [84]_{FS}) \oplus \nonumber\\
&([21]_{C}, [\bar{6}]_{FS}).
\label{eq-1-6}
\end{align}
By using the Young tableau, we can find that the multiplets in the right hand side of Eq.~(\ref{eq-1-6}) is fully antisymmetric.

\begin{center}
\begin{tabular}{c}
$([\bar{3}]_{C}, [420]_{FS})$=
\end{tabular}
\begin{tabular}{|c|c|}
\hline
$\quad$ & $\quad$   \\
\cline{1-2}
$\quad$ &  $\quad$  \\
\cline{1-2}
\multicolumn{1}{|c|}{$\quad$}  \\
\cline{1-1}
\end{tabular}
$\otimes$
\begin{tabular}{|c|c|c|}
\hline
$\quad$ & $\quad$ & $\quad$    \\
\cline{1-3} \multicolumn{1}{|c|}{$\quad$}    & \multicolumn{1}{|c|}{$\quad$}    \\
\cline{1-2}
\end{tabular}
,
\end{center}

\begin{center}
\begin{tabular}{c}
$([6]_{C}, [336]_{FS})$=
\end{tabular}
\begin{tabular}{|c|c|c|}
\hline
$\quad$ & $\quad$ & $\quad$   \\
\hline
\multicolumn{1}{|c|}{$\quad$}  \\
\cline{1-1}
\multicolumn{1}{|c|}{$\quad$}   \\
\cline{1-1}
\end{tabular}
$\otimes$
\begin{tabular}{|c|c|c|}
\hline
$\quad$ & $\quad$ & $\quad$   \\
\hline
\multicolumn{1}{|c|}{$\quad$}  \\
\cline{1-1}
\multicolumn{1}{|c|}{$\quad$}   \\
\cline{1-1}
\end{tabular}
,
\end{center}

\begin{center}
\begin{tabular}{c}
$([\bar{15}]_{C}, [210]_{FS})$=
\end{tabular}
\begin{tabular}{|c|c|c|}
\hline
$\quad$ & $\quad$ & $\quad$    \\
\cline{1-3}
\multicolumn{1}{|c|}{$\quad$}    & \multicolumn{1}{|c|}{$\quad$}    \\
\cline{1-2}
\end{tabular}
$\otimes$
\begin{tabular}{|c|c|}
\hline
$\quad$ & $\quad$   \\
\cline{1-2}
$\quad$ &  $\quad$  \\
\cline{1-2}
\multicolumn{1}{|c|}{$\quad$}   \\
\cline{1-1}
\end{tabular}
,
\end{center}


\begin{center}
\begin{tabular}{c}
$([24]_{C}, [84]_{FS})$=
\end{tabular}
\begin{tabular}{|c|c|c|c|}
\hline
$\quad$ & $\quad$ & $\quad$ & $\quad$   \\
\cline{1-4}
\multicolumn{1}{|c|}{$\quad$}   \\
\cline{1-1}
\end{tabular}
$\otimes$
\begin{tabular}{|c|c|}
\hline
$\quad$ & $\quad$    \\
\cline{1-2}
\multicolumn{1}{|c|}{$\quad$}   \\
\cline{1-1}
\multicolumn{1}{|c|}{$\quad$}   \\
\cline{1-1}
\multicolumn{1}{|c|}{$\quad$}   \\
\cline{1-1}
\end{tabular}
,
\end{center}



\begin{center}
\begin{tabular}{c}
$([21]_{C}, [6]_{FS})$=
\end{tabular}
\begin{tabular}{|c|c|c|c|c|}
\hline
$\quad$ & $\quad$ & $\quad$ & $\quad$ & $\quad$  \\
\hline
\end{tabular}
$\otimes$
\begin{tabular}{|c|}
\hline
$\quad$  \\
\hline
$\quad$   \\
\hline
$\quad$   \\
\hline
$\quad$   \\
\hline
$\quad$   \\
\hline
\end{tabular}
.
\end{center}

\begin{figure}[htbp]
  \begin{center}
         \includegraphics[scale=0.13]{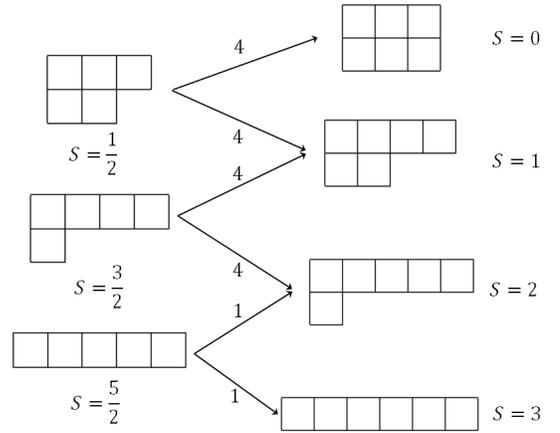}
\caption[]{Spin Young tableau of $q^5$ and $q^5Q$.}
\label{fig1}
  \end{center}
\end{figure}

Since the dibaryon is a color singlet, the color state of $q^5$ should be anti-triplet. So flavor and spin state of $q^5$ should be $[420]_{FS}$.  The $[420]_{FS}$ multiplet  can be decomposed as

\begin{align}
[420]_{FS}=&([\bar{3}]_{F}, [2]_{S}) \oplus ([6]_{F}, [2]_{S}) \oplus ([\bar{15}]_{F}, [2]_{S}) \oplus \nonumber\\
& ([21]_{F}, [2]_{S}) \oplus ([24]_{F}, [2]_{S}) \oplus ([\bar{3}]_{F}, [4]_{S}) \oplus \nonumber\\
&([6]_{F}, [4]_{S})  \oplus ([\bar{15}]_{F}, [4]_{S}) \oplus ([24]_{F}, [4]_{S}) \oplus \nonumber\\
&([\bar{15}]_{F}, [6]_{S}).
\label{eq-50is}
\end{align}
The corresponding Young tableau of each states are given as follows.
\begin{center}
\begin{tabular}{c}
$([\bar{3}]_F,[2]_S)$=
\end{tabular}
\begin{tabular}{|c|c|}
\hline
$\quad$ & $\quad$   \\
\cline{1-2}
$\quad$ &  $\quad$  \\
\cline{1-2}
\multicolumn{1}{|c|}{$\quad$}  \\
\cline{1-1}
\end{tabular}
$\otimes$
\begin{tabular}{|c|c|c|}
\hline
$\quad$ & $\quad$ & $\quad$  \\
\cline{1-3} \multicolumn{1}{|c|}{$\quad$} & \multicolumn{1}{|c|}{$\quad$}    \\
\cline{1-2}
\end{tabular}
,
\end{center}
\begin{center}
\begin{tabular}{c}
$([6]_F,[2]_S)$=
\end{tabular}
\begin{tabular}{|c|c|c|}
\hline
$\quad$ & $\quad$ & $\quad$   \\
\cline{1-3}
\multicolumn{1}{|c|}{$\quad$}  \\
\cline{1-1}
\multicolumn{1}{|c|}{$\quad$}  \\
\cline{1-1}
\end{tabular}
$\otimes$
\begin{tabular}{|c|c|c|}
\hline
$\quad$ & $\quad$ & $\quad$  \\
\cline{1-3} \multicolumn{1}{|c|}{$\quad$} & \multicolumn{1}{|c|}{$\quad$}    \\
\cline{1-2}
\end{tabular}
,
\end{center}
\begin{center}
\begin{tabular}{c}
$([\bar{15}]_F,[2]_S)$=
\end{tabular}
\begin{tabular}{|c|c|c|}
\hline
$\quad$ & $\quad$ & $\quad$   \\
\cline{1-3}
\multicolumn{1}{|c|}{$\quad$} & \multicolumn{1}{|c|}{$\quad$} \\
\cline{1-2}
\end{tabular}
$\otimes$
\begin{tabular}{|c|c|c|}
\hline
$\quad$ & $\quad$ & $\quad$  \\
\cline{1-3} \multicolumn{1}{|c|}{$\quad$} & \multicolumn{1}{|c|}{$\quad$}    \\
\cline{1-2}
\end{tabular}
,
\end{center}
\begin{center}
\begin{tabular}{c}
$([24]_F,[2]_S)$=
\end{tabular}
\begin{tabular}{|c|c|c|c|}
\hline
$\quad$ & $\quad$ & $\quad$ & $\quad$  \\
\cline{1-4}
\multicolumn{1}{|c|}{$\quad$} \\
\cline{1-1}
\end{tabular}
$\otimes$
\begin{tabular}{|c|c|c|}
\hline
$\quad$ & $\quad$ & $\quad$  \\
\cline{1-3} \multicolumn{1}{|c|}{$\quad$} & \multicolumn{1}{|c|}{$\quad$}    \\
\cline{1-2}
\end{tabular}
,
\end{center}
\begin{center}
\begin{tabular}{c}
$([\bar{3}]_F,[4]_S)$=
\end{tabular}
\begin{tabular}{|c|c|}
\hline
$\quad$ & $\quad$   \\
\cline{1-2}
$\quad$ &  $\quad$  \\
\cline{1-2}
\multicolumn{1}{|c|}{$\quad$}  \\
\cline{1-1}
\end{tabular}
$\otimes$
\begin{tabular}{|c|c|c|c|}
\hline
$\quad$ & $\quad$ & $\quad$ & $\quad$   \\
\cline{1-4} \multicolumn{1}{|c|}{$\quad$}     \\
\cline{1-1}
\end{tabular}
,
\end{center}
\begin{center}
\begin{tabular}{c}
$([6]_F,[4]_S)$=
\end{tabular}
\begin{tabular}{|c|c|c|}
\hline
$\quad$ & $\quad$ & $\quad$   \\
\cline{1-3}
\multicolumn{1}{|c|}{$\quad$}  \\
\cline{1-1}
\multicolumn{1}{|c|}{$\quad$}  \\
\cline{1-1}
\end{tabular}
$\otimes$
\begin{tabular}{|c|c|c|c|}
\hline
$\quad$ & $\quad$ & $\quad$ & $\quad$   \\
\cline{1-4} \multicolumn{1}{|c|}{$\quad$}     \\
\cline{1-1}
\end{tabular}
,
\end{center}
\begin{center}
\begin{tabular}{c}
$([\bar{15}]_F,[4]_S)$=
\end{tabular}
\begin{tabular}{|c|c|c|}
\hline
$\quad$ & $\quad$ & $\quad$   \\
\cline{1-3}
\multicolumn{1}{|c|}{$\quad$} & \multicolumn{1}{|c|}{$\quad$} \\
\cline{1-2}
\end{tabular}
$\otimes$
\begin{tabular}{|c|c|c|c|}
\hline
$\quad$ & $\quad$ & $\quad$ & $\quad$   \\
\cline{1-4} \multicolumn{1}{|c|}{$\quad$}     \\
\cline{1-1}
\end{tabular}
,
\end{center}
\begin{center}
\begin{tabular}{c}
$([24]_F,[4]_S)$=
\end{tabular}
\begin{tabular}{|c|c|c|c|}
\hline
$\quad$ & $\quad$ & $\quad$ & $\quad$  \\
\cline{1-4}
\multicolumn{1}{|c|}{$\quad$} \\
\cline{1-1}
\end{tabular}
$\otimes$
\begin{tabular}{|c|c|c|c|}
\hline
$\quad$ & $\quad$ & $\quad$ & $\quad$   \\
\cline{1-4} \multicolumn{1}{|c|}{$\quad$}     \\
\cline{1-1}
\end{tabular}
,
\end{center}
\begin{center}
\begin{tabular}{c}
$([\bar{15}]_F,[6]_S)$=
\end{tabular}
\begin{tabular}{|c|c|c|}
\hline
$\quad$ & $\quad$ & $\quad$   \\
\cline{1-3}
\multicolumn{1}{|c|}{$\quad$} & \multicolumn{1}{|c|}{$\quad$} \\
\cline{1-2}
\end{tabular}
$\otimes$
\begin{tabular}{|c|c|c|c|c|}
\hline
$\quad$ & $\quad$ & $\quad$ & $\quad$ & $\quad$  \\
\hline
\end{tabular}
.
\end{center}
In this article, we consider only $I=\frac{1}{2}$, so we exclude $[21]_F$ case. Therefore, there are four flavor states for each $S=\frac{1}{2}$ and $S=\frac{3}{2}$, and one state for $S=\frac{5}{2}$. Hence, as we can see in Figure ~\ref{fig1}, we can determine the number of possible states of $q^5Q$ when adding one heavy quark. For the dibaryons, there can be four spin states S=0,1,2,3. From the above decomposition, for S=0 case, there are 4 possible states. For S=1, there are 8 possible states. For S=2, there are 5 possible states.  For S=3, there is only one flavor that is $[\bar{15}]$.
\subsection{Flavor, color and spin state of $q^5$}
\label{subsec:section5b}
After $SU(3)_F$ breaking, there are only two flavor bases for $q^5$.
\begin{eqnarray}
|F_1 \rangle=(
\begin{tabular}{|c|c|}
\hline
1 & 2 \\
\hline
3 \\
\cline{1-1}
\end{tabular}
,~
\begin{tabular}{|c|c|}
\hline
4 & 5 \\
\hline
\end{tabular}
),~
|F_2 \rangle=(
\begin{tabular}{|c|c|}
\hline
1 & 3 \\
\hline
2 \\
\cline{1-1}
\end{tabular}
,~
\begin{tabular}{|c|c|}
\hline
4 & 5 \\
\hline
\end{tabular}
).
\end{eqnarray}
We can construct the wave function of $q^5$ using the same method as in Sec.~\ref{subsec:section4b}

\begin{itemize}
\item F=$[\bar{3}]$ : Young tableau of color and spin state is [3,2].\\
\begin{align}
\psi_1=\frac{1}{\sqrt{2}}|F_1 \rangle
\otimes
\Big( \frac{1}{2}
\begin{tabular}{|c|c|c|}
\hline
1 & 3 & 5 \\
\hline
2 & 4  \\
\cline{1-2}
\end{tabular}
-\frac{\sqrt{3}}{2}
\begin{tabular}{|c|c|c|}
\hline
1 & 3 & 4 \\
\hline
2 & 5  \\
\cline{1-2}
\end{tabular}
\Big)_{CS}
\nonumber\\
-\frac{1}{\sqrt{2}}|F_2 \rangle
\otimes
\Big( \frac{1}{2}
\begin{tabular}{|c|c|c|}
\hline
1 & 2 & 5 \\
\hline
3 & 4  \\
\cline{1-2}
\end{tabular}
-\frac{\sqrt{3}}{2}
\begin{tabular}{|c|c|c|}
\hline
1 & 2 & 4 \\
\hline
3 & 5  \\
\cline{1-2}
\end{tabular}
\Big)_{CS}.\nonumber
\\
\label{eq-q5-1}
\end{align}
\item F=[6] : Young tableau of color and spin state is [3,1,1].
\begin{align}
\psi_2=\frac{1}{\sqrt{2}}|F_1 \rangle
\otimes
\Big( \frac{\sqrt{3}}{\sqrt{8}}
\begin{tabular}{|c|c|c|}
\hline
1 & 3 & 5 \\
\hline
2   \\
\cline{1-1}
4   \\
\cline{1-1}
\end{tabular}
-\frac{\sqrt{5}}{\sqrt{8}}
\begin{tabular}{|c|c|c|}
\hline
1 & 3 & 4 \\
\hline
2   \\
\cline{1-1}
5   \\
\cline{1-1}
\end{tabular}
\Big)_{CS}
\nonumber\\
-\frac{1}{\sqrt{2}}|F_2 \rangle
\otimes
\Big( \frac{\sqrt{3}}{\sqrt{8}}
\begin{tabular}{|c|c|c|}
\hline
1 & 2 & 5 \\
\hline
3   \\
\cline{1-1}
4   \\
\cline{1-1}
\end{tabular}
-\frac{\sqrt{5}}{\sqrt{8}}
\begin{tabular}{|c|c|c|}
\hline
1 & 2 & 4 \\
\hline
3   \\
\cline{1-1}
5   \\
\cline{1-1}
\end{tabular}
\Big)_{CS}.\nonumber
\\
\label{eq-q5-2}
\end{align}
\item F=$[\bar{15}]$ : Young tableau of color and spin state is [2,2,1].
\begin{align}
\psi_3=\frac{1}{\sqrt{2}}|F_1 \rangle
\otimes
\Big( \frac{1}{2}
\begin{tabular}{|c|c|}
\hline
1 & 3  \\
\hline
2 & 5  \\
\cline{1-2}
4   \\
\cline{1-1}
\end{tabular}
-\frac{\sqrt{3}}{2}
\begin{tabular}{|c|c|}
\hline
1 & 3 \\
\hline
2 & 4  \\
\cline{1-2}
5   \\
\cline{1-1}
\end{tabular}
\Big)_{CS}
\nonumber\\
-\frac{1}{\sqrt{2}}|F_2 \rangle
\otimes
\Big( \frac{1}{2}
\begin{tabular}{|c|c|}
\hline
1 & 2  \\
\hline
3 & 5  \\
\cline{1-2}
4   \\
\cline{1-1}
\end{tabular}
-\frac{\sqrt{3}}{2}
\begin{tabular}{|c|c|}
\hline
1 & 2  \\
\hline
3 & 4  \\
\cline{1-2}
5   \\
\cline{1-1}
\end{tabular}
\Big)_{CS}.\nonumber
\\
\label{eq-q5-3}
\end{align}
\item F=[24] : Young tableau of color and spin state is [2,1,1,1].
\begin{align}
\psi_4=\frac{1}{\sqrt{2}}|F_1 \rangle
\otimes
\begin{tabular}{|c|c|}
\hline
1 & 3  \\
\hline
2   \\
\cline{1-1}
4   \\
\cline{1-1}
5   \\
\cline{1-1}
\end{tabular}
_{CS}
-\frac{1}{\sqrt{2}}|F_2 \rangle
\otimes
\begin{tabular}{|c|c|}
\hline
1 & 2  \\
\hline
3   \\
\cline{1-1}
4   \\
\cline{1-1}
5   \\
\cline{1-1}
\end{tabular}
_{CS}.\nonumber
\\
\label{eq-q5-4}
\end{align}
\end{itemize}
\section{Flavor, color and spin state of $q^5Q$}
\label{section6}
In the appendix, we represent the color and spin state of $q^5$ and Clebsch-Gordan coefficients, from which we can construct the color and spin state of $q^5Q$ using the following basis functions.
\subsection{Flavor basis function}
Since the isospin of the dibaryon is $\frac{1}{2}$, there are two flavor basis functions.\\
\begin{eqnarray}
|F_1 \rangle=(
\begin{tabular}{|c|c|}
\hline
1 & 2 \\
\hline
3 \\
\cline{1-1}
\end{tabular}
,~
\begin{tabular}{|c|c|}
\hline
4 & 5 \\
\hline
\end{tabular}
,~
\begin{tabular}{|c|}
\hline
6 \\
\hline
\end{tabular}
),~
|F_2 \rangle=(
\begin{tabular}{|c|c|}
\hline
1 & 3 \\
\hline
2 \\
\cline{1-1}
\end{tabular}
,~
\begin{tabular}{|c|c|}
\hline
4 & 5 \\
\hline
\end{tabular}
,~
\begin{tabular}{|c|}
\hline
6 \\
\hline
\end{tabular}
).
\end{eqnarray}

\subsection{Color basis function}

Color singlet : five basis functions with Young tableau [2,2,2]
\begin{align}
&\begin{tabular}{c}
$\vert C_1 \rangle$=
\end{tabular}
\begin{tabular}{|c|c|}
\hline
1 & 2   \\
\cline{1-2}
3 &  4  \\
\cline{1-2}
5 &  6  \\
\hline
\end{tabular}
\quad
\begin{tabular}{c}
$\vert C_2 \rangle$=
\end{tabular}
\begin{tabular}{|c|c|}
\hline
1 & 3   \\
\cline{1-2}
2 &  4  \\
\cline{1-2}
5 &  6  \\
\hline
\end{tabular}
\quad
\begin{tabular}{c}
$\vert C_3 \rangle$=
\end{tabular}
\begin{tabular}{|c|c|}
\hline
1 & 2   \\
\cline{1-2}
3 &  5  \\
\cline{1-2}
4 &  6  \\
\hline
\end{tabular}
\quad
\begin{tabular}{c}
$\vert C_4 \rangle$=
\end{tabular}
\begin{tabular}{|c|c|}
\hline
1 & 3   \\
\cline{1-2}
2 &  5  \\
\cline{1-2}
4 &  6  \\
\hline
\end{tabular}
\nonumber \\
&\begin{tabular}{c}
$\vert C_5 \rangle$=
\end{tabular}
\begin{tabular}{|c|c|}
\hline
1 & 4   \\
\cline{1-2}
2 &  5  \\
\cline{1-2}
3 &  6  \\
\hline
\end{tabular}
\label{eq-color}
\end{align}

\subsection{Spin basis function}

\begin{itemize}
\item S=0 : five basis functions with Young tableau [3,3]
\begin{tabular}{c}
$\vert S_1^0 \rangle$=
\end{tabular}
\begin{tabular}{|c|c|c|}
\hline
                1   & 2   & 3    \\
\cline{1-3} 4  &  5 & 6  \\
\hline
\end{tabular}
\begin{tabular}{c}
$\vert S_2^0 \rangle$=
\end{tabular}
\begin{tabular}{|c|c|c|}
\hline
                1   & 2   & 4    \\
\cline{1-3} 3  &  5 & 6  \\
\hline
\end{tabular}
\begin{tabular}{c}
$\vert S_3^0 \rangle$=
\end{tabular}
\begin{tabular}{|c|c|c|}
\hline
                1   & 3   & 4    \\
\cline{1-3} 2  &  5 & 6  \\
\hline
\end{tabular}

\begin{tabular}{c}
$\vert S_4^0 \rangle$=
\end{tabular}
\begin{tabular}{|c|c|c|}
\hline
                1   & 2   & 5    \\
\cline{1-3} 3  &  4 & 6  \\
\hline
\end{tabular}
\begin{tabular}{c}
$\vert S_5^0 \rangle$=
\end{tabular}
\begin{tabular}{|c|c|c|}
\hline
                1   & 3   & 5    \\
\cline{1-3} 2  &  4 & 6  \\
\hline
\end{tabular}
\item S=1 : nine basis functions with Young tableau [4,2]
\begin{tabular}{c}
$\vert S_1^1 \rangle$=
\end{tabular}
\begin{tabular}{|c|c|c|c|}
\hline
                  1 & 2 & 3 & 4   \\
\cline{1-4}
\multicolumn{1}{|c|}{5} & \multicolumn{1}{c|}{6}  \\
\cline{1-2}
\end{tabular}
\begin{tabular}{c}
$\vert S_2^1 \rangle$=
\end{tabular}
\begin{tabular}{|c|c|c|c|}
\hline
                  1 & 2 & 3 & 5   \\
\cline{1-4}
\multicolumn{1}{|c|}{4} & \multicolumn{1}{c|}{6}  \\
\cline{1-2}
\end{tabular}
\begin{tabular}{c}
$\vert S_3^1 \rangle$=
\end{tabular}
\begin{tabular}{|c|c|c|c|}
\hline
                  1 & 2 & 4 & 5   \\
\cline{1-4}
\multicolumn{1}{|c|}{3} & \multicolumn{1}{c|}{6}  \\
\cline{1-2}
\end{tabular}

\begin{tabular}{c}
$\vert S_4^1 \rangle$=
\end{tabular}
\begin{tabular}{|c|c|c|c|}
\hline
                  1 & 3 & 4 & 5   \\
\cline{1-4}
\multicolumn{1}{|c|}{2} & \multicolumn{1}{c|}{6}  \\
\cline{1-2}
\end{tabular}
\begin{tabular}{c}
$\vert S_5^1 \rangle$=
\end{tabular}
\begin{tabular}{|c|c|c|c|}
\hline
                  1 & 2 & 3 & 6   \\
\cline{1-4}
\multicolumn{1}{|c|}{4} & \multicolumn{1}{c|}{5}  \\
\cline{1-2}
\end{tabular}
\begin{tabular}{c}
$\vert S_6^1 \rangle$=
\end{tabular}
\begin{tabular}{|c|c|c|c|}
\hline
                  1 & 2 & 4 & 6   \\
\cline{1-4}
\multicolumn{1}{|c|}{3} & \multicolumn{1}{c|}{5}  \\
\cline{1-2}
\end{tabular}

\begin{tabular}{c}
$\vert S_7^1 \rangle$=
\end{tabular}
\begin{tabular}{|c|c|c|c|}
\hline
                  1 & 3 & 4 & 6   \\
\cline{1-4}
\multicolumn{1}{|c|}{2} & \multicolumn{1}{c|}{5}  \\
\cline{1-2}
\end{tabular}
\begin{tabular}{c}
$\vert S_8^1 \rangle$=
\end{tabular}
\begin{tabular}{|c|c|c|c|}
\hline
                  1 & 2 & 5 & 6   \\
\cline{1-4}
\multicolumn{1}{|c|}{3} & \multicolumn{1}{c|}{4}  \\
\cline{1-2}
\end{tabular}
\begin{tabular}{c}
$\vert S_9^1 \rangle$=
\end{tabular}
\begin{tabular}{|c|c|c|c|}
\hline
                  1 & 3 & 5 & 6   \\
\cline{1-4}
\multicolumn{1}{|c|}{2} & \multicolumn{1}{c|}{4}  \\
\cline{1-2}
\end{tabular}
\item S=2 : five basis functions with Young tableau [5,1]
\begin{tabular}{c}
$\vert S_1^2 \rangle$=
\end{tabular}
\begin{tabular}{|c|c|c|c|c|}
\hline
                  1 & 2 & 3 & 4 & 5   \\
\cline{1-5}
\multicolumn{1}{|c|}{6}  \\
\cline{1-1}
\end{tabular}
\begin{tabular}{c}
$\vert S_2^2 \rangle$=
\end{tabular}
\begin{tabular}{|c|c|c|c|c|}
\hline
                  1 & 2 & 3 & 4 & 6   \\
\cline{1-5}
\multicolumn{1}{|c|}{5}  \\
\cline{1-1}
\end{tabular}

\begin{tabular}{c}
$\vert S_3^2 \rangle$=
\end{tabular}
\begin{tabular}{|c|c|c|c|c|}
\hline
                  1 & 2 & 3 & 5 & 6   \\
\cline{1-5}
\multicolumn{1}{|c|}{4}  \\
\cline{1-1}
\end{tabular}
\begin{tabular}{c}
$\vert S_4^2 \rangle$=
\end{tabular}
\begin{tabular}{|c|c|c|c|c|}
\hline
                  1 & 2 & 4 & 5 & 6   \\
\cline{1-5}
\multicolumn{1}{|c|}{3}  \\
\cline{1-1}
\end{tabular}

\begin{tabular}{c}
$\vert S_5^2 \rangle$=
\end{tabular}
\begin{tabular}{|c|c|c|c|c|}
\hline
                  1 & 3 & 4 & 5 & 6   \\
\cline{1-5}
\multicolumn{1}{|c|}{2}  \\
\cline{1-1}
\end{tabular}
\item S=3 : one basis function with Young tableau [6]
\begin{tabular}{c}
$\vert S^3 \rangle$=
\end{tabular}
\begin{tabular}{|c|c|c|c|c|c|}
\hline
                  1 & 2 & 3 & 4 & 5 & 6   \\
\hline
\end{tabular}
\end{itemize}

\subsection{Flavor, color and spin state of $q^5Q$}

Here, we can construct the wave function of $q^5Q$ from the state of $q^5$. Since the color state of $q^5$ is $[\bar{3}]$, there is only one way to construct color singlet state by adding one heavy flavor. There is no change in color basis function from $q^5$ to $q^5Q$. However, we should treat the spin state transformation carefully. The spin basis function of $q^5$ for S=$\frac{1}{2}$ can be either S=0 or S=1 of $q^5Q$ state. We represent the spin basis transformation for each cases in the appendix.
\begin{itemize}
\item S=0 : 4 possible states\\
$\psi^5_{1,S=\frac{1}{2}} \rightarrow \psi^6_{1,S=0} $, $\psi^5_{2,S=\frac{1}{2}} \rightarrow \psi^6_{2,S=0} $,\\
$\psi^5_{3,S=\frac{1}{2}} \rightarrow \psi^6_{3,S=0} $, $\psi^5_{4,S=\frac{1}{2}} \rightarrow \psi^6_{4,S=0} $.
\item S=1 : 8 possible states\\
$\psi^5_{1,S=\frac{1}{2}} \rightarrow \psi^6_{1,S=1} $, $\psi^5_{2,S=\frac{1}{2}} \rightarrow \psi^6_{2,S=1} $,\\
$\psi^5_{3,S=\frac{1}{2}} \rightarrow \psi^6_{3,S=1} $, $\psi^5_{4,S=\frac{1}{2}} \rightarrow \psi^6_{4,S=1} $,\\
$\psi^5_{1,S=\frac{3}{2}} \rightarrow \psi^6_{5,S=1} $, $\psi^5_{2,S=\frac{3}{2}} \rightarrow \psi^6_{6,S=1} $,\\
$\psi^5_{3,S=\frac{3}{2}} \rightarrow \psi^6_{7,S=1} $, $\psi^5_{4,S=\frac{3}{2}} \rightarrow \psi^6_{8,S=1} $.
\item S=2 : 5 possible states\\
$\psi^5_{1,S=\frac{3}{2}} \rightarrow \psi^6_{1,S=2} $, $\psi^5_{2,S=\frac{3}{2}} \rightarrow \psi^6_{2,S=2} $,\\
$\psi^5_{3,S=\frac{3}{2}} \rightarrow \psi^6_{3,S=2} $, $\psi^5_{4,S=\frac{3}{2}} \rightarrow \psi^6_{4,S=2} $,\\
$\psi^5_{3,S=\frac{5}{2}} \rightarrow \psi^6_{5,S=2} $.
\item S=3 : 1 possible state\\
$\psi^5_{3,S=\frac{5}{2}} \rightarrow \psi^6_{1,S=3} $.
\end{itemize}

We find that there is an orthogonal transformation between $\{\psi_i\}$ and $\{\phi_i\}$.
\begin{itemize}
  \item S=0
  \begin{align}
    \phi_1=\psi_1,~ \phi_2=-\psi_2,~ \phi_3=\psi_3,~ \phi_4=\psi_4. \nonumber
  \end{align}
  \item S=1
  \begin{align}
  \frac{2}{\sqrt{5}}\phi_1-\frac{1}{\sqrt{5}}\phi_5=\psi_1,~ \frac{1}{\sqrt{5}}\phi_1+\frac{2}{\sqrt{5}}\phi_5=\psi_5, \nonumber\\
  -\frac{\sqrt{5}}{3}\phi_2+\frac{2}{3}\phi_6=\psi_2,~~~~ \frac{2}{3}\phi_2+\frac{\sqrt{5}}{3}\phi_6=\psi_6, \nonumber\\
  \frac{\sqrt{5}}{3}\phi_3+\frac{2}{3}\phi_7=\psi_3,~~~~ \frac{2}{3}\phi_3-\frac{\sqrt{5}}{3}\phi_7=\psi_7, \nonumber\\
  -\frac{2}{3}\phi_4-\frac{\sqrt{5}}{3}\phi_8=\psi_4,~ -\frac{\sqrt{5}}{3}\phi_4+\frac{2}{3}\phi_8=\psi_8. \nonumber
\end{align}
  \item S=2
  \begin{align}
    \phi_1=-\psi_1,~ \phi_2=\psi_2,~ \phi_4=-\psi_4, \nonumber\\
    -\frac{4}{5}\phi_3+\frac{3}{5}\phi_5=\psi_3,~ -\frac{3}{5}\phi_3-\frac{4}{5}\phi_5=\psi_5. \nonumber
  \end{align}
  \item S=3
  \begin{align}
    \phi_1=\psi_1. \nonumber
  \end{align}
\end{itemize}

Since we calculate the expectation values using these wave functions, the results from both methods are the same as expected.

\subsection{Baryon-baryon configuration of $q^5Q$}

Considering the decay channel, we construct the baryon-baryon wave function of $q^5Q$. Among the five color basis functions, $|C_5\rangle$ is color singlet for 1,2,3 and at the same time color singlet for 4,5,6. Therefore, we have to construct the wave function only by using $|C_5\rangle$ as color state if we want to investigate decay channel through $uud$ baryon and $ssQ$ baryon. As done in Sec.\ref{subsec:section5b}, we can construct the color, flavor and spin state of $q^5Q$ which has the specific symmetry property for S=0,1,2.
\begin{align}
  \psi^{BB}_{S=0}=\frac{1}{\sqrt{2}}|C_5\rangle \otimes |F_1\rangle \otimes (\frac{1}{2}|S^0_2\rangle + \frac{\sqrt{3}}{2}|S^0_4\rangle ) \nonumber\\
  +\frac{1}{\sqrt{2}}|C_5\rangle \otimes |F_2\rangle \otimes (\frac{1}{2}|S^0_3\rangle + \frac{\sqrt{3}}{2}|S^0_5\rangle ).
\end{align}
\begin{align}
\psi^{BB}_{S=1}=\frac{1}{\sqrt{2}}|C_5\rangle \otimes |F_1\rangle \otimes (A_1|S^1_3\rangle + A_2(\frac{1}{2}|S^1_6\rangle +\frac{\sqrt{3}}{2}|S^1_8\rangle )) \nonumber\\
+\frac{1}{\sqrt{2}}|C_5\rangle \otimes |F_2\rangle \otimes (A_1|S^1_4\rangle + A_2(\frac{1}{2}|S^1_7\rangle +\frac{\sqrt{3}}{2}|S^1_9\rangle )). \nonumber\\
\end{align}
\begin{align}
\psi^{BB}_{S=2}=\frac{1}{\sqrt{2}}|C_5\rangle \otimes |F_1\rangle \otimes |S^2_4\rangle ) + \frac{1}{\sqrt{2}}|C_5\rangle \otimes |F_2\rangle \otimes |S^2_5\rangle . \nonumber\\
\end{align}
\label{subsec:section6e}
In the above expressions, $A_1$ and $A_2$ are undetermined constants. However, to construct S=1 dibaryon, the spin of $ssQ$ baryon can be either $\frac{1}{2}$ or $\frac{3}{2}$. When the spin of $ssQ$ baryon is $\frac{3}{2}$, it should have the symmetry property \{456\}. Using this symmetry property, we can decide $A_1$ and $A_2$. Once this state is determined, we can obtain the other state which is consist of $ssQ$ baryon($S=\frac{1}{2}$) by using the orthogonality of wave function.
\begin{itemize}
  \item $uud(S=\frac{1}{2})+ssQ(S=\frac{1}{2})$
\begin{align}
\psi^{BB}_{S=1}=\frac{1}{\sqrt{2}}|C_5\rangle \otimes |F_1\rangle \otimes (-\frac{2\sqrt{2}}{3}|S^1_3\rangle + \frac{1}{6}|S^1_6\rangle +\frac{\sqrt{3}}{6}|S^1_8\rangle ) \nonumber\\
+\frac{1}{\sqrt{2}}|C_5\rangle \otimes |F_2\rangle \otimes (-\frac{2\sqrt{2}}{3}|S^1_4\rangle + \frac{1}{6}|S^1_7\rangle +\frac{\sqrt{3}}{6}|S^1_9\rangle ). \nonumber\\
\end{align}
  \item $uud(S=\frac{1}{2})+ssQ(S=\frac{3}{2})$
\begin{align}
\psi^{BB}_{S=1}=\frac{1}{\sqrt{2}}|C_5\rangle \otimes |F_1\rangle \otimes (\frac{1}{3}|S^1_3\rangle + \frac{\sqrt{2}}{3}|S^1_6\rangle +\frac{\sqrt{6}}{3}|S^1_8\rangle ) \nonumber\\
+\frac{1}{\sqrt{2}}|C_5\rangle \otimes |F_2\rangle \otimes (\frac{1}{3}|S^1_4\rangle + \frac{\sqrt{2}}{3}|S^1_7\rangle +\frac{\sqrt{6}}{3}|S^1_9\rangle ). \nonumber\\
\end{align}
\end{itemize}
We can not construct the baryon\{123\}-baryon\{456\} wave function with S=3, because there is no $uud$ baryon with $I=\frac{1}{2}$ and $S=\frac{3}{2}$.

\section{Numerical Results}
\label{section7}
In this section, we present our numerical results.
 Before using variational method, we can estimate the stability condition only by using the simple color-spin interaction formula given as $-\sum_{i<j}^N 1/(m_i m_j)\langle \lambda_i^c \lambda_j^c \sigma_i \cdot \sigma_j \rangle$ without taking into account the r-dependence of the hyperfine potential.   In the upper part of Table~\ref{table:hyperfine2}, we show the matrix elements without the masses, which reflects the attraction in the SU(4) flavor symmetric limit.  As one can see here, the $\psi_{5,S=1}$ is the most attractive channel.  The lower part of Table~\ref{table:hyperfine2} is the realistic case where all the relevant constituent quark masses are included.   One notes that for S=0,1, $\Lambda \Xi_c$ system is now more attractive than anti-triplet dibaryon state, but for S=2, the dibaryon state is more attractive. Therefore, we can expect to have a compact stable dibaryon state for S=2 configuration  when considering the color-spin interaction only.

\begin{table}
\caption{The expectation values of $-\sum_{i<j}^N \langle \lambda_i^c \lambda_j^c \sigma_i \cdot \sigma_j \rangle$ and $-\sum_{i<j}^N 1/(m_i m_j)\langle \lambda_i^c \lambda_j^c \sigma_i \cdot \sigma_j \rangle$ for each anti-triplet flavor states and corresponding decay channel}
\begin{center}
\begin{tabular}{c|c|c|c|c}
\hline \hline
\multicolumn{5}{c}{$-\sum_{i<j}^N \langle \lambda_i^c \lambda_j^c \sigma_i \cdot \sigma_j \rangle$} \\
\hline
$uudssc$ & $\psi_{1,S=0}$ & $\psi_{1,S=1}$ & $\psi_{5,S=1}$ & $\psi_{1,S=2}$ \\
\hline
& -24 & $-\frac{40}{3}$ & $-\frac{76}{3}$ & -4 \\
\hline
decay channel & $N\Omega_c$ & $N\Omega_c$ & $N\Omega_c$ & $N\Omega_c^*$  \\
\hline
& -16 & -16 & -16 & 0 \\
\hline
decay channel & $\Lambda \Xi_c$ & $\Lambda \Xi_c$ & $\Lambda \Xi_c$ & $\Lambda \Xi_c^*$  \\
\hline
& -16 & -16 & -16 & 0 \\
\hline \hline
\end{tabular}
\end{center}
\begin{center}
\begin{tabular}{c|c|c|c|c}
\hline \hline
\multicolumn{5}{c}{$-\sum_{i<j}^N 1/(m_i m_j)\langle \lambda_i^c \lambda_j^c \sigma_i \cdot \sigma_j \rangle$} \\
\hline
$uudssc$ & $\psi_{1,S=0}$ & $\psi_{1,S=1}$ & $\psi_{5,S=1}$ & $\psi_{1,S=2}$ \\
\hline
& -98.1 & -87.1 & -91.2 & -64.9 \\
\hline
decay channel & $N\Omega_c$ & $N\Omega_c$ & $N\Omega_c$ & $N\Omega_c^*$  \\
\hline
& -70.1 & -70.1 & -70.1 & -57 \\
\hline
decay channel & $\Lambda \Xi_c$ & $\Lambda \Xi_c$ & $\Lambda \Xi_c$ & $\Lambda \Xi_c^*$  \\
\hline
& -104.9 & -104.9 & -104.9 & -49.5 \\
\hline \hline
\end{tabular}
\end{center}
\label{table:hyperfine2}
\end{table}

We now analyse the numerical results obtained from the variational method. Varying the parameters of spatial function, we find the ground state of the dibaryon. The parameters and masses are in Table V.
Here, we represent the color, flavor and spin wave function of ground state dibaryon obtained from variational method for S=0,1,2,3.

\begin{align}
  &\psi^{\mathrm{ground}}_{S=0}& \nonumber\\
  &=-0.5\psi^6_1+0.71\psi^6_2-0.17\psi^6_3-0.47\psi^6_4& \nonumber\\
  &=0.35|C_5 \rangle \otimes |F_1 \rangle \otimes |S^0_2 \rangle +0.35|C_5 \rangle \otimes |F_2 \rangle \otimes |S^0_3 \rangle & \nonumber\\
  &+0.61|C_5 \rangle \otimes |F_1 \rangle \otimes |S^0_4 \rangle+0.61|C_5 \rangle \otimes |F_2 \rangle \otimes |S^0_5 \rangle. & \nonumber
\\
\end{align}
\begin{align}
  \psi^{\mathrm{ground}}_{S=1}& \nonumber\\
  =&-0.17\psi^6_1+0.24\psi^6_2-0.06\psi^6_3-0.16\psi^6_4 \nonumber\\
  &+0.67\psi^6_5+0.42\psi^6_6-0.5\psi^6_7+0.14\psi^6_8 \nonumber\\
  =&-0.67|C_5 \rangle \otimes |F_1 \rangle \otimes |S^1_3 \rangle-0.67|C_5 \rangle \otimes |F_2 \rangle \otimes |S^1_4 \rangle \nonumber\\
  &+0.12|C_5 \rangle \otimes |F_1 \rangle \otimes |S^1_6 \rangle+0.12|C_5 \rangle \otimes |F_2 \rangle \otimes |S^1_7 \rangle \nonumber\\
  &+0.2|C_5 \rangle \otimes |F_1 \rangle \otimes |S^1_8 \rangle+0.2|C_5 \rangle \otimes |F_2 \rangle \otimes |S^1_9 \rangle. \nonumber
\\
\label{G-1}
\end{align}
\begin{align}
  \psi^{\mathrm{ground}}_{S=2}& \nonumber\\
  =&-0.71\psi^6_1-0.45\psi^6_2+0.53\psi^6_3-0.15\psi^6_4 \nonumber\\
  =&0.71|C_5 \rangle \otimes |F_1 \rangle \otimes |S^2_4 \rangle+0.71|C_5 \rangle \otimes |F_2 \rangle \otimes |S^2_5 \rangle. \nonumber
\\
\end{align}
\begin{align}
  \psi^{\mathrm{ground}}_{S=3}&=\psi^6_1 \nonumber\\
  =&\frac{1}{2\sqrt{2}}|C_4\rangle\otimes|F_1\rangle\otimes|S^3_1\rangle-\frac{\sqrt{3}}{2\sqrt{2}}|C_2\rangle\otimes|F_1\rangle\otimes|S^3_1\rangle \nonumber\\
  -&\frac{1}{2\sqrt{2}}|C_3\rangle\otimes|F_2\rangle\otimes|S^3_1\rangle+\frac{\sqrt{3}}{2\sqrt{2}}|C_1\rangle\otimes|F_2\rangle\otimes|S^3_1\rangle. \nonumber
\\
\end{align}
\begin{table}
\caption{Variational parameters fitted to the ground state of the dibaryon using the variational method. The other parameters are the same as baryon. The mass unit is GeV.}
\begin{center}
\begin{tabular}{|c|c|c|c|c|c|c|}
\hline \hline
uudssc & a & b  & c & d & Mass & Decay mode   \\
\hline
S=0 & 2.5 & 3.5 & 5.7  & 0 & 3.666 & N$\Omega_c$  \\
\hline
S=1 & 2.5 & 3.5 & 5.7  & 0 & 3.666 & N$\Omega_c$  \\
\hline
S=2 & 2.5 & 3.2 & 4.8  & 0 & 3.737 & N$\Omega_c^*$  \\
\hline
S=3 & 1.6 & 3.3 & 2.8  & 2.4 & 4.404 &$\Sigma^* \Xi_c^*$  \\
\hline \hline
\end{tabular}
\end{center}
\begin{center}
\begin{tabular}{|c|c|c|c|c|c|c|}
\hline \hline
uudssb & a & b  & c & d & Mass & Decay mode   \\
\hline
S=0 & 2.5 & 3.5 & 6.5  & 0 & 6.997 & N$\Omega_b$  \\
\hline
S=1 & 2.5 & 3.5 & 6.5  & 0 & 6.997 & N$\Omega_b$  \\
\hline
S=2 & 2.5 & 3.3 & 6.1  & 0 & 7.028 & N$\Omega_b^*$  \\
\hline
S=3 & 1.6 & 3.3 & 4.5  & 2.5 & 7.803 &$\Sigma^* \Xi_b^*$  \\
\hline \hline
\end{tabular}
\end{center}
\label{parameter-02}
\end{table}

As we can see, the color wave function of the ground state for S=0,1,2 is $|C_5 \rangle$ and the coefficients are almost the same as in Sec.~\ref{subsec:section6e}. Therefore, the ground state of $q^5Q$ for S=0,1,2 is the sum of two baryon states. However, for S=3, the color state is not $|C_5 \rangle$ because there is no baryon state with I=1/2 and S=3/2 for $uud$. Therefore, the ground state for S=3 is not the sum of two baryon states.  Also, as can be seen in Table~\ref{parameter-02}, $d=0$ for the ground state other than the S=3.  This parameter corresponds to a well separated baryon baryon configuration.
As $uudssQ$ has the \{123\}\{45\}6 symmetry property, even with a single Guassian trial wave function given in Eq.~\ref{spatial function}, one  can express the wave function for a well separated baryon(123)-baryon(456) configuration.

Additionally, we have to notice that there are two cases to make S=1 dibaryon from the two baryon states. The first one is two S=1/2 baryons and second is one S=1/2 baryon and one S=3/2 baryon. The ground state in Eq.~(\ref{G-1}) corresponds to the  first case.  We can also calculate the excited state for S=1 by using variational parameters obtained from the ground state. We find that the flavor color spin wave function for the first excited state is given as
\begin{align}
  \psi^{\mathrm{excited}}_{S=1}& \nonumber\\
  =&0.24|C_5 \rangle \otimes |F_1 \rangle \otimes |S^1_3 \rangle+0.24|C_5 \rangle \otimes |F_2 \rangle \otimes |S^1_4 \rangle \nonumber\\
  +&0.33|C_5 \rangle \otimes |F_1 \rangle \otimes |S^1_6 \rangle+0.33|C_5 \rangle \otimes |F_2 \rangle \otimes |S^1_7 \rangle \nonumber\\
  +&0.58|C_5 \rangle \otimes |F_1 \rangle \otimes |S^1_8 \rangle+0.58|C_5 \rangle \otimes |F_2 \rangle \otimes |S^1_9 \rangle. \nonumber
\\
\end{align}

This form turns out to be the same as the second case for S=1 given in Sec.\ref{subsec:section6e}, which represents flavor color spin state of the $N\Omega_c^*$ system.

\section{Summary}
\label{section8}
In this work, we investigate the symmetry property and the stability of the dibaryon containing two strange quarks and one heavy flavor with $I=\frac{1}{2}$. To obtain the wave function of the dibaryon with required symmetry property, we utilize two methods; the first one is to construct the color and spin wave function of the dibaryon from each flavor state directly, and the second one is to form the color and spin state of $q^5$ first, and then construct the wave function of the dibaryon by adding one heavy quark. We verify that their results are the same.

In the SU(3) flavor limit, by using Eq.~(\ref{eq:hyperfine-formula}), we expect the flavor anti-triplet state to be the most attractive channel when five light quarks are considered. Additionally, if we increase the mass of sixth quark, then its additional kinetic term will become smaller so that it may lead more chance to form the stable and compact dibaryon state. Hence, we calculate the mass of the dibaryon with two strange quarks and one heavy flavor.

We first estimate the stability condition by using only the color-spin interaction without the r-dependence. In the SU(4) flavor symmetric limit, all the anti-triplet flavor states of the dibaryon with S=0,1,2 are more attractive than the decay channel. However, when we consider the relevant constituent quark masses obtained from baryon fitting results, only the anti-triplet state for S=2 is more attractive.

Finally, we calculate the masses of the dibaryons in a nonrelativistic Hamiltonian by using variational method. We find that the ground state of the dibaryon for S=0,1,2 is the well separated two baryon states. As for S=3, with our single Gaussian form, the mass of the ground state is found to be more repulsive than the decay channel. Hence, we conclude that there are no stable and compact $uudssQ$ dibaryon state with $I=\frac{1}{2}$ within the given potential. If we consider the dibaryon containing more than one heavy quark, then their additional kinetic terms will decrease further than the dibaryon with one heavy quark so that it may lead more chance to form the stable dibaryon.
\\
\section*{Acknowledgements}

This work was supported by the Korea
National Research Foundation under the grant number
KRF-2011-0020333 and KRF-2011-0030621.
\\
\appendix
\section{Color and Spin state of $q^5$}
Here, we represent the color and spin basis function of $q^5$.
\subsection{Color basis function}
\begin{align}
&\begin{tabular}{c}
$\vert C_1 \rangle$=
\end{tabular}
\begin{tabular}{|c|c|}
\hline
1 & 2   \\
\cline{1-2}
3 &  4  \\
\cline{1-2}
5   \\
\cline{1-1}
\end{tabular}
\quad
\begin{tabular}{c}
$\vert C_2 \rangle$=
\end{tabular}
\begin{tabular}{|c|c|}
\hline
1 & 3   \\
\cline{1-2}
2 &  4  \\
\cline{1-2}
5   \\
\cline{1-1}
\end{tabular}
\quad
\begin{tabular}{c}
$\vert C_3 \rangle$=
\end{tabular}
\begin{tabular}{|c|c|}
\hline
1 & 2   \\
\cline{1-2}
3 &  5  \\
\cline{1-2}
4   \\
\cline{1-1}
\end{tabular}
\quad
\begin{tabular}{c}
$\vert C_4 \rangle$=
\end{tabular}
\begin{tabular}{|c|c|}
\hline
1 & 3   \\
\cline{1-2}
2 &  5  \\
\cline{1-2}
4   \\
\cline{1-1}
\end{tabular}
\nonumber \\
&\begin{tabular}{c}
$\vert C_5 \rangle$=
\end{tabular}
\begin{tabular}{|c|c|}
\hline
1 & 4   \\
\cline{1-2}
2 &  5  \\
\cline{1-2}
3   \\
\cline{1-1}
\end{tabular}
\label{eq-color}
\end{align}

\subsection{Spin basis function}

\begin{itemize}
\item S=1/2 : 5 basis functions with Young tableau [3,3]
\begin{tabular}{c}
$\vert S_1^{\frac{1}{2}} \rangle$=
\end{tabular}
\begin{tabular}{|c|c|c|}
\hline
                1   & 2   & 3    \\
\cline{1-3} 4  &  5   \\
\cline{1-2}
\end{tabular}
\begin{tabular}{c}
$\vert S_2^{\frac{1}{2}} \rangle$=
\end{tabular}
\begin{tabular}{|c|c|c|}
\hline
                1   & 2   & 4    \\
\cline{1-3} 3  &  5  \\
\cline{1-2}
\end{tabular}
\begin{tabular}{c}
$\vert S_3^{\frac{1}{2}} \rangle$=
\end{tabular}
\begin{tabular}{|c|c|c|}
\hline
                1   & 3   & 4    \\
\cline{1-3} 2  &  5  \\
\cline{1-2}
\end{tabular}
\\
\\
\begin{tabular}{c}
$\vert S_4^{\frac{1}{2}} \rangle$=
\end{tabular}
\begin{tabular}{|c|c|c|}
\hline
                1   & 2   & 5    \\
\cline{1-3} 3  &  4   \\
\cline{1-2}
\end{tabular}
\begin{tabular}{c}
$\vert S_5^{\frac{1}{2}} \rangle$=
\end{tabular}
\begin{tabular}{|c|c|c|}
\hline
                1   & 3   & 5    \\
\cline{1-3} 2  &  4   \\
\cline{1-2}
\end{tabular}
\\
\item S=3/2 : 4 basis functions with Young tableau [4,2]
\begin{tabular}{c}
$\vert S_1^{\frac{3}{2}} \rangle$=
\end{tabular}
\begin{tabular}{|c|c|c|c|}
\hline
                  1 & 2 & 3 & 4   \\
\cline{1-4}
5   \\
\cline{1-1}
\end{tabular}
\begin{tabular}{c}
$\vert S_2^{\frac{3}{2}} \rangle$=
\end{tabular}
\begin{tabular}{|c|c|c|c|}
\hline
                  1 & 2 & 3 & 5   \\
\cline{1-4}
4  \\
\cline{1-1}
\end{tabular}
\begin{tabular}{c}
$\vert S_3^{\frac{3}{2}} \rangle$=
\end{tabular}
\begin{tabular}{|c|c|c|c|}
\hline
                  1 & 2 & 4 & 5   \\
\cline{1-4}
3  \\
\cline{1-1}
\end{tabular}

\begin{tabular}{c}
$\vert S_4^{\frac{3}{2}} \rangle$=
\end{tabular}
\begin{tabular}{|c|c|c|c|}
\hline
                  1 & 3 & 4 & 5   \\
\cline{1-4}
2  \\
\cline{1-1}
\end{tabular}

\item S=5/2 : 1 basis function with Young tableau [6]
\begin{tabular}{c}
$\vert S^{\frac{5}{2}}_1 \rangle$=
\end{tabular}
\begin{tabular}{|c|c|c|c|c|}
\hline
                  1 & 2 & 3 & 4 & 5    \\
\hline
\end{tabular}
\end{itemize}

\begin{widetext}\allowdisplaybreaks
\section{CS coupling of $q^5Q$}

In this section, we present the color $\otimes$ spin basis. The Clebsch-Gordon coefficients of combining the color and spin basis are calculated by using K matrix~\cite{Stancu:1991rc, Stancu:1999qr}.
\subsection{S=0}
\begin{eqnarray}
\begin{tabular}{|c|c|c|}
\hline
1 & 2 & 3 \\
\hline
4 & 5 & 6  \\
\hline
\end{tabular}_{~CS}
&=&\frac{1}{2}|C_1 \rangle \otimes |S_2^0 \rangle + \frac{1}{2}|C_2 \rangle \otimes |S_3^0\rangle + \frac{1}{2}|C_3 \rangle \otimes |S_4^0\rangle + \frac{1}{2}|C_4 \rangle \otimes |S_5^0\rangle
\nonumber\\
\begin{tabular}{|c|c|c|}
\hline
1 & 2 & 4 \\
\hline
3 & 5 & 6 \\
\hline
\end{tabular}_{~CS}
&=&\frac{1}{2}|C_1 \rangle \otimes |S_1^0 \rangle + \frac{1}{2\sqrt{2}}|C_1 \rangle \otimes |S_2^0\rangle - \frac{1}{2\sqrt{2}}|C_2 \rangle \otimes |S_3^0\rangle - \frac{1}{2\sqrt{2}}|C_3 \rangle \otimes |S_4^0\rangle + \frac{1}{2\sqrt{2}}|C_4 \rangle \otimes |S_5^0\rangle - \frac{1}{2}|C_5 \rangle \otimes |S_5^0\rangle
\nonumber\\
\begin{tabular}{|c|c|c|}
\hline
1 & 3 & 4 \\
\hline
2 & 5 & 6 \\
\hline
\end{tabular}_{~CS}
&=&\frac{1}{2}|C_2 \rangle \otimes |S_1^0 \rangle - \frac{1}{2\sqrt{2}}|C_2 \rangle \otimes |S_2^0\rangle - \frac{1}{2\sqrt{2}}|C_1 \rangle \otimes |S_3^0\rangle + \frac{1}{2\sqrt{2}}|C_4 \rangle \otimes |S_4^0\rangle + \frac{1}{2}|C_5 \rangle \otimes |S_4^0\rangle + \frac{1}{2\sqrt{2}}|C_3 \rangle \otimes |S_5^0\rangle
\nonumber\\
\begin{tabular}{|c|c|c|}
\hline
1 & 2 & 5 \\
\hline
3 & 4 & 6 \\
\hline
\end{tabular}_{~CS}
&=&\frac{1}{2}|C_3 \rangle \otimes |S_1^0 \rangle - \frac{1}{2\sqrt{2}}|C_3 \rangle \otimes |S_2^0\rangle + \frac{1}{2\sqrt{2}}|C_4 \rangle \otimes |S_3^0\rangle + \frac{1}{2}|C_5 \rangle \otimes |S_3^0\rangle - \frac{1}{2\sqrt{2}}|C_1 \rangle \otimes |S_4^0\rangle + \frac{1}{2\sqrt{2}}|C_2 \rangle \otimes |S_5^0\rangle
\nonumber\\
\begin{tabular}{|c|c|c|}
\hline
1 & 3 & 5 \\
\hline
2 & 4 & 6 \\
\hline
\end{tabular}_{~CS}
&=&\frac{1}{2}|C_4 \rangle \otimes |S_1^0 \rangle + \frac{1}{2\sqrt{2}}|C_4 \rangle \otimes |S_2^0\rangle - \frac{1}{2}|C_5 \rangle \otimes |S_2^0\rangle + \frac{1}{2\sqrt{2}}|C_3 \rangle \otimes |S_3^0\rangle + \frac{1}{2\sqrt{2}}|C_2 \rangle \otimes |S_4^0\rangle + \frac{1}{2\sqrt{2}}|C_1 \rangle \otimes |S_5^0\rangle
\nonumber
\end{eqnarray}
\begin{eqnarray}
  \begin{tabular}{|c|c|c|c|}
    \hline
    1 & 2 & 3 & 4 \\
    \hline
    5 \\
    \cline{1-1}
    6 \\
    \cline{1-1}
  \end{tabular}_{CS}
  &=&\frac{1}{\sqrt{2}}|C_1 \rangle \otimes |S_4^0 \rangle + \frac{1}{\sqrt{2}}|C_2 \rangle \otimes |S_5^0 \rangle \nonumber\\
  \begin{tabular}{|c|c|c|c|}
    \hline
    1 & 2 & 3 & 5 \\
    \hline
    4 \\
    \cline{1-1}
    6 \\
    \cline{1-1}
  \end{tabular}_{CS}
  &=&-\frac{1}{\sqrt{10}}|C_1 \rangle \otimes |S_2^0 \rangle + \frac{\sqrt{3}}{\sqrt{10}}|C_3 \rangle \otimes |S_2^0 \rangle - \frac{1}{\sqrt{10}}|C_2 \rangle \otimes |S_3^0 \rangle +\frac{\sqrt{3}}{\sqrt{10}}|C_4 \rangle \otimes |S_3^0 \rangle +\frac{1}{\sqrt{10}}|C_3 \rangle \otimes |S_4^0 \rangle \nonumber\\
  &&+ \frac{1}{\sqrt{10}}|C_4 \rangle \otimes |S_5^0 \rangle \nonumber\\
  \begin{tabular}{|c|c|c|c|}
    \hline
    1 & 2 & 4 & 5 \\
    \hline
    3 \\
    \cline{1-1}
    6 \\
    \cline{1-1}
  \end{tabular}_{CS}
  &=&-\frac{1}{\sqrt{10}}|C_1 \rangle \otimes |S_1^0 \rangle - \frac{\sqrt{3}}{\sqrt{10}}|C_3 \rangle \otimes |S_1^0 \rangle -\frac{1}{2\sqrt{5}}|C_1 \rangle \otimes |S_2^0 \rangle +\frac{1}{2\sqrt{5}}|C_2 \rangle \otimes |S_3^0 \rangle +\frac{\sqrt{3}}{\sqrt{10}}|C_5 \rangle \otimes |S_3^0 \rangle \nonumber\\
  && -\frac{1}{2\sqrt{5}}|C_3 \rangle \otimes |S_4^0 \rangle +\frac{1}{2\sqrt{5}}|C_4 \rangle \otimes |S_5^0 \rangle -\frac{1}{\sqrt{10}}|C_5 \rangle \otimes |S_5^0 \rangle \nonumber\\
   \begin{tabular}{|c|c|c|c|}
    \hline
    1 & 3 & 4 & 5 \\
    \hline
    2 \\
    \cline{1-1}
    6 \\
    \cline{1-1}
  \end{tabular}_{CS}
  &=&-\frac{1}{\sqrt{10}}|C_2 \rangle \otimes |S_1^0 \rangle - \frac{\sqrt{3}}{\sqrt{10}}|C_4 \rangle \otimes |S_1^0 \rangle +\frac{1}{2\sqrt{5}}|C_2 \rangle \otimes |S_2^0 \rangle -\frac{\sqrt{3}}{\sqrt{10}}|C_5 \rangle \otimes |S_2^0 \rangle +\frac{1}{2\sqrt{5}}|C_1 \rangle \otimes |S_3^0 \rangle \nonumber\\
  && +\frac{1}{2\sqrt{5}}|C_4 \rangle \otimes |S_4^0 \rangle +\frac{1}{\sqrt{10}}|C_5 \rangle \otimes |S_4^0 \rangle +\frac{1}{2\sqrt{5}}|C_3 \rangle \otimes |S_5^0 \rangle \nonumber\\
  \begin{tabular}{|c|c|c|c|}
    \hline
    1 & 2 & 3 & 6 \\
    \hline
    4 \\
    \cline{1-1}
    5 \\
    \cline{1-1}
  \end{tabular}_{CS}
  &=&-\frac{\sqrt{3}}{2\sqrt{5}}|C_1 \rangle \otimes |S_2^0 \rangle - \frac{1}{\sqrt{5}}|C_3 \rangle \otimes |S_2^0 \rangle -\frac{\sqrt{3}}{2\sqrt{5}}|C_2 \rangle \otimes |S_3^0 \rangle -\frac{1}{\sqrt{5}}|C_4 \rangle \otimes |S_3^0 \rangle +\frac{\sqrt{3}}{2\sqrt{5}}|C_3 \rangle \otimes |S_4^0 \rangle \nonumber\\
  && +\frac{\sqrt{3}}{2\sqrt{5}}|C_4 \rangle \otimes |S_5^0 \rangle \nonumber\\
  \begin{tabular}{|c|c|c|c|}
    \hline
    1 & 2 & 4 & 6 \\
    \hline
    3 \\
    \cline{1-1}
    5 \\
    \cline{1-1}
  \end{tabular}_{CS}
  &=&-\frac{\sqrt{3}}{2\sqrt{5}}|C_1 \rangle \otimes |S_1^0 \rangle +\frac{1}{\sqrt{5}}|C_3 \rangle \otimes |S_1^0 \rangle -\frac{\sqrt{3}}{2\sqrt{10}}|C_1 \rangle \otimes |S_2^0 \rangle +\frac{\sqrt{3}}{2\sqrt{10}}|C_2 \rangle \otimes |S_3^0 \rangle -\frac{1}{\sqrt{5}}|C_5 \rangle \otimes |S_3^0 \rangle \nonumber\\
  && -\frac{\sqrt{3}}{2\sqrt{10}}|C_3 \rangle \otimes |S_4^0 \rangle +\frac{\sqrt{3}}{2\sqrt{10}}|C_4 \rangle \otimes |S_5^0 \rangle -\frac{\sqrt{3}}{2\sqrt{5}}|C_5 \rangle \otimes |S_5^0 \rangle \nonumber\\
  \begin{tabular}{|c|c|c|c|}
    \hline
    1 & 3 & 4 & 6 \\
    \hline
    2 \\
    \cline{1-1}
    5 \\
    \cline{1-1}
  \end{tabular}_{CS}
  &=&-\frac{\sqrt{3}}{2\sqrt{5}}|C_2 \rangle \otimes |S_1^0 \rangle +\frac{1}{\sqrt{5}}|C_4 \rangle \otimes |S_1^0 \rangle +\frac{\sqrt{3}}{2\sqrt{10}}|C_2 \rangle \otimes |S_2^0 \rangle +\frac{1}{\sqrt{5}}|C_5 \rangle \otimes |S_2^0 \rangle +\frac{\sqrt{3}}{2\sqrt{10}}|C_1 \rangle \otimes |S_3^0 \rangle \nonumber\\
  && +\frac{\sqrt{3}}{2\sqrt{10}}|C_4 \rangle \otimes |S_4^0 \rangle +\frac{\sqrt{3}}{2\sqrt{5}}|C_5 \rangle \otimes |S_4^0 \rangle +\frac{\sqrt{3}}{2\sqrt{10}}|C_3 \rangle \otimes |S_5^0 \rangle \nonumber\\
  \begin{tabular}{|c|c|c|c|}
    \hline
    1 & 2 & 5 & 6 \\
    \hline
    3 \\
    \cline{1-1}
    4 \\
    \cline{1-1}
  \end{tabular}_{CS}
  &=&\frac{1}{2}|C_1 \rangle \otimes |S_1^0 \rangle - \frac{1}{2\sqrt{2}}|C_1 \rangle \otimes |S_2^0 \rangle +\frac{1}{2\sqrt{2}}|C_2 \rangle \otimes |S_3^0 \rangle -\frac{1}{2\sqrt{2}}|C_3 \rangle \otimes |S_4^0 \rangle +\frac{1}{2\sqrt{2}}|C_4 \rangle \otimes |S_5^0 \rangle \nonumber\\
  && +\frac{1}{2}|C_5 \rangle \otimes |S_5^0 \rangle \nonumber\\
  \begin{tabular}{|c|c|c|c|}
    \hline
    1 & 3 & 5 & 6 \\
    \hline
    2 \\
    \cline{1-1}
    4 \\
    \cline{1-1}
  \end{tabular}_{CS}
  &=&\frac{1}{2}|C_2 \rangle \otimes |S_1^0 \rangle +\frac{1}{2\sqrt{2}}|C_2 \rangle \otimes |S_2^0 \rangle +\frac{1}{2\sqrt{2}}|C_1 \rangle \otimes |S_3^0 \rangle +\frac{1}{2\sqrt{2}}|C_4 \rangle \otimes |S_4^0 \rangle -\frac{1}{2}|C_5 \rangle \otimes |S_4^0 \rangle \nonumber\\
  && +\frac{1}{2\sqrt{2}}|C_3 \rangle \otimes |S_5^0 \rangle \nonumber\\
  \begin{tabular}{|c|c|c|c|}
    \hline
    1 & 4 & 5 & 6 \\
    \hline
    2 \\
    \cline{1-1}
    3 \\
    \cline{1-1}
  \end{tabular}_{CS}
  &=&-\frac{1}{2}|C_2 \rangle \otimes |S_2^0 \rangle +\frac{1}{3}|C_1 \rangle \otimes |S_3^0 \rangle -\frac{1}{2}|C_4 \rangle \otimes |S_4^0 \rangle +\frac{1}{2}|C_3 \rangle \otimes |S_5^0 \rangle \nonumber
\end{eqnarray}
\begin{eqnarray}
  \begin{tabular}{|c|c|}
    \hline
    1 & 2 \\
    \hline
    3 & 4 \\
    \hline
    5 \\
    \cline{1-1}
    6 \\
    \cline{1-1}
  \end{tabular}_{CS}
  &=&-\frac{1}{2\sqrt{3}}|C_3 \rangle \otimes |S_1^0 \rangle +\frac{1}{2\sqrt{6}}|C_3 \rangle \otimes |S_2^0 \rangle -\frac{1}{2\sqrt{6}}|C_4 \rangle \otimes |S_3^0 \rangle -\frac{1}{2\sqrt{3}}|C_5 \rangle \otimes |S_3^0 \rangle -\frac{\sqrt{3}}{2\sqrt{2}}|C_1 \rangle \otimes |S_4^0 \rangle \nonumber\\
  &&+ \frac{\sqrt{3}}{2\sqrt{2}}|C_2 \rangle \otimes |S_5^0 \rangle\nonumber\\
  \begin{tabular}{|c|c|}
    \hline
    1 & 3 \\
    \hline
    2 & 4 \\
    \hline
    5 \\
    \cline{1-1}
    6 \\
    \cline{1-1}
  \end{tabular}_{CS}
  &=&-\frac{1}{2\sqrt{3}}|C_4 \rangle \otimes |S_1^0 \rangle -\frac{1}{2\sqrt{6}}|C_4 \rangle \otimes |S_2^0 \rangle +\frac{1}{2\sqrt{3}}|C_5 \rangle \otimes |S_2^0 \rangle -\frac{1}{2\sqrt{6}}|C_3 \rangle \otimes |S_3^0 \rangle +\frac{\sqrt{3}}{2\sqrt{2}}|C_2 \rangle \otimes |S_4^0 \rangle \nonumber\\
  &&+\frac{\sqrt{3}}{2\sqrt{2}}|C_1 \rangle \otimes |S_5^0 \rangle\nonumber\\
  \begin{tabular}{|c|c|}
    \hline
    1 & 2 \\
    \hline
    3 & 5 \\
    \hline
    4 \\
    \cline{1-1}
    6 \\
    \cline{1-1}
  \end{tabular}_{CS}
  &=&-\frac{1}{2\sqrt{3}}|C_1 \rangle \otimes |S_1^0 \rangle -\frac{1}{3}|C_3 \rangle \otimes |S_1^0 \rangle +\frac{1}{2\sqrt{6}}|C_1 \rangle \otimes |S_2^0 \rangle -\frac{\sqrt{2}}{3}|C_3 \rangle \otimes |S_2^0 \rangle -\frac{1}{2\sqrt{6}}|C_2 \rangle \otimes |S_3^0 \rangle \nonumber\\
  &&+\frac{\sqrt{2}}{3}|C_4 \rangle \otimes |S_3^0 \rangle -\frac{1}{3}|C_5 \rangle \otimes |S_3^0 -\frac{1}{2\sqrt{6}}|C_3 \rangle \otimes |S_4^0 \rangle \rangle +\frac{1}{2\sqrt{6}}|C_4 \rangle \otimes |S_5^0 \rangle +\frac{1}{2\sqrt{3}}|C_5 \rangle \otimes |S_5^0 \rangle \nonumber\\
  \begin{tabular}{|c|c|}
    \hline
    1 & 3 \\
    \hline
    2 & 5 \\
    \hline
    4 \\
    \cline{1-1}
    6 \\
    \cline{1-1}
  \end{tabular}_{CS}
  &=&-\frac{1}{2\sqrt{3}}|C_2 \rangle \otimes |S_1^0 \rangle -\frac{1}{3}|C_4 \rangle \otimes |S_1^0 \rangle -\frac{1}{2\sqrt{6}}|C_2 \rangle \otimes |S_2^0 \rangle +\frac{\sqrt{2}}{3}|C_4 \rangle \otimes |S_2^0 \rangle +\frac{1}{3}|C_5 \rangle \otimes |S_2^0 \rangle \nonumber\\
  &&-\frac{1}{2\sqrt{6}}|C_1 \rangle \otimes |S_3^0 \rangle +\frac{\sqrt{2}}{3}|C_3 \rangle \otimes |S_3^0 +\frac{1}{2\sqrt{6}}|C_4 \rangle \otimes |S_4^0 \rangle \rangle -\frac{1}{2\sqrt{3}}|C_5 \rangle \otimes |S_4^0 \rangle +\frac{1}{2\sqrt{6}}|C_3 \rangle \otimes |S_5^0 \rangle \nonumber\\
  \begin{tabular}{|c|c|}
    \hline
    1 & 4 \\
    \hline
    2 & 5 \\
    \hline
    3 \\
    \cline{1-1}
    6 \\
    \cline{1-1}
  \end{tabular}_{CS}
  &=&\frac{2}{3}|C_5 \rangle \otimes |S_1^0 \rangle +\frac{1}{2\sqrt{3}}|C_2 \rangle \otimes |S_2^0 \rangle +\frac{1}{3}|C_4 \rangle \otimes |S_2^0 \rangle -\frac{1}{2\sqrt{3}}|C_1 \rangle \otimes |S_3^0 \rangle -\frac{1}{3}|C_3 \rangle \otimes |S_3^0 \rangle \nonumber\\
  &&-\frac{1}{2\sqrt{3}}|C_4 \rangle \otimes |S_4^0 \rangle +\frac{1}{2\sqrt{3}}|C_3 \rangle \otimes |S_5^0 \rangle \nonumber\\
  \begin{tabular}{|c|c|}
    \hline
    1 & 2 \\
    \hline
    3 & 6 \\
    \hline
    4 \\
    \cline{1-1}
    5 \\
    \cline{1-1}
  \end{tabular}_{CS}
  &=&-\frac{1}{\sqrt{6}}|C_1 \rangle \otimes |S_1^0 \rangle +\frac{1}{3\sqrt{2}}|C_3 \rangle \otimes |S_1^0 \rangle +\frac{1}{2\sqrt{3}}|C_1 \rangle \otimes |S_2^0 \rangle +\frac{1}{3}|C_3 \rangle \otimes |S_2^0 \rangle -\frac{1}{2\sqrt{3}}|C_2 \rangle \otimes |S_3^0 \rangle \nonumber\\
  &&-\frac{1}{3}|C_4 \rangle \otimes |S_3^0 \rangle +\frac{1}{3\sqrt{2}}|C_5 \rangle \otimes |S_3^0 \rangle -\frac{1}{2\sqrt{3}}|C_3 \rangle \otimes |S_4^0 \rangle +\frac{1}{2\sqrt{3}}|C_4 \rangle \otimes |S_5^0 \rangle +\frac{1}{\sqrt{6}}|C_5 \rangle \otimes |S_5^0 \rangle \nonumber\\
  \begin{tabular}{|c|c|}
    \hline
    1 & 3 \\
    \hline
    2 & 6 \\
    \hline
    4 \\
    \cline{1-1}
    5 \\
    \cline{1-1}
  \end{tabular}_{CS}
  &=&-\frac{1}{\sqrt{6}}|C_2 \rangle \otimes |S_1^0 \rangle +\frac{1}{3\sqrt{2}}|C_4 \rangle \otimes |S_1^0 \rangle -\frac{1}{2\sqrt{3}}|C_2 \rangle \otimes |S_2^0 \rangle -\frac{1}{3}|C_4 \rangle \otimes |S_2^0 \rangle -\frac{1}{3\sqrt{2}}|C_5 \rangle \otimes |S_2^0 \rangle \nonumber\\
  &&-\frac{1}{2\sqrt{3}}|C_1 \rangle \otimes |S_3^0 \rangle -\frac{1}{3}|C_3 \rangle \otimes |S_3^0 \rangle +\frac{1}{2\sqrt{3}}|C_4 \rangle \otimes |S_4^0 \rangle -\frac{1}{\sqrt{6}}|C_5 \rangle \otimes |S_4^0 \rangle +\frac{1}{2\sqrt{3}}|C_3 \rangle \otimes |S_5^0 \rangle \nonumber\\
  \begin{tabular}{|c|c|}
    \hline
    1 & 4 \\
    \hline
    2 & 6 \\
    \hline
    3 \\
    \cline{1-1}
    5 \\
    \cline{1-1}
  \end{tabular}_{CS}
  &=&-\frac{\sqrt{2}}{3}|C_5 \rangle \otimes |S_1^0 \rangle +\frac{1}{\sqrt{6}}|C_2 \rangle \otimes |S_2^0 \rangle -\frac{1}{3\sqrt{2}}|C_4 \rangle \otimes |S_2^0 \rangle -\frac{1}{\sqrt{6}}|C_1 \rangle \otimes |S_3^0 \rangle +\frac{1}{3\sqrt{2}}|C_3 \rangle \otimes |S_3^0 \rangle \nonumber\\
  &&-\frac{1}{\sqrt{6}}|C_4 \rangle \otimes |S_4^0 \rangle +\frac{1}{\sqrt{6}}|C_3 \rangle \otimes |S_5^0 \rangle \nonumber\\
  \begin{tabular}{|c|c|}
    \hline
    1 & 5 \\
    \hline
    2 & 6 \\
    \hline
    3 \\
    \cline{1-1}
    4 \\
    \cline{1-1}
  \end{tabular}_{CS}
  &=&\frac{\sqrt{2}}{\sqrt{15}}|C_5 \rangle \otimes |S_1^0 \rangle -\frac{\sqrt{2}}{\sqrt{15}}|C_4 \rangle \otimes |S_2^0 \rangle +\frac{\sqrt{2}}{\sqrt{15}}|C_3 \rangle \otimes |S_3^0 \rangle -\frac{\sqrt{3}}{\sqrt{10}}|C_2 \rangle \otimes |S_4^0 \rangle +\frac{\sqrt{3}}{\sqrt{10}}|C_1 \rangle \otimes |S_5^0 \rangle \nonumber
\end{eqnarray}
\subsection{S=1}
\begin{eqnarray}
  \begin{tabular}{|c|c|c|}
    \hline
    1 & 2 & 3 \\
    \hline
    4 & 5 \\
    \cline{1-2}
    6 \\
    \cline{1-1}
  \end{tabular}_{CS}
  &=&-\frac{1}{2\sqrt{10}}|C_1 \rangle \otimes |S_3^1 \rangle -\frac{\sqrt{3}}{2\sqrt{10}}|C_3 \rangle \otimes |S_3^1 \rangle -\frac{1}{2\sqrt{10}}|C_2 \rangle \otimes |S_4^1 \rangle -\frac{\sqrt{3}}{2\sqrt{10}}|C_4 \rangle \otimes |S_4^1 \rangle +\frac{1}{\sqrt{5}}|C_1 \rangle \otimes |S_6^1 \rangle \nonumber\\
  &&+\frac{1}{\sqrt{5}}|C_2 \rangle \otimes |S_7^1 \rangle +\frac{1}{\sqrt{5}}|C_3 \rangle \otimes |S_8^1 \rangle +\frac{1}{\sqrt{5}}|C_4 \rangle \otimes |S_9^1 \rangle \nonumber\\
  \begin{tabular}{|c|c|c|}
    \hline
    1 & 2 & 4 \\
    \hline
    3 & 5 \\
    \cline{1-2}
    6 \\
    \cline{1-1}
  \end{tabular}_{CS}
  &=&-\frac{1}{2\sqrt{10}}|C_1 \rangle \otimes |S_2^1 \rangle +\frac{\sqrt{3}}{2\sqrt{10}}|C_3 \rangle \otimes |S_2^1 \rangle -\frac{1}{4\sqrt{5}}|C_1 \rangle \otimes |S_3^1 \rangle +\frac{1}{4\sqrt{5}}|C_2 \rangle \otimes |S_4^1 \rangle -\frac{\sqrt{3}}{2\sqrt{10}}|C_5 \rangle \otimes |S_4^1 \rangle \nonumber\\
  &&+\frac{1}{\sqrt{5}}|C_1 \rangle \otimes |S_5^1 \rangle +\frac{1}{\sqrt{10}}|C_1 \rangle \otimes |S_6^1 \rangle -\frac{1}{\sqrt{10}}|C_2 \rangle \otimes |S_7^1 \rangle -\frac{1}{\sqrt{10}}|C_3 \rangle \otimes |S_8^1 \rangle +\frac{1}{\sqrt{10}}|C_4 \rangle \otimes |S_9^1 \rangle \nonumber\\
  &&-\frac{1}{\sqrt{5}}|C_5 \rangle \otimes |S_9^1 \rangle \nonumber\\
  \begin{tabular}{|c|c|c|}
    \hline
    1 & 3 & 4 \\
    \hline
    2 & 5 \\
    \cline{1-2}
    6 \\
    \cline{1-1}
  \end{tabular}_{CS}
  &=&-\frac{1}{2\sqrt{10}}|C_2 \rangle \otimes |S_2^1 \rangle +\frac{\sqrt{3}}{2\sqrt{10}}|C_4 \rangle \otimes |S_2^1 \rangle +\frac{1}{4\sqrt{5}}|C_2 \rangle \otimes |S_3^1 \rangle +\frac{\sqrt{3}}{2\sqrt{10}}|C_5 \rangle \otimes |S_3^1 \rangle +\frac{1}{4\sqrt{5}}|C_1 \rangle \otimes |S_4^1 \rangle \nonumber\\
  &&+\frac{1}{\sqrt{5}}|C_2 \rangle \otimes |S_5^1 \rangle -\frac{1}{\sqrt{10}}|C_2 \rangle \otimes |S_6^1 \rangle -\frac{1}{\sqrt{10}}|C_1 \rangle \otimes |S_7^1 \rangle +\frac{1}{\sqrt{10}}|C_4 \rangle \otimes |S_8^1 \rangle +\frac{1}{\sqrt{5}}|C_5 \rangle \otimes |S_8^1 \rangle \nonumber\\
  &&+\frac{1}{\sqrt{10}}|C_3 \rangle \otimes |S_9^1 \rangle \nonumber\\
  \begin{tabular}{|c|c|c|}
    \hline
    1 & 2 & 5 \\
    \hline
    3 & 4 \\
    \cline{1-2}
    6 \\
    \cline{1-1}
  \end{tabular}_{CS}
  &=&\frac{1}{2\sqrt{2}}|C_1 \rangle \otimes |S_1^1 \rangle +\frac{1}{2\sqrt{10}}|C_3 \rangle \otimes |S_2^1 \rangle -\frac{1}{4\sqrt{5}}|C_3 \rangle \otimes |S_3^1 \rangle +\frac{1}{4\sqrt{5}}|C_4 \rangle \otimes |S_4^1 \rangle +\frac{1}{2\sqrt{10}}|C_5 \rangle \otimes |S_4^1 \rangle \nonumber\\
  &&+\frac{1}{\sqrt{5}}|C_3 \rangle \otimes |S_5^1 \rangle -\frac{1}{\sqrt{10}}|C_3 \rangle \otimes |S_6^1 \rangle +\frac{1}{\sqrt{10}}|C_4 \rangle \otimes |S_7^1 \rangle +\frac{1}{\sqrt{5}}|C_5 \rangle \otimes |S_7^1 \rangle -\frac{1}{\sqrt{10}}|C_1 \rangle \otimes |S_8^1 \rangle \nonumber\\
  &&+\frac{1}{\sqrt{10}}|C_2 \rangle \otimes |S_9^1 \rangle \nonumber\\
  \begin{tabular}{|c|c|c|}
    \hline
    1 & 3 & 5 \\
    \hline
    2 & 4 \\
    \cline{1-2}
    6 \\
    \cline{1-1}
  \end{tabular}_{CS}
  &=&\frac{1}{2\sqrt{2}}|C_2 \rangle \otimes |S_1^1 \rangle +\frac{1}{2\sqrt{10}}|C_4 \rangle \otimes |S_2^1 \rangle +\frac{1}{4\sqrt{5}}|C_4 \rangle \otimes |S_3^1 \rangle -\frac{1}{2\sqrt{10}}|C_5 \rangle \otimes |S_3^1 \rangle +\frac{1}{4\sqrt{5}}|C_3 \rangle \otimes |S_4^1 \rangle \nonumber\\
  &&+\frac{1}{\sqrt{5}}|C_4 \rangle \otimes |S_5^1 \rangle +\frac{1}{\sqrt{10}}|C_4 \rangle \otimes |S_6^1 \rangle -\frac{1}{\sqrt{5}}|C_5 \rangle \otimes |S_6^1 \rangle +\frac{1}{\sqrt{10}}|C_3 \rangle \otimes |S_7^1 \rangle +\frac{1}{\sqrt{10}}|C_2 \rangle \otimes |S_8^1 \rangle \nonumber\\
  &&+\frac{1}{\sqrt{10}}|C_1 \rangle \otimes |S_9^1 \rangle \nonumber\\
  \begin{tabular}{|c|c|c|}
    \hline
    1 & 2 & 3 \\
    \hline
    4 & 6 \\
    \cline{1-2}
    5 \\
    \cline{1-1}
  \end{tabular}_{CS}
  &=&\frac{\sqrt{5}}{2\sqrt{6}}|C_1 \rangle \otimes |S_3^1 \rangle -\frac{\sqrt{5}}{6\sqrt{2}}|C_3 \rangle \otimes |S_3^1 \rangle +\frac{\sqrt{5}}{2\sqrt{6}}|C_2 \rangle \otimes |S_4^1 \rangle -\frac{\sqrt{5}}{6\sqrt{2}}|C_4 \rangle \otimes |S_4^1 \rangle +\frac{1}{\sqrt{15}}|C_1 \rangle \otimes |S_6^1 \rangle \nonumber\\
  &&+\frac{2}{3\sqrt{5}}|C_3 \rangle \otimes |S_6^1 \rangle +\frac{1}{\sqrt{15}}|C_2 \rangle \otimes |S_6^1 \rangle +\frac{2}{3\sqrt{5}}|C_4 \rangle \otimes |S_7^1 \rangle -\frac{1}{\sqrt{15}}|C_3 \rangle \otimes |S_8^1 \rangle -\frac{1}{\sqrt{15}}|C_4 \rangle \otimes |S_9^1 \rangle \nonumber\\
  \begin{tabular}{|c|c|c|}
    \hline
    1 & 2 & 4 \\
    \hline
    3 & 6 \\
    \cline{1-2}
    5 \\
    \cline{1-1}
  \end{tabular}_{CS}
  &=&\frac{\sqrt{5}}{2\sqrt{6}}|C_1 \rangle \otimes |S_2^1 \rangle +\frac{\sqrt{5}}{6\sqrt{2}}|C_3 \rangle \otimes |S_2^1 \rangle +\frac{\sqrt{5}}{4\sqrt{3}}|C_1 \rangle \otimes |S_3^1 \rangle -\frac{\sqrt{5}}{4\sqrt{3}}|C_2 \rangle \otimes |S_4^1 \rangle -\frac{\sqrt{5}}{6\sqrt{2}}|C_5 \rangle \otimes |S_4^1 \rangle \nonumber\\
  &&+\frac{1}{\sqrt{15}}|C_1 \rangle \otimes |S_5^1 \rangle -\frac{2}{3\sqrt{5}}|C_3 \rangle \otimes |S_5^1 \rangle +\frac{1}{\sqrt{30}}|C_1 \rangle \otimes |S_6^1 \rangle -\frac{1}{\sqrt{30}}|C_2 \rangle \otimes |S_7^1 \rangle +\frac{2}{3\sqrt{5}}|C_5 \rangle \otimes |S_7^1 \rangle \nonumber\\
  &&+\frac{1}{\sqrt{30}}|C_3 \rangle \otimes |S_8^1 \rangle -\frac{1}{\sqrt{30}}|C_4 \rangle \otimes |S_9^1 \rangle +\frac{1}{\sqrt{15}}|C_5 \rangle \otimes |S_9^1 \rangle \nonumber\\
  \begin{tabular}{|c|c|c|}
    \hline
    1 & 3 & 4 \\
    \hline
    2 & 6 \\
    \cline{1-2}
    5 \\
    \cline{1-1}
  \end{tabular}_{CS}
  &=&\frac{\sqrt{5}}{2\sqrt{6}}|C_2 \rangle \otimes |S_2^1 \rangle +\frac{\sqrt{5}}{6\sqrt{2}}|C_4 \rangle \otimes |S_2^1 \rangle -\frac{\sqrt{5}}{4\sqrt{3}}|C_2 \rangle \otimes |S_3^1 \rangle +\frac{\sqrt{5}}{6\sqrt{2}}|C_5 \rangle \otimes |S_3^1 \rangle -\frac{\sqrt{5}}{4\sqrt{3}}|C_1 \rangle \otimes |S_4^1 \rangle \nonumber\\
  &&+\frac{1}{\sqrt{15}}|C_2 \rangle \otimes |S_5^1 \rangle -\frac{2}{3\sqrt{5}}|C_4 \rangle \otimes |S_5^1 \rangle -\frac{1}{\sqrt{30}}|C_2 \rangle \otimes |S_6^1 \rangle -\frac{2}{3\sqrt{5}}|C_5 \rangle \otimes |S_6^1 \rangle -\frac{1}{\sqrt{30}}|C_1 \rangle \otimes |S_7^1 \rangle \nonumber\\
  &&-\frac{1}{\sqrt{30}}|C_4 \rangle \otimes |S_8^1 \rangle -\frac{1}{\sqrt{15}}|C_5 \rangle \otimes |S_8^1 \rangle -\frac{1}{\sqrt{30}}|C_3 \rangle \otimes |S_9^1 \rangle \nonumber\\
  \begin{tabular}{|c|c|c|}
    \hline
    1 & 2 & 5 \\
    \hline
    3 & 6 \\
    \cline{1-2}
    4 \\
    \cline{1-1}
  \end{tabular}_{CS}
  &=&\frac{\sqrt{5}}{3\sqrt{2}}|C_3 \rangle \otimes |S_1^1 \rangle +\frac{1}{6\sqrt{2}}|C_1 \rangle \otimes |S_2^1 \rangle +\frac{1}{2\sqrt{6}}|C_3 \rangle \otimes |S_2^1 \rangle -\frac{1}{12}|C_1 \rangle \otimes |S_3^1 \rangle +\frac{1}{2\sqrt{3}}|C_3 \rangle \otimes |S_3^1 \rangle \nonumber\\
  &&+\frac{1}{12}|C_2 \rangle \otimes |S_4^1 \rangle -\frac{1}{2\sqrt{3}}|C_4 \rangle \otimes |S_4^1 \rangle +\frac{1}{2\sqrt{6}}|C_5 \rangle \otimes |S_4^1 \rangle -\frac{1}{3}|C_1 \rangle \otimes |S_5^1 \rangle +\frac{1}{3\sqrt{2}}|C_1 \rangle \otimes |S_6^1 \rangle \nonumber\\
  &&-\frac{1}{3\sqrt{2}}|C_2 \rangle \otimes |S_7^1 \rangle +\frac{1}{3\sqrt{2}}|C_3 \rangle \otimes |S_8^1 \rangle -\frac{1}{3\sqrt{2}}|C_4 \rangle \otimes |S_9^1 \rangle -\frac{1}{3}|C_5 \rangle \otimes |S_9^1 \rangle \nonumber\\
  \begin{tabular}{|c|c|c|}
    \hline
    1 & 3 & 5 \\
    \hline
    2 & 6 \\
    \cline{1-2}
    4 \\
    \cline{1-1}
  \end{tabular}_{CS}
  &=&\frac{\sqrt{5}}{3\sqrt{2}}|C_4 \rangle \otimes |S_1^1 \rangle +\frac{1}{6\sqrt{2}}|C_2 \rangle \otimes |S_2^1 \rangle +\frac{1}{2\sqrt{6}}|C_4 \rangle \otimes |S_2^1 \rangle +\frac{1}{12}|C_2 \rangle \otimes |S_3^1 \rangle -\frac{1}{2\sqrt{3}}|C_4 \rangle \otimes |S_3^1 \rangle \nonumber\\
  &&-\frac{1}{2\sqrt{6}}|C_5 \rangle \otimes |S_3^1 \rangle +\frac{1}{12}|C_1 \rangle \otimes |S_4^1 \rangle -\frac{1}{2\sqrt{3}}|C_3 \rangle \otimes |S_4^1 \rangle -\frac{1}{3}|C_2 \rangle \otimes |S_5^1 \rangle -\frac{1}{3\sqrt{2}}|C_2 \rangle \otimes |S_6^1 \rangle \nonumber\\
  &&-\frac{1}{3\sqrt{2}}|C_1 \rangle \otimes |S_7^1 \rangle -\frac{1}{3\sqrt{2}}|C_4 \rangle \otimes |S_8^1 \rangle +\frac{1}{3}|C_5 \rangle \otimes |S_8^1 \rangle -\frac{1}{3\sqrt{2}}|C_3 \rangle \otimes |S_9^1 \rangle \nonumber\\
  \begin{tabular}{|c|c|c|}
    \hline
    1 & 4 & 5 \\
    \hline
    2 & 6 \\
    \cline{1-2}
    3 \\
    \cline{1-1}
  \end{tabular}_{CS}
  &=&\frac{\sqrt{5}}{3\sqrt{2}}|C_5 \rangle \otimes |S_1^1 \rangle -\frac{1}{\sqrt{6}}|C_5 \rangle \otimes |S_2^1 \rangle -\frac{1}{6\sqrt{2}}|C_2 \rangle \otimes |S_3^1 \rangle -\frac{1}{2\sqrt{6}}|C_4 \rangle \otimes |S_3^1 \rangle +\frac{1}{6\sqrt{2}}|C_1 \rangle \otimes |S_4^1 \rangle \nonumber\\
  &&+\frac{1}{2\sqrt{6}}|C_3 \rangle \otimes |S_4^1 \rangle +\frac{1}{3}|C_2 \rangle \otimes |S_6^1 \rangle -\frac{1}{3}|C_1 \rangle \otimes |S_7^1 \rangle +\frac{1}{3}|C_4 \rangle \otimes |S_8^1 \rangle -\frac{1}{3}|C_3 \rangle \otimes |S_9^1 \rangle \nonumber\\
  \begin{tabular}{|c|c|c|}
    \hline
    1 & 2 & 6 \\
    \hline
    3 & 4 \\
    \cline{1-2}
    5 \\
    \cline{1-1}
  \end{tabular}_{CS}
  &=&\frac{\sqrt{5}}{2\sqrt{6}}|C_1 \rangle \otimes |S_1^1 \rangle -\frac{5}{6\sqrt{6}}|C_3 \rangle \otimes |S_2^1 \rangle +\frac{5}{12\sqrt{3}}|C_3 \rangle \otimes |S_3^1 \rangle -\frac{5}{12\sqrt{3}}|C_4 \rangle \otimes |S_4^1 \rangle -\frac{5}{6\sqrt{6}}|C_5 \rangle \otimes |S_4^1 \rangle \nonumber\\
  &&-\frac{1}{3\sqrt{3}}|C_3 \rangle \otimes |S_5^1 \rangle +\frac{1}{3\sqrt{6}}|C_3 \rangle \otimes |S_6^1 \rangle -\frac{1}{3\sqrt{6}}|C_4 \rangle \otimes |S_7^1 \rangle -\frac{1}{3\sqrt{3}}|C_5 \rangle \otimes |S_7^1 \rangle -\frac{1}{\sqrt{6}}|C_1 \rangle \otimes |S_8^1 \rangle \nonumber\\
  &&+\frac{1}{\sqrt{6}}|C_2 \rangle \otimes |S_9^1 \rangle \nonumber\\
  \begin{tabular}{|c|c|c|}
    \hline
    1 & 3 & 6 \\
    \hline
    2 & 4 \\
    \cline{1-2}
    5 \\
    \cline{1-1}
  \end{tabular}_{CS}
  &=&\frac{\sqrt{5}}{2\sqrt{6}}|C_2 \rangle \otimes |S_1^1 \rangle -\frac{5}{6\sqrt{6}}|C_4 \rangle \otimes |S_2^1 \rangle -\frac{5}{12\sqrt{3}}|C_4 \rangle \otimes |S_3^1 \rangle +\frac{5}{6\sqrt{6}}|C_5 \rangle \otimes |S_3^1 \rangle -\frac{5}{12\sqrt{3}}|C_3 \rangle \otimes |S_4^1 \rangle \nonumber\\
  &&-\frac{1}{3\sqrt{3}}|C_4 \rangle \otimes |S_5^1 \rangle -\frac{1}{3\sqrt{6}}|C_4 \rangle \otimes |S_6^1 \rangle +\frac{1}{3\sqrt{3}}|C_5 \rangle \otimes |S_6^1 \rangle -\frac{1}{3\sqrt{6}}|C_3 \rangle \otimes |S_7^1 \rangle +\frac{1}{\sqrt{6}}|C_2 \rangle \otimes |S_8^1 \rangle \nonumber\\
  &&+\frac{1}{\sqrt{6}}|C_1 \rangle \otimes |S_9^1 \rangle \nonumber\\
  \begin{tabular}{|c|c|c|}
    \hline
    1 & 2 & 6 \\
    \hline
    3 & 5 \\
    \cline{1-2}
    4 \\
    \cline{1-1}
  \end{tabular}_{CS}
  &=&-\frac{\sqrt{5}}{3\sqrt{6}}|C_3 \rangle \otimes |S_1^1 \rangle -\frac{5}{6\sqrt{6}}|C_1 \rangle \otimes |S_2^1 \rangle +\frac{5}{18\sqrt{2}}|C_3 \rangle \otimes |S_2^1 \rangle +\frac{5}{12\sqrt{3}}|C_1 \rangle \otimes |S_3^1 \rangle +\frac{5}{18}|C_3 \rangle \otimes |S_3^1 \rangle \nonumber\\
  &&-\frac{5}{12\sqrt{3}}|C_2 \rangle \otimes |S_4^1 \rangle -\frac{5}{18}|C_4 \rangle \otimes |S_4^1 \rangle +\frac{5}{18\sqrt{2}}|C_5 \rangle \otimes |S_4^1 \rangle -\frac{1}{3\sqrt{3}}|C_1 \rangle \otimes |S_5^1 \rangle -\frac{2}{9}|C_3 \rangle \otimes |S_5^1 \rangle \nonumber\\
  &&+\frac{1}{3\sqrt{6}}|C_1 \rangle \otimes |S_6^1 \rangle -\frac{2\sqrt{2}}{9}|C_3 \rangle \otimes |S_6^1 \rangle -\frac{1}{3\sqrt{6}}|C_2 \rangle \otimes |S_7^1 \rangle +\frac{2\sqrt{2}}{9}|C_4 \rangle \otimes |S_7^1 \rangle -\frac{2}{9}|C_5 \rangle \otimes |S_7^1 \rangle \nonumber\\
  &&-\frac{1}{3\sqrt{6}}|C_3 \rangle \otimes |S_8^1 \rangle +\frac{1}{3\sqrt{6}}|C_4 \rangle \otimes |S_9^1 \rangle +\frac{1}{3\sqrt{3}}|C_5 \rangle \otimes |S_9^1 \rangle \nonumber\\
  \begin{tabular}{|c|c|c|}
    \hline
    1 & 3 & 6 \\
    \hline
    2 & 5 \\
    \cline{1-2}
    4 \\
    \cline{1-1}
  \end{tabular}_{CS}
  &=&-\frac{\sqrt{5}}{3\sqrt{6}}|C_4 \rangle \otimes |S_1^1 \rangle -\frac{5}{6\sqrt{6}}|C_2 \rangle \otimes |S_2^1 \rangle +\frac{5}{18\sqrt{2}}|C_4 \rangle \otimes |S_2^1 \rangle -\frac{5}{12\sqrt{3}}|C_2 \rangle \otimes |S_3^1 \rangle -\frac{5}{18}|C_4 \rangle \otimes |S_3^1 \rangle \nonumber\\
  &&-\frac{5}{18\sqrt{2}}|C_5 \rangle \otimes |S_3^1 \rangle -\frac{5}{12\sqrt{3}}|C_1 \rangle \otimes |S_4^1 \rangle -\frac{5}{18}|C_3 \rangle \otimes |S_4^1 \rangle -\frac{1}{3\sqrt{3}}|C_2 \rangle \otimes |S_5^1 \rangle -\frac{2}{9}|C_4 \rangle \otimes |S_5^1 \rangle \nonumber\\
  &&-\frac{1}{3\sqrt{6}}|C_2 \rangle \otimes |S_6^1 \rangle +\frac{2\sqrt{2}}{9}|C_4 \rangle \otimes |S_6^1 \rangle +\frac{2}{9}|C_5 \rangle \otimes |S_6^1 \rangle -\frac{1}{3\sqrt{6}}|C_1 \rangle \otimes |S_7^1 \rangle +\frac{2\sqrt{2}}{9}|C_3 \rangle \otimes |S_7^1 \rangle \nonumber\\
  &&+\frac{1}{3\sqrt{6}}|C_4 \rangle \otimes |S_8^1 \rangle -\frac{1}{3\sqrt{3}}|C_5 \rangle \otimes |S_8^1 \rangle +\frac{1}{3\sqrt{6}}|C_3 \rangle \otimes |S_9^1 \rangle \nonumber\\
  \begin{tabular}{|c|c|c|}
    \hline
    1 & 4 & 6 \\
    \hline
    2 & 5 \\
    \cline{1-2}
    3 \\
    \cline{1-1}
  \end{tabular}_{CS}
  &=&-\frac{\sqrt{5}}{3\sqrt{6}}|C_5 \rangle \otimes |S_1^1 \rangle -\frac{5}{9\sqrt{2}}|C_5 \rangle \otimes |S_2^1 \rangle +\frac{5}{6\sqrt{6}}|C_2 \rangle \otimes |S_3^1 \rangle -\frac{5}{18\sqrt{2}}|C_4 \rangle \otimes |S_3^1 \rangle -\frac{5}{6\sqrt{6}}|C_1 \rangle \otimes |S_4^1 \rangle \nonumber\\
  &&+\frac{5}{18\sqrt{2}}|C_3 \rangle \otimes |S_4^1 \rangle +\frac{4}{9}|C_5 \rangle \otimes |S_5^1 \rangle +\frac{1}{3\sqrt{3}}|C_2 \rangle \otimes |S_6^1 \rangle +\frac{2}{9}|C_4 \rangle \otimes |S_6^1 \rangle -\frac{1}{3\sqrt{3}}|C_1 \rangle \otimes |S_7^1 \rangle \nonumber\\
  &&-\frac{2}{9}|C_3 \rangle \otimes |S_7^1 \rangle -\frac{1}{3\sqrt{3}}|C_4 \rangle \otimes |S_8^1 \rangle +\frac{1}{3\sqrt{3}}|C_3 \rangle \otimes |S_9^1 \rangle \nonumber
\end{eqnarray}
\begin{eqnarray}
  \begin{tabular}{|c|c|c|}
    \hline
    1 & 2 & 3 \\
    \hline
    4 \\
    \cline{1-1}
    5 \\
    \cline{1-1}
    6 \\
    \cline{1-1}
  \end{tabular}_{CS}
  &=&\frac{1}{\sqrt{6}}|C_1 \rangle \otimes |S_3^1 \rangle -\frac{1}{3\sqrt{2}}|C_3 \rangle \otimes |S_3^1 \rangle +\frac{1}{\sqrt{6}}|C_2 \rangle \otimes |S_4^1 \rangle -\frac{1}{3\sqrt{2}}|C_4 \rangle \otimes |S_4^1 \rangle -\frac{1}{2\sqrt{3}}|C_1 \rangle \otimes |S_6^1 \rangle \nonumber\\
  &&-\frac{1}{3}|C_3 \rangle \otimes |S_6^1 \rangle -\frac{1}{2\sqrt{3}}|C_2 \rangle \otimes |S_7^1 \rangle -\frac{1}{3}|C_4 \rangle \otimes |S_7^1 \rangle +\frac{1}{2\sqrt{3}}|C_3 \rangle \otimes |S_8^1 \rangle +\frac{1}{2\sqrt{3}}|C_4 \rangle \otimes |S_9^1 \rangle \nonumber\\
  \begin{tabular}{|c|c|c|}
    \hline
    1 & 2 & 4 \\
    \hline
    3 \\
    \cline{1-1}
    5 \\
    \cline{1-1}
    6 \\
    \cline{1-1}
  \end{tabular}_{CS}
  &=&\frac{1}{\sqrt{6}}|C_1 \rangle \otimes |S_2^1 \rangle +\frac{1}{3\sqrt{2}}|C_3 \rangle \otimes |S_2^1 \rangle +\frac{1}{2\sqrt{3}}|C_1 \rangle \otimes |S_3^1 \rangle -\frac{1}{2\sqrt{3}}|C_2 \rangle \otimes |S_4^1 \rangle -\frac{1}{3\sqrt{2}}|C_5 \rangle \otimes |S_4^1 \rangle \nonumber\\
  &&-\frac{1}{2\sqrt{3}}|C_1 \rangle \otimes |S_5^1 \rangle +\frac{1}{3}|C_3 \rangle \otimes |S_5^1 \rangle -\frac{1}{2\sqrt{6}}|C_1 \rangle \otimes |S_6^1 \rangle +\frac{1}{2\sqrt{6}}|C_2 \rangle \otimes |S_7^1 \rangle -\frac{1}{3}|C_5 \rangle \otimes |S_7^1 \rangle \nonumber\\
  &&-\frac{1}{2\sqrt{6}}|C_3 \rangle \otimes |S_8^1 \rangle +\frac{1}{2\sqrt{6}}|C_4 \rangle \otimes |S_9^1 \rangle -\frac{1}{2\sqrt{3}}|C_5 \rangle \otimes |S_9^1 \rangle \nonumber\\
  \begin{tabular}{|c|c|c|}
    \hline
    1 & 3 & 4 \\
    \hline
    2 \\
    \cline{1-1}
    5 \\
    \cline{1-1}
    6 \\
    \cline{1-1}
  \end{tabular}_{CS}
  &=&\frac{1}{\sqrt{6}}|C_2 \rangle \otimes |S_2^1 \rangle +\frac{1}{3\sqrt{2}}|C_4 \rangle \otimes |S_2^1 \rangle -\frac{1}{2\sqrt{3}}|C_2 \rangle \otimes |S_3^1 \rangle +\frac{1}{3\sqrt{2}}|C_5 \rangle \otimes |S_3^1 \rangle -\frac{1}{2\sqrt{3}}|C_1 \rangle \otimes |S_4^1 \rangle \nonumber\\
  &&-\frac{1}{2\sqrt{3}}|C_2 \rangle \otimes |S_5^1 \rangle +\frac{1}{3}|C_4 \rangle \otimes |S_5^1 \rangle +\frac{1}{2\sqrt{6}}|C_2 \rangle \otimes |S_6^1 \rangle +\frac{1}{3}|C_5 \rangle \otimes |S_6^1 \rangle +\frac{1}{2\sqrt{6}}|C_1 \rangle \otimes |S_7^1 \rangle \nonumber\\
  &&+\frac{1}{2\sqrt{6}}|C_4 \rangle \otimes |S_8^1 \rangle +\frac{1}{2\sqrt{3}}|C_5 \rangle \otimes |S_8^1 \rangle +\frac{1}{2\sqrt{6}}|C_3 \rangle \otimes |S_9^1 \rangle \nonumber\\
  \begin{tabular}{|c|c|c|}
    \hline
    1 & 2 & 5 \\
    \hline
    3 \\
    \cline{1-1}
    4 \\
    \cline{1-1}
    6 \\
    \cline{1-1}
  \end{tabular}_{CS}
  &=&\frac{\sqrt{2}}{3}|C_3 \rangle \otimes |S_1^1 \rangle +\frac{1}{3\sqrt{10}}|C_1 \rangle \otimes |S_2^1 \rangle +\frac{1}{\sqrt{30}}|C_3 \rangle \otimes |S_2^1 \rangle -\frac{1}{6\sqrt{5}}|C_1 \rangle \otimes |S_3^1 \rangle -\frac{1}{\sqrt{15}}|C_3 \rangle \otimes |S_3^1 \rangle \nonumber\\
  &&+\frac{1}{6\sqrt{5}}|C_2 \rangle \otimes |S_4^1 \rangle -\frac{1}{\sqrt{15}}|C_4 \rangle \otimes |S_4^1 \rangle +\frac{1}{\sqrt{30}}|C_5 \rangle \otimes |S_4^1 \rangle +\frac{\sqrt{5}}{6}|C_1 \rangle \otimes |S_5^1 \rangle -\frac{\sqrt{5}}{6\sqrt{2}}|C_1 \rangle \otimes |S_6^1 \rangle \nonumber\\
  &&+\frac{\sqrt{5}}{6\sqrt{2}}|C_2 \rangle \otimes |S_7^1 \rangle -\frac{\sqrt{5}}{6\sqrt{2}}|C_3 \rangle \otimes |S_8^1 \rangle +\frac{\sqrt{5}}{6\sqrt{2}}|C_4 \rangle \otimes |S_9^1 \rangle +\frac{\sqrt{5}}{6}|C_5 \rangle \otimes |S_9^1 \rangle \nonumber\\
  \begin{tabular}{|c|c|c|}
    \hline
    1 & 3 & 5 \\
    \hline
    2 \\
    \cline{1-1}
    4 \\
    \cline{1-1}
    6 \\
    \cline{1-1}
  \end{tabular}_{CS}
  &=&\frac{\sqrt{2}}{3}|C_4 \rangle \otimes |S_1^1 \rangle +\frac{1}{3\sqrt{10}}|C_2 \rangle \otimes |S_2^1 \rangle +\frac{1}{\sqrt{30}}|C_4 \rangle \otimes |S_2^1 \rangle +\frac{1}{6\sqrt{5}}|C_2 \rangle \otimes |S_3^1 \rangle -\frac{1}{\sqrt{15}}|C_4 \rangle \otimes |S_3^1 \rangle \nonumber\\
  &&-\frac{1}{\sqrt{30}}|C_5 \rangle \otimes |S_3^1 \rangle +\frac{1}{6\sqrt{5}}|C_1 \rangle \otimes |S_4^1 \rangle -\frac{1}{\sqrt{15}}|C_3 \rangle \otimes |S_4^1 \rangle +\frac{\sqrt{5}}{6}|C_2 \rangle \otimes |S_5^1 \rangle +\frac{\sqrt{5}}{6\sqrt{2}}|C_2 \rangle \otimes |S_6^1 \rangle \nonumber\\
  &&+\frac{\sqrt{5}}{6\sqrt{2}}|C_1 \rangle \otimes |S_7^1 \rangle +\frac{\sqrt{5}}{6\sqrt{2}}|C_4 \rangle \otimes |S_8^1 \rangle -\frac{\sqrt{5}}{6}|C_5 \rangle \otimes |S_8^1 \rangle +\frac{\sqrt{5}}{6\sqrt{2}}|C_3 \rangle \otimes |S_9^1 \rangle \nonumber\\
  \begin{tabular}{|c|c|c|}
    \hline
    1 & 4 & 5 \\
    \hline
    2 \\
    \cline{1-1}
    3 \\
    \cline{1-1}
    6 \\
    \cline{1-1}
  \end{tabular}_{CS}
  &=&\frac{\sqrt{2}}{3}|C_5 \rangle \otimes |S_1^1 \rangle -\frac{\sqrt{2}}{\sqrt{15}}|C_5 \rangle \otimes |S_2^1 \rangle -\frac{1}{3\sqrt{10}}|C_2 \rangle \otimes |S_3^1 \rangle -\frac{1}{\sqrt{30}}|C_4 \rangle \otimes |S_3^1 \rangle +\frac{1}{3\sqrt{10}}|C_1 \rangle \otimes |S_4^1 \rangle \nonumber\\
  &&+\frac{1}{\sqrt{30}}|C_3 \rangle \otimes |S_4^1 \rangle -\frac{\sqrt{5}}{6}|C_2 \rangle \otimes |S_6^1 \rangle +\frac{\sqrt{5}}{6}|C_1 \rangle \otimes |S_7^1 \rangle -\frac{\sqrt{5}}{6}|C_4 \rangle \otimes |S_8^1 \rangle +\frac{\sqrt{5}}{6}|C_3 \rangle \otimes |S_9^1 \rangle \nonumber\\
  \begin{tabular}{|c|c|c|}
    \hline
    1 & 2 & 6 \\
    \hline
    3 \\
    \cline{1-1}
    4 \\
    \cline{1-1}
    5 \\
    \cline{1-1}
  \end{tabular}_{CS}
  &=&-\frac{2}{3\sqrt{3}}|C_3 \rangle \otimes |S_1^1 \rangle +\frac{4}{3\sqrt{15}}|C_1 \rangle \otimes |S_2^1 \rangle +\frac{2}{9\sqrt{5}}|C_3 \rangle \otimes |S_2^1 \rangle -\frac{2\sqrt{2}}{3\sqrt{15}}|C_1 \rangle \otimes |S_3^1 \rangle +\frac{2\sqrt{2}}{9\sqrt{5}}|C_3 \rangle \otimes |S_3^1 \rangle \nonumber\\
  &&+\frac{2\sqrt{2}}{3\sqrt{15}}|C_2 \rangle \otimes |S_4^1 \rangle -\frac{2\sqrt{2}}{9\sqrt{5}}|C_4 \rangle \otimes |S_4^1 \rangle +\frac{2}{9\sqrt{5}}|C_5 \rangle \otimes |S_4^1 \rangle +\frac{\sqrt{5}}{3\sqrt{6}}|C_1 \rangle \otimes |S_5^1 \rangle -\frac{\sqrt{5}}{9\sqrt{2}}|C_3 \rangle \otimes |S_5^1 \rangle \nonumber\\
  &&-\frac{\sqrt{5}}{6\sqrt{3}}|C_1 \rangle \otimes |S_6^1 \rangle -\frac{\sqrt{5}}{9}|C_3 \rangle \otimes |S_6^1 \rangle +\frac{\sqrt{5}}{6\sqrt{3}}|C_2 \rangle \otimes |S_7^1 \rangle +\frac{\sqrt{5}}{9}|C_4 \rangle \otimes |S_7^1 \rangle -\frac{\sqrt{5}}{9\sqrt{2}}|C_5 \rangle \otimes |S_7^1 \rangle \nonumber\\
  &&+\frac{\sqrt{5}}{6\sqrt{3}}|C_3 \rangle \otimes |S_8^1 \rangle -\frac{\sqrt{5}}{6\sqrt{3}}|C_4 \rangle \otimes |S_9^1 \rangle -\frac{\sqrt{5}}{3\sqrt{6}}|C_5 \rangle \otimes |S_9^1 \rangle \nonumber\\
  \begin{tabular}{|c|c|c|}
    \hline
    1 & 3 & 6 \\
    \hline
    2 \\
    \cline{1-1}
    4 \\
    \cline{1-1}
    5 \\
    \cline{1-1}
  \end{tabular}_{CS}
  &=&-\frac{2}{3\sqrt{3}}|C_4 \rangle \otimes |S_1^1 \rangle +\frac{4}{3\sqrt{15}}|C_2 \rangle \otimes |S_2^1 \rangle +\frac{2}{9\sqrt{5}}|C_4 \rangle \otimes |S_2^1 \rangle +\frac{2\sqrt{2}}{3\sqrt{15}}|C_2 \rangle \otimes |S_3^1 \rangle -\frac{2\sqrt{2}}{9\sqrt{5}}|C_4 \rangle \otimes |S_3^1 \rangle \nonumber\\
  &&-\frac{2}{9\sqrt{5}}|C_5 \rangle \otimes |S_3^1 \rangle +\frac{2\sqrt{2}}{3\sqrt{15}}|C_1 \rangle \otimes |S_4^1 \rangle -\frac{2\sqrt{2}}{9\sqrt{5}}|C_3 \rangle \otimes |S_4^1 \rangle +\frac{\sqrt{5}}{3\sqrt{6}}|C_2 \rangle \otimes |S_5^1 \rangle -\frac{\sqrt{5}}{9\sqrt{2}}|C_4 \rangle \otimes |S_5^1 \rangle \nonumber\\
  &&+\frac{\sqrt{5}}{6\sqrt{3}}|C_2 \rangle \otimes |S_6^1 \rangle +\frac{\sqrt{5}}{9}|C_4 \rangle \otimes |S_6^1 \rangle +\frac{\sqrt{5}}{9\sqrt{2}}|C_5 \rangle \otimes |S_6^1 \rangle +\frac{\sqrt{5}}{6\sqrt{3}}|C_1 \rangle \otimes |S_7^1 \rangle +\frac{\sqrt{5}}{9}|C_3 \rangle \otimes |S_7^1 \rangle \nonumber\\
  &&-\frac{\sqrt{5}}{6\sqrt{3}}|C_4 \rangle \otimes |S_8^1 \rangle +\frac{\sqrt{5}}{3\sqrt{6}}|C_5 \rangle \otimes |S_8^1 \rangle -\frac{\sqrt{5}}{6\sqrt{3}}|C_3 \rangle \otimes |S_9^1 \rangle \nonumber\\
  \begin{tabular}{|c|c|c|}
    \hline
    1 & 4 & 6 \\
    \hline
    2 \\
    \cline{1-1}
    3 \\
    \cline{1-1}
    5 \\
    \cline{1-1}
  \end{tabular}_{CS}
  &=&-\frac{2}{3\sqrt{3}}|C_5 \rangle \otimes |S_1^1 \rangle -\frac{4}{9\sqrt{5}}|C_5 \rangle \otimes |S_2^1 \rangle -\frac{4}{3\sqrt{15}}|C_2 \rangle \otimes |S_3^1 \rangle -\frac{2}{9\sqrt{5}}|C_4 \rangle \otimes |S_3^1 \rangle +\frac{4}{3\sqrt{15}}|C_1 \rangle \otimes |S_4^1 \rangle \nonumber\\
  &&+\frac{2}{9\sqrt{5}}|C_3 \rangle \otimes |S_4^1 \rangle +\frac{\sqrt{10}}{9}|C_5 \rangle \otimes |S_5^1 \rangle -\frac{\sqrt{5}}{3\sqrt{6}}|C_2 \rangle \otimes |S_6^1 \rangle +\frac{\sqrt{5}}{9\sqrt{2}}|C_4 \rangle \otimes |S_6^1 \rangle +\frac{\sqrt{5}}{3\sqrt{6}}|C_1 \rangle \otimes |S_7^1 \rangle \nonumber\\
  &&-\frac{\sqrt{5}}{9\sqrt{2}}|C_3 \rangle \otimes |S_7^1 \rangle +\frac{\sqrt{5}}{3\sqrt{6}}|C_4 \rangle \otimes |S_8^1 \rangle -\frac{\sqrt{5}}{3\sqrt{6}}|C_3 \rangle \otimes |S_9^1 \rangle \nonumber\\
  \begin{tabular}{|c|c|c|}
    \hline
    1 & 5 & 6 \\
    \hline
    2 \\
    \cline{1-1}
    3 \\
    \cline{1-1}
    4 \\
    \cline{1-1}
  \end{tabular}_{CS}
  &=&\frac{2}{3\sqrt{3}}|C_5 \rangle \otimes |S_2^1 \rangle -\frac{2}{3\sqrt{3}}|C_4 \rangle \otimes |S_3^1 \rangle +\frac{2}{3\sqrt{3}}|C_3 \rangle \otimes |S_4^1 \rangle -\frac{\sqrt{2}}{3\sqrt{3}}|C_5 \rangle \otimes |S_5^1 \rangle +\frac{\sqrt{2}}{3\sqrt{3}}|C_4 \rangle \otimes |S_6^1 \rangle \nonumber\\
  &&-\frac{\sqrt{2}}{3\sqrt{3}}|C_3 \rangle \otimes |S_7^1 \rangle +\frac{1}{\sqrt{6}}|C_2 \rangle \otimes |S_8^1 \rangle -\frac{1}{\sqrt{6}}|C_1 \rangle \otimes |S_9^1 \rangle \nonumber
\end{eqnarray}
\begin{eqnarray}
  \begin{tabular}{|c|c|c|}
    \hline
    1 & 2 \\
    \hline
    3 & 4 \\
    \hline
    5 & 6 \\
    \hline
  \end{tabular}_{~CS}
  &=&-\frac{1}{\sqrt{6}}|C_1 \rangle \otimes |S_1^1 \rangle +\frac{\sqrt{5}}{3\sqrt{6}}|C_3 \rangle \otimes |S_2^1 \rangle -\frac{\sqrt{5}}{6\sqrt{3}}|C_3 \rangle \otimes |S_3^1 \rangle +\frac{\sqrt{5}}{6\sqrt{3}}|C_4 \rangle \otimes |S_4^1 \rangle +\frac{\sqrt{5}}{3\sqrt{6}}|C_5 \rangle \otimes |S_4^1 \rangle \nonumber\\
  &&-\frac{\sqrt{5}}{6\sqrt{3}}|C_3 \rangle \otimes |S_5^1 \rangle +\frac{\sqrt{5}}{6\sqrt{6}}|C_3 \rangle \otimes |S_6^1 \rangle -\frac{\sqrt{5}}{6\sqrt{6}}|C_4 \rangle \otimes |S_7^1 \rangle -\frac{\sqrt{5}}{6\sqrt{3}}|C_5 \rangle \otimes |S_7^1 \rangle -\frac{\sqrt{5}}{2\sqrt{6}}|C_1 \rangle \otimes |S_8^1 \rangle \nonumber\\ &&+\frac{\sqrt{5}}{2\sqrt{6}}|C_2 \rangle \otimes |S_9^1 \rangle \nonumber\\
  \begin{tabular}{|c|c|c|}
    \hline
    1 & 3 \\
    \hline
    2 & 4 \\
    \hline
    5 & 6 \\
    \hline
  \end{tabular}_{~CS}
  &=&-\frac{1}{\sqrt{6}}|C_2 \rangle \otimes |S_1^1 \rangle +\frac{\sqrt{5}}{3\sqrt{6}}|C_4 \rangle \otimes |S_2^1 \rangle +\frac{\sqrt{5}}{6\sqrt{3}}|C_4 \rangle \otimes |S_3^1 \rangle -\frac{\sqrt{5}}{3\sqrt{6}}|C_5 \rangle \otimes |S_3^1 \rangle +\frac{\sqrt{5}}{6\sqrt{3}}|C_3 \rangle \otimes |S_4^1 \rangle \nonumber\\
  &&-\frac{\sqrt{5}}{6\sqrt{3}}|C_4 \rangle \otimes |S_5^1 \rangle -\frac{\sqrt{5}}{6\sqrt{6}}|C_4 \rangle \otimes |S_6^1 \rangle +\frac{\sqrt{5}}{6\sqrt{3}}|C_5 \rangle \otimes |S_6^1 \rangle -\frac{\sqrt{5}}{6\sqrt{6}}|C_3 \rangle \otimes |S_7^1 \rangle +\frac{\sqrt{5}}{2\sqrt{6}}|C_2 \rangle \otimes |S_8^1 \rangle \nonumber\\ &&+\frac{\sqrt{5}}{2\sqrt{6}}|C_1 \rangle \otimes |S_9^1 \rangle \nonumber\\
  \begin{tabular}{|c|c|c|}
    \hline
    1 & 2 \\
    \hline
    3 & 5 \\
    \hline
    4 & 6 \\
    \hline
  \end{tabular}_{~CS}
  &=&\frac{\sqrt{2}}{3\sqrt{3}}|C_3 \rangle \otimes |S_1^1 \rangle +\frac{\sqrt{5}}{3\sqrt{6}}|C_1 \rangle \otimes |S_2^1 \rangle -\frac{\sqrt{5}}{9\sqrt{2}}|C_3 \rangle \otimes |S_2^1 \rangle -\frac{\sqrt{5}}{6\sqrt{3}}|C_1 \rangle \otimes |S_3^1 \rangle -\frac{\sqrt{5}}{9}|C_3 \rangle \otimes |S_3^1 \rangle \nonumber\\
  &&+\frac{\sqrt{5}}{6\sqrt{3}}|C_2 \rangle \otimes |S_4^1 \rangle +\frac{\sqrt{5}}{9}|C_4 \rangle \otimes |S_4^1 \rangle -\frac{\sqrt{5}}{9\sqrt{2}}|C_5 \rangle \otimes |S_4^1 \rangle -\frac{\sqrt{5}}{6\sqrt{3}}|C_1 \rangle \otimes |S_5^1 \rangle -\frac{\sqrt{5}}{9}|C_3 \rangle \otimes |S_5^1 \rangle \nonumber\\ &&+\frac{\sqrt{5}}{6\sqrt{6}}|C_1 \rangle \otimes |S_6^1 \rangle -\frac{\sqrt{10}}{9}|C_3 \rangle \otimes |S_6^1 \rangle -\frac{\sqrt{5}}{6\sqrt{6}}|C_2 \rangle \otimes |S_7^1 \rangle +\frac{\sqrt{10}}{9}|C_4 \rangle \otimes |S_5^1 \rangle -\frac{\sqrt{5}}{9}|C_5 \rangle \otimes |S_7^1 \rangle \nonumber\\
  &&-\frac{\sqrt{5}}{6\sqrt{6}}|C_3 \rangle \otimes |S_8^1 \rangle +\frac{\sqrt{5}}{6\sqrt{6}}|C_4 \rangle \otimes |S_9^1 \rangle +\frac{\sqrt{5}}{6\sqrt{3}}|C_5 \rangle \otimes |S_9^1 \rangle \nonumber\\
  \begin{tabular}{|c|c|c|}
    \hline
    1 & 3 \\
    \hline
    2 & 5 \\
    \hline
    4 & 6 \\
    \hline
  \end{tabular}_{~CS}
  &=&\frac{\sqrt{2}}{3\sqrt{3}}|C_4 \rangle \otimes |S_1^1 \rangle +\frac{\sqrt{5}}{3\sqrt{6}}|C_2 \rangle \otimes |S_2^1 \rangle -\frac{\sqrt{5}}{9\sqrt{2}}|C_4 \rangle \otimes |S_2^1 \rangle +\frac{\sqrt{5}}{6\sqrt{3}}|C_2 \rangle \otimes |S_3^1 \rangle +\frac{\sqrt{5}}{9}|C_4 \rangle \otimes |S_3^1 \rangle \nonumber\\
  &&+\frac{\sqrt{5}}{9\sqrt{2}}|C_5 \rangle \otimes |S_3^1 \rangle +\frac{\sqrt{5}}{6\sqrt{3}}|C_1 \rangle \otimes |S_4^1 \rangle +\frac{\sqrt{5}}{9}|C_3 \rangle \otimes |S_4^1 \rangle -\frac{\sqrt{5}}{6\sqrt{3}}|C_2 \rangle \otimes |S_5^1 \rangle -\frac{\sqrt{5}}{9}|C_4 \rangle \otimes |S_5^1 \rangle \nonumber\\ &&-\frac{\sqrt{5}}{6\sqrt{6}}|C_2 \rangle \otimes |S_6^1 \rangle +\frac{\sqrt{10}}{9}|C_4 \rangle \otimes |S_6^1 \rangle +\frac{\sqrt{5}}{9}|C_5 \rangle \otimes |S_6^1 \rangle -\frac{\sqrt{5}}{6\sqrt{6}}|C_1 \rangle \otimes |S_7^1 \rangle +\frac{\sqrt{10}}{9}|C_3 \rangle \otimes |S_7^1 \rangle \nonumber\\
  &&+\frac{\sqrt{5}}{6\sqrt{6}}|C_4 \rangle \otimes |S_8^1 \rangle -\frac{\sqrt{5}}{6\sqrt{3}}|C_5 \rangle \otimes |S_8^1 \rangle +\frac{\sqrt{5}}{6\sqrt{6}}|C_3 \rangle \otimes |S_9^1 \rangle \nonumber\\
  \begin{tabular}{|c|c|c|}
    \hline
    1 & 4 \\
    \hline
    2 & 5 \\
    \hline
    3 & 6 \\
    \hline
  \end{tabular}_{~CS}
  &=&\frac{\sqrt{2}}{3\sqrt{3}}|C_5 \rangle \otimes |S_1^1 \rangle +\frac{\sqrt{10}}{9}|C_5 \rangle \otimes |S_2^1 \rangle -\frac{\sqrt{5}}{3\sqrt{6}}|C_2 \rangle \otimes |S_3^1 \rangle +\frac{\sqrt{5}}{9\sqrt{2}}|C_4 \rangle \otimes |S_3^1 \rangle +\frac{\sqrt{5}}{3\sqrt{6}}|C_1 \rangle \otimes |S_4^1 \rangle \nonumber\\
  &&-\frac{\sqrt{5}}{9\sqrt{2}}|C_3 \rangle \otimes |S_4^1 \rangle +\frac{2\sqrt{5}}{9}|C_5 \rangle \otimes |S_5^1 \rangle +\frac{\sqrt{5}}{6\sqrt{3}}|C_2 \rangle \otimes |S_6^1 \rangle +\frac{\sqrt{5}}{9}|C_4 \rangle \otimes |S_6^1 \rangle -\frac{\sqrt{5}}{6\sqrt{3}}|C_1 \rangle \otimes |S_7^1 \rangle \nonumber\\ &&-\frac{\sqrt{5}}{9}|C_3 \rangle \otimes |S_7^1 \rangle -\frac{\sqrt{5}}{6\sqrt{3}}|C_4 \rangle \otimes |S_8^1 \rangle +\frac{\sqrt{5}}{6\sqrt{3}}|C_3 \rangle \otimes |S_9^1 \rangle \nonumber
\end{eqnarray}
\begin{eqnarray}
  \begin{tabular}{|c|c|}
    \hline
    1 & 2 \\
    \hline
    3 \\
    \cline{1-1}
    4 \\
    \cline{1-1}
    5 \\
    \cline{1-1}
    6 \\
    \cline{1-1}
  \end{tabular}_{CS}
  &=&\frac{\sqrt{5}}{3\sqrt{3}}|C_3 \rangle \otimes |S_1^1 \rangle -\frac{2}{3\sqrt{3}}|C_1 \rangle \otimes |S_2^1 \rangle -\frac{1}{9}|C_3 \rangle \otimes |S_2^1 \rangle +\frac{\sqrt{2}}{3\sqrt{3}}|C_1 \rangle \otimes |S_3^1 \rangle -\frac{\sqrt{2}}{9}|C_3 \rangle \otimes |S_3^1 \rangle \nonumber\\
  &&-\frac{\sqrt{2}}{3\sqrt{3}}|C_2 \rangle \otimes |S_4^1 \rangle +\frac{\sqrt{2}}{9}|C_4 \rangle \otimes |S_4^1 \rangle -\frac{1}{9}|C_5 \rangle \otimes |S_4^1 \rangle +\frac{\sqrt{2}}{3\sqrt{3}}|C_1 \rangle \otimes |S_5^1 \rangle -\frac{\sqrt{2}}{9}|C_3 \rangle \otimes |S_5^1 \rangle \nonumber\\
  &&-\frac{1}{3\sqrt{3}}|C_1 \rangle \otimes |S_6^1 \rangle -\frac{2}{9}|C_3 \rangle \otimes |S_6^1 \rangle +\frac{1}{3\sqrt{3}}|C_2 \rangle \otimes |S_7^1 \rangle +\frac{2}{9}|C_4 \rangle \otimes |S_7^1 \rangle -\frac{\sqrt{2}}{9}|C_5 \rangle \otimes |S_7^1 \rangle \nonumber\\
  &&+\frac{1}{3\sqrt{3}}|C_3 \rangle \otimes |S_8^1 \rangle -\frac{1}{3\sqrt{3}}|C_4 \rangle \otimes |S_9^1 \rangle -\frac{\sqrt{2}}{3\sqrt{3}}|C_5 \rangle \otimes |S_9^1 \rangle \nonumber\\
  \begin{tabular}{|c|c|}
    \hline
    1 & 3 \\
    \hline
    2 \\
    \cline{1-1}
    4 \\
    \cline{1-1}
    5 \\
    \cline{1-1}
    6 \\
    \cline{1-1}
  \end{tabular}_{CS}
  &=&\frac{\sqrt{5}}{3\sqrt{3}}|C_4 \rangle \otimes |S_1^1 \rangle -\frac{2}{3\sqrt{3}}|C_2 \rangle \otimes |S_2^1 \rangle -\frac{1}{9}|C_4 \rangle \otimes |S_2^1 \rangle -\frac{\sqrt{2}}{3\sqrt{3}}|C_2 \rangle \otimes |S_3^1 \rangle +\frac{\sqrt{2}}{9}|C_4 \rangle \otimes |S_3^1 \rangle \nonumber\\
  &&+\frac{1}{9}|C_5 \rangle \otimes |S_3^1 \rangle -\frac{\sqrt{2}}{3\sqrt{3}}|C_1 \rangle \otimes |S_4^1 \rangle +\frac{\sqrt{2}}{9}|C_3 \rangle \otimes |S_4^1 \rangle +\frac{\sqrt{2}}{3\sqrt{3}}|C_2 \rangle \otimes |S_5^1 \rangle -\frac{\sqrt{2}}{9}|C_4 \rangle \otimes |S_5^1 \rangle \nonumber\\
  &&+\frac{1}{3\sqrt{3}}|C_2 \rangle \otimes |S_6^1 \rangle +\frac{2}{9}|C_4 \rangle \otimes |S_6^1 \rangle +\frac{\sqrt{2}}{9}|C_5 \rangle \otimes |S_6^1 \rangle +\frac{1}{3\sqrt{3}}|C_1 \rangle \otimes |S_7^1 \rangle +\frac{2}{9}|C_3 \rangle \otimes |S_7^1 \rangle \nonumber\\
  &&-\frac{1}{3\sqrt{3}}|C_4 \rangle \otimes |S_8^1 \rangle +\frac{\sqrt{2}}{3\sqrt{3}}|C_5 \rangle \otimes |S_8^1 \rangle -\frac{1}{3\sqrt{3}}|C_3 \rangle \otimes |S_9^1 \rangle \nonumber\\
  \begin{tabular}{|c|c|}
    \hline
    1 & 4 \\
    \hline
    2 \\
    \cline{1-1}
    3 \\
    \cline{1-1}
    5 \\
    \cline{1-1}
    6 \\
    \cline{1-1}
  \end{tabular}_{CS}
  &=&\frac{\sqrt{5}}{3\sqrt{3}}|C_5 \rangle \otimes |S_1^1 \rangle +\frac{2}{9}|C_5 \rangle \otimes |S_2^1 \rangle +\frac{2}{3\sqrt{3}}|C_2 \rangle \otimes |S_3^1 \rangle +\frac{1}{9}|C_4 \rangle \otimes |S_3^1 \rangle -\frac{2}{3\sqrt{3}}|C_1 \rangle \otimes |S_4^1 \rangle \nonumber\\
  &&-\frac{1}{9}|C_3 \rangle \otimes |S_4^1 \rangle +\frac{2\sqrt{2}}{9}|C_5 \rangle \otimes |S_5^1 \rangle -\frac{\sqrt{2}}{3\sqrt{3}}|C_2 \rangle \otimes |S_6^1 \rangle +\frac{\sqrt{2}}{9}|C_4 \rangle \otimes |S_6^1 \rangle +\frac{\sqrt{2}}{3\sqrt{3}}|C_1 \rangle \otimes |S_7^1 \rangle \nonumber\\
  &&-\frac{\sqrt{2}}{9}|C_3 \rangle \otimes |S_7^1 \rangle +\frac{\sqrt{2}}{3\sqrt{3}}|C_4 \rangle \otimes |S_8^1 \rangle -\frac{\sqrt{2}}{3\sqrt{3}}|C_3 \rangle \otimes |S_9^1 \rangle \nonumber\\
  \begin{tabular}{|c|c|}
    \hline
    1 & 5 \\
    \hline
    2 \\
    \cline{1-1}
    3 \\
    \cline{1-1}
    4 \\
    \cline{1-1}
    6 \\
    \cline{1-1}
  \end{tabular}_{CS}
  &=&-\frac{\sqrt{5}}{3\sqrt{3}}|C_5 \rangle \otimes |S_2^1 \rangle +\frac{\sqrt{5}}{3\sqrt{3}}|C_4 \rangle \otimes |S_3^1 \rangle -\frac{\sqrt{5}}{3\sqrt{3}}|C_3 \rangle \otimes |S_4^1 \rangle -\frac{2\sqrt{2}}{3\sqrt{15}}|C_5 \rangle \otimes |S_5^1 \rangle +\frac{2\sqrt{2}}{3\sqrt{15}}|C_4 \rangle \otimes |S_6^1 \rangle \nonumber\\
  &&-\frac{2\sqrt{2}}{3\sqrt{15}}|C_3 \rangle \otimes |S_7^1 \rangle +\frac{\sqrt{2}}{\sqrt{15}}|C_2 \rangle \otimes |S_8^1 \rangle -\frac{\sqrt{2}}{\sqrt{15}}|C_1 \rangle \otimes |S_9^1 \rangle \nonumber\\
  \begin{tabular}{|c|c|}
    \hline
    1 & 6 \\
    \hline
    2 \\
    \cline{1-1}
    3 \\
    \cline{1-1}
    4 \\
    \cline{1-1}
    5 \\
    \cline{1-1}
  \end{tabular}_{CS}
  &=&\frac{1}{\sqrt{5}}|C_5 \rangle \otimes |S_5^1 \rangle -\frac{1}{\sqrt{5}}|C_4 \rangle \otimes |S_6^1 \rangle +\frac{1}{\sqrt{5}}|C_3 \rangle \otimes |S_7^1 \rangle +\frac{1}{\sqrt{5}}|C_2 \rangle \otimes |S_8^1 \rangle -\frac{1}{\sqrt{5}}|C_1 \rangle \otimes |S_9^1 \rangle \nonumber
\end{eqnarray}
\begin{eqnarray}
  \begin{tabular}{|c|c|c|c|}
    \hline
    1 & 2 & 3 & 4 \\
    \hline
    5 & 6 \\
    \cline{1-2}
  \end{tabular}_{CS}
  &=&\frac{1}{\sqrt{2}}|C_1 \rangle \otimes |S_8^1 \rangle +\frac{1}{\sqrt{2}}|C_2 \rangle \otimes |S_9^1 \rangle \nonumber\\
  \begin{tabular}{|c|c|c|c|}
    \hline
    1 & 2 & 3 & 5 \\
    \hline
    4 & 6 \\
    \cline{1-2}
  \end{tabular}_{CS}
  &=&-\frac{1}{\sqrt{10}}|C_1 \rangle \otimes |S_6^1 \rangle +\frac{\sqrt{3}}{\sqrt{10}}|C_3 \rangle \otimes |S_6^1 \rangle -\frac{1}{\sqrt{10}}|C_2 \rangle \otimes |S_7^1 \rangle +\frac{\sqrt{3}}{\sqrt{10}}|C_4 \rangle \otimes |S_7^1 \rangle +\frac{1}{\sqrt{10}}|C_3 \rangle \otimes |S_8^1 \rangle \nonumber\\
  &&+\frac{1}{\sqrt{10}}|C_4 \rangle \otimes |S_9^1 \rangle \nonumber\\
  \begin{tabular}{|c|c|c|c|}
    \hline
    1 & 2 & 4 & 5 \\
    \hline
    3 & 6 \\
    \cline{1-2}
  \end{tabular}_{CS}
  &=&-\frac{1}{\sqrt{10}}|C_1 \rangle \otimes |S_5^1 \rangle -\frac{\sqrt{3}}{\sqrt{10}}|C_3 \rangle \otimes |S_5^1 \rangle -\frac{1}{2\sqrt{5}}|C_1 \rangle \otimes |S_6^1 \rangle +\frac{1}{2\sqrt{5}}|C_2 \rangle \otimes |S_7^1 \rangle +\frac{\sqrt{3}}{\sqrt{10}}|C_5 \rangle \otimes |S_7^1 \rangle \nonumber\\
  &&-\frac{1}{2\sqrt{5}}|C_3 \rangle \otimes |S_8^1 \rangle +\frac{1}{2\sqrt{5}}|C_4 \rangle \otimes |S_9^1 \rangle -\frac{1}{\sqrt{10}}|C_5 \rangle \otimes |S_9^1 \rangle \nonumber\\
  \begin{tabular}{|c|c|c|c|}
    \hline
    1 & 3 & 4 & 5 \\
    \hline
    2 & 6 \\
    \cline{1-2}
  \end{tabular}_{CS}
  &=&-\frac{1}{\sqrt{10}}|C_2 \rangle \otimes |S_5^1 \rangle -\frac{\sqrt{3}}{\sqrt{10}}|C_4 \rangle \otimes |S_5^1 \rangle +\frac{1}{2\sqrt{5}}|C_2 \rangle \otimes |S_6^1 \rangle -\frac{\sqrt{3}}{\sqrt{10}}|C_5 \rangle \otimes |S_6^1 \rangle +\frac{1}{2\sqrt{5}}|C_1 \rangle \otimes |S_7^1 \rangle \nonumber\\
  &&+\frac{1}{2\sqrt{5}}|C_4 \rangle \otimes |S_8^1 \rangle +\frac{1}{\sqrt{10}}|C_5 \rangle \otimes |S_8^1 \rangle +\frac{1}{2\sqrt{5}}|C_3 \rangle \otimes |S_9^1 \rangle \nonumber\\
  \begin{tabular}{|c|c|c|c|}
    \hline
    1 & 2 & 3 & 6 \\
    \hline
    4 & 5 \\
    \cline{1-2}
  \end{tabular}_{CS}
  &=&-\frac{1}{\sqrt{10}}|C_1 \rangle \otimes |S_3^1 \rangle -\frac{\sqrt{3}}{\sqrt{10}}|C_3 \rangle \otimes |S_3^1 \rangle -\frac{1}{\sqrt{10}}|C_2 \rangle \otimes |S_4^1 \rangle -\frac{\sqrt{3}}{\sqrt{10}}|C_4 \rangle \otimes |S_4^1 \rangle -\frac{1}{2\sqrt{5}}|C_1 \rangle \otimes |S_6^1 \rangle \nonumber\\
  &&-\frac{1}{2\sqrt{5}}|C_2 \rangle \otimes |S_7^1 \rangle -\frac{1}{2\sqrt{5}}|C_3 \rangle \otimes |S_8^1 \rangle -\frac{1}{2\sqrt{5}}|C_4 \rangle \otimes |S_9^1 \rangle \nonumber\\
  \begin{tabular}{|c|c|c|c|}
    \hline
    1 & 2 & 4 & 6 \\
    \hline
    3 & 5 \\
    \cline{1-2}
  \end{tabular}_{CS}
  &=&-\frac{1}{\sqrt{10}}|C_1 \rangle \otimes |S_2^1 \rangle +\frac{\sqrt{3}}{\sqrt{10}}|C_3 \rangle \otimes |S_2^1 \rangle -\frac{1}{2\sqrt{5}}|C_1 \rangle \otimes |S_3^1 \rangle +\frac{1}{2\sqrt{5}}|C_2 \rangle \otimes |S_4^1 \rangle -\frac{\sqrt{3}}{\sqrt{10}}|C_5 \rangle \otimes |S_4^1 \rangle \nonumber\\
  &&-\frac{1}{2\sqrt{5}}|C_1 \rangle \otimes |S_5^1 \rangle -\frac{1}{2\sqrt{10}}|C_1 \rangle \otimes |S_6^1 \rangle +\frac{1}{2\sqrt{10}}|C_2 \rangle \otimes |S_7^1 \rangle +\frac{1}{2\sqrt{10}}|C_3 \rangle \otimes |S_8^1 \rangle -\frac{1}{2\sqrt{10}}|C_4 \rangle \otimes |S_9^1 \rangle \nonumber\\
  &&+\frac{1}{2\sqrt{5}}|C_5 \rangle \otimes |S_9^1 \rangle \nonumber\\
  \begin{tabular}{|c|c|c|c|}
    \hline
    1 & 3 & 4 & 6 \\
    \hline
    2 & 5 \\
    \cline{1-2}
  \end{tabular}_{CS}
  &=&-\frac{1}{\sqrt{10}}|C_2 \rangle \otimes |S_2^1 \rangle +\frac{\sqrt{3}}{\sqrt{10}}|C_4 \rangle \otimes |S_2^1 \rangle +\frac{1}{2\sqrt{5}}|C_2 \rangle \otimes |S_3^1 \rangle +\frac{\sqrt{3}}{\sqrt{10}}|C_5 \rangle \otimes |S_3^1 \rangle +\frac{1}{2\sqrt{5}}|C_1 \rangle \otimes |S_4^1 \rangle \nonumber\\
  &&-\frac{1}{2\sqrt{5}}|C_2 \rangle \otimes |S_5^1 \rangle +\frac{1}{2\sqrt{10}}|C_2 \rangle \otimes |S_6^1 \rangle +\frac{1}{2\sqrt{10}}|C_1 \rangle \otimes |S_7^1 \rangle -\frac{1}{2\sqrt{10}}|C_4 \rangle \otimes |S_8^1 \rangle -\frac{1}{2\sqrt{5}}|C_5 \rangle \otimes |S_8^1 \rangle \nonumber\\
  &&-\frac{1}{2\sqrt{10}}|C_3 \rangle \otimes |S_9^1 \rangle \nonumber\\
  \begin{tabular}{|c|c|c|c|}
    \hline
    1 & 2 & 5 & 6 \\
    \hline
    3 & 4 \\
    \cline{1-2}
  \end{tabular}_{CS}
  &=&\frac{1}{\sqrt{2}}|C_1 \rangle \otimes |S_1^1 \rangle +\frac{1}{\sqrt{10}}|C_3 \rangle \otimes |S_2^1 \rangle -\frac{1}{2\sqrt{5}}|C_3 \rangle \otimes |S_3^1 \rangle +\frac{1}{2\sqrt{5}}|C_4 \rangle \otimes |S_4^1 \rangle +\frac{1}{\sqrt{10}}|C_5 \rangle \otimes |S_4^1 \rangle \nonumber\\
  &&-\frac{1}{2\sqrt{5}}|C_3 \rangle \otimes |S_5^1 \rangle +\frac{1}{2\sqrt{10}}|C_3 \rangle \otimes |S_6^1 \rangle -\frac{1}{2\sqrt{10}}|C_4 \rangle \otimes |S_7^1 \rangle -\frac{1}{2\sqrt{5}}|C_5 \rangle \otimes |S_7^1 \rangle +\frac{1}{2\sqrt{10}}|C_1 \rangle \otimes |S_8^1 \rangle \nonumber\\
  &&-\frac{1}{2\sqrt{10}}|C_2 \rangle \otimes |S_9^1 \rangle \nonumber\\
  \begin{tabular}{|c|c|c|c|}
    \hline
    1 & 3 & 5 & 6 \\
    \hline
    2 & 4 \\
    \cline{1-2}
  \end{tabular}_{CS}
  &=&\frac{1}{\sqrt{2}}|C_2 \rangle \otimes |S_1^1 \rangle +\frac{1}{\sqrt{10}}|C_4 \rangle \otimes |S_2^1 \rangle +\frac{1}{2\sqrt{5}}|C_4 \rangle \otimes |S_3^1 \rangle -\frac{1}{\sqrt{10}}|C_5 \rangle \otimes |S_3^1 \rangle +\frac{1}{2\sqrt{5}}|C_3 \rangle \otimes |S_4^1 \rangle \nonumber\\
  &&-\frac{1}{2\sqrt{5}}|C_4 \rangle \otimes |S_5^1 \rangle -\frac{1}{2\sqrt{10}}|C_4 \rangle \otimes |S_6^1 \rangle +\frac{1}{2\sqrt{5}}|C_5 \rangle \otimes |S_6^1 \rangle -\frac{1}{2\sqrt{10}}|C_3 \rangle \otimes |S_7^1 \rangle -\frac{1}{2\sqrt{10}}|C_2 \rangle \otimes |S_8^1 \rangle \nonumber\\
  &&-\frac{1}{2\sqrt{10}}|C_1 \rangle \otimes |S_9^1 \rangle \nonumber
\end{eqnarray}
\subsection{S=2}
\begin{eqnarray}
  \begin{tabular}{|c|c|c|}
    \hline
    1 & 2 & 3 \\
    \hline
    4 & 5 \\
    \cline{1-2}
    6 \\
    \cline{1-1}
  \end{tabular}_{CS}
  &=&\frac{1}{2\sqrt{2}}|C_1 \rangle \otimes |S_4^2 \rangle +\frac{\sqrt{3}}{2\sqrt{2}}|C_3 \rangle \otimes |S_4^2 \rangle +\frac{1}{2\sqrt{2}}|C_2 \rangle \otimes |S_5^2 \rangle +\frac{\sqrt{3}}{2\sqrt{2}}|C_4 \rangle \otimes |S_5^2 \rangle \nonumber\\
  \begin{tabular}{|c|c|c|}
    \hline
    1 & 2 & 4 \\
    \hline
    3 & 5 \\
    \cline{1-2}
    6 \\
    \cline{1-1}
  \end{tabular}_{CS}
  &=&\frac{1}{2\sqrt{2}}|C_1 \rangle \otimes |S_3^2 \rangle -\frac{\sqrt{3}}{2\sqrt{2}}|C_3 \rangle \otimes |S_3^2 \rangle +\frac{1}{4}|C_1 \rangle \otimes |S_4^2 \rangle -\frac{1}{4}|C_2 \rangle \otimes |S_5^2 \rangle +\frac{\sqrt{3}}{2\sqrt{2}}|C_5 \rangle \otimes |S_5^2 \rangle \nonumber\\
  \begin{tabular}{|c|c|c|}
    \hline
    1 & 3 & 4 \\
    \hline
    2 & 5 \\
    \cline{1-2}
    6 \\
    \cline{1-1}
  \end{tabular}_{CS}
  &=&\frac{1}{2\sqrt{2}}|C_2 \rangle \otimes |S_3^2 \rangle -\frac{\sqrt{3}}{2\sqrt{2}}|C_4 \rangle \otimes |S_3^2 \rangle -\frac{1}{4}|C_2 \rangle \otimes |S_4^2 \rangle -\frac{\sqrt{3}}{2\sqrt{2}}|C_5 \rangle \otimes |S_4^2 \rangle -\frac{1}{4}|C_1 \rangle \otimes |S_5^2 \rangle \nonumber\\
  \begin{tabular}{|c|c|c|}
    \hline
    1 & 2 & 5 \\
    \hline
    3 & 4 \\
    \cline{1-2}
    6 \\
    \cline{1-1}
  \end{tabular}_{CS}
  &=&-\frac{\sqrt{5}}{2\sqrt{2}}|C_1 \rangle \otimes |S_2^2 \rangle -\frac{1}{2\sqrt{2}}|C_3 \rangle \otimes |S_3^2 \rangle +\frac{1}{4}|C_3 \rangle \otimes |S_4^2 \rangle -\frac{1}{4}|C_4 \rangle \otimes |S_5^2 \rangle -\frac{1}{2\sqrt{2}}|C_5 \rangle \otimes |S_5^2 \rangle \nonumber\\
  \begin{tabular}{|c|c|c|}
    \hline
    1 & 3 & 5 \\
    \hline
    2 & 4 \\
    \cline{1-2}
    6 \\
    \cline{1-1}
  \end{tabular}_{CS}
  &=&-\frac{\sqrt{5}}{2\sqrt{2}}|C_2 \rangle \otimes |S_2^2 \rangle -\frac{1}{2\sqrt{2}}|C_4 \rangle \otimes |S_3^2 \rangle -\frac{1}{4}|C_4 \rangle \otimes |S_4^2 \rangle +\frac{1}{2\sqrt{2}}|C_5 \rangle \otimes |S_4^2 \rangle -\frac{1}{4}|C_3 \rangle \otimes |S_5^2 \rangle \nonumber\\
  \begin{tabular}{|c|c|c|}
    \hline
    1 & 2 & 3 \\
    \hline
    4 & 6 \\
    \cline{1-2}
    5 \\
    \cline{1-1}
  \end{tabular}_{CS}
  &=&\frac{\sqrt{3}}{2\sqrt{2}}|C_1 \rangle \otimes |S_4^2 \rangle -\frac{1}{2\sqrt{2}}|C_3 \rangle \otimes |S_4^2 \rangle +\frac{\sqrt{3}}{2\sqrt{2}}|C_2 \rangle \otimes |S_5^2 \rangle -\frac{1}{2\sqrt{2}}|C_4 \rangle \otimes |S_5^2 \rangle \nonumber\\
  \begin{tabular}{|c|c|c|}
    \hline
    1 & 2 & 4 \\
    \hline
    3 & 6 \\
    \cline{1-2}
    5 \\
    \cline{1-1}
  \end{tabular}_{CS}
  &=&\frac{\sqrt{3}}{2\sqrt{2}}|C_1 \rangle \otimes |S_3^2 \rangle +\frac{1}{2\sqrt{2}}|C_3 \rangle \otimes |S_3^2 \rangle +\frac{\sqrt{3}}{4}|C_1 \rangle \otimes |S_4^2 \rangle -\frac{\sqrt{3}}{4}|C_2 \rangle \otimes |S_5^2 \rangle -\frac{1}{2\sqrt{2}}|C_5 \rangle \otimes |S_5^2 \rangle \nonumber\\
  \begin{tabular}{|c|c|c|}
    \hline
    1 & 3 & 4 \\
    \hline
    2 & 6 \\
    \cline{1-2}
    5 \\
    \cline{1-1}
  \end{tabular}_{CS}
  &=&\frac{\sqrt{3}}{2\sqrt{2}}|C_2 \rangle \otimes |S_3^2 \rangle +\frac{1}{2\sqrt{2}}|C_4 \rangle \otimes |S_3^2 \rangle -\frac{\sqrt{3}}{4}|C_2 \rangle \otimes |S_4^2 \rangle +\frac{1}{2\sqrt{2}}|C_5 \rangle \otimes |S_4^2 \rangle -\frac{\sqrt{3}}{4}|C_1 \rangle \otimes |S_5^2 \rangle \nonumber\\
  \begin{tabular}{|c|c|c|}
    \hline
    1 & 2 & 5 \\
    \hline
    3 & 6 \\
    \cline{1-2}
    4 \\
    \cline{1-1}
  \end{tabular}_{CS}
  &=&\frac{1}{\sqrt{2}}|C_3 \rangle \otimes |S_2^2 \rangle +\frac{1}{2\sqrt{10}}|C_1 \rangle \otimes |S_3^2 \rangle +\frac{\sqrt{3}}{2\sqrt{10}}|C_3 \rangle \otimes |S_3^2 \rangle -\frac{1}{4\sqrt{5}}|C_1 \rangle \otimes |S_4^2 \rangle +\frac{\sqrt{3}}{2\sqrt{5}}|C_3 \rangle \otimes |S_4^2 \rangle \nonumber\\
  &&+\frac{1}{4\sqrt{5}}|C_2 \rangle \otimes |S_5^2 \rangle -\frac{\sqrt{3}}{2\sqrt{5}}|C_4 \rangle \otimes |S_5^2 \rangle +\frac{\sqrt{3}}{2\sqrt{10}}|C_5 \rangle \otimes |S_5^2 \rangle \nonumber\\
  \begin{tabular}{|c|c|c|}
    \hline
    1 & 3 & 5 \\
    \hline
    2 & 6 \\
    \cline{1-2}
    4 \\
    \cline{1-1}
  \end{tabular}_{CS}
  &=&\frac{1}{\sqrt{2}}|C_4 \rangle \otimes |S_2^2 \rangle +\frac{1}{2\sqrt{10}}|C_2 \rangle \otimes |S_3^2 \rangle +\frac{\sqrt{3}}{2\sqrt{10}}|C_4 \rangle \otimes |S_3^2 \rangle +\frac{1}{4\sqrt{5}}|C_2 \rangle \otimes |S_4^2 \rangle -\frac{\sqrt{3}}{2\sqrt{5}}|C_4 \rangle \otimes |S_4^2 \rangle \nonumber\\
  &&-\frac{\sqrt{3}}{2\sqrt{10}}|C_5 \rangle \otimes |S_4^2 \rangle +\frac{1}{4\sqrt{5}}|C_1 \rangle \otimes |S_5^2 \rangle -\frac{\sqrt{3}}{2\sqrt{5}}|C_3 \rangle \otimes |S_5^2 \rangle \nonumber\\
  \begin{tabular}{|c|c|c|}
    \hline
    1 & 4 & 5 \\
    \hline
    2 & 6 \\
    \cline{1-2}
    3 \\
    \cline{1-1}
  \end{tabular}_{CS}
  &=&\frac{1}{\sqrt{2}}|C_5 \rangle \otimes |S_2^2 \rangle -\frac{\sqrt{3}}{\sqrt{10}}|C_5 \rangle \otimes |S_3^2 \rangle -\frac{1}{2\sqrt{10}}|C_2 \rangle \otimes |S_4^2 \rangle -\frac{\sqrt{3}}{2\sqrt{10}}|C_4 \rangle \otimes |S_4^2 \rangle +\frac{1}{2\sqrt{10}}|C_1 \rangle \otimes |S_5^2 \rangle \nonumber\\
  &&+\frac{\sqrt{3}}{2\sqrt{10}}|C_3 \rangle \otimes |S_5^2 \rangle \nonumber\\
  \begin{tabular}{|c|c|c|}
    \hline
    1 & 2 & 6 \\
    \hline
    3 & 4 \\
    \cline{1-2}
    5 \\
    \cline{1-1}
  \end{tabular}_{CS}
  &=&-\frac{4}{5}|C_1 \rangle \otimes |S_1^2 \rangle -\frac{3\sqrt{3}}{10\sqrt{2}}|C_1 \rangle \otimes |S_2^2 \rangle +\frac{\sqrt{3}}{2\sqrt{10}}|C_3 \rangle \otimes |S_3^2 \rangle -\frac{\sqrt{3}}{4\sqrt{5}}|C_3 \rangle \otimes |S_4^2 \rangle +\frac{\sqrt{3}}{4\sqrt{5}}|C_4 \rangle \otimes |S_5^2 \rangle \nonumber\\
  &&+\frac{\sqrt{3}}{2\sqrt{10}}|C_5 \rangle \otimes |S_5^2 \rangle \nonumber\\
  \begin{tabular}{|c|c|c|}
    \hline
    1 & 3 & 6 \\
    \hline
    2 & 4 \\
    \cline{1-2}
    5 \\
    \cline{1-1}
  \end{tabular}_{CS}
  &=&-\frac{4}{5}|C_2 \rangle \otimes |S_1^2 \rangle -\frac{3\sqrt{3}}{10\sqrt{2}}|C_2 \rangle \otimes |S_2^2 \rangle +\frac{\sqrt{3}}{2\sqrt{10}}|C_4 \rangle \otimes |S_3^2 \rangle +\frac{\sqrt{3}}{4\sqrt{5}}|C_4 \rangle \otimes |S_4^2 \rangle -\frac{\sqrt{3}}{2\sqrt{10}}|C_5 \rangle \otimes |S_4^2 \rangle \nonumber\\
  &&+\frac{\sqrt{3}}{4\sqrt{5}}|C_3 \rangle \otimes |S_5^2 \rangle \nonumber\\
  \begin{tabular}{|c|c|c|}
    \hline
    1 & 2 & 6 \\
    \hline
    3 & 5 \\
    \cline{1-2}
    4 \\
    \cline{1-1}
  \end{tabular}_{CS}
  &=&-\frac{4}{5}|C_3 \rangle \otimes |S_1^2 \rangle +\frac{\sqrt{3}}{5\sqrt{2}}|C_3 \rangle \otimes |S_2^2 \rangle +\frac{\sqrt{3}}{2\sqrt{10}}|C_1 \rangle \otimes |S_3^2 \rangle -\frac{1}{2\sqrt{10}}|C_3 \rangle \otimes |S_3^2 \rangle -\frac{\sqrt{3}}{4\sqrt{5}}|C_1 \rangle \otimes |S_4^2 \rangle \nonumber\\
  &&-\frac{1}{2\sqrt{5}}|C_3 \rangle \otimes |S_4^2 \rangle +\frac{\sqrt{3}}{4\sqrt{5}}|C_2 \rangle \otimes |S_5^2 \rangle +\frac{1}{2\sqrt{5}}|C_4 \rangle \otimes |S_5^2 \rangle -\frac{1}{2\sqrt{10}}|C_5 \rangle \otimes |S_5^2 \rangle \nonumber\\
  \begin{tabular}{|c|c|c|}
    \hline
    1 & 3 & 6 \\
    \hline
    2 & 5 \\
    \cline{1-2}
    4 \\
    \cline{1-1}
  \end{tabular}_{CS}
  &=&-\frac{4}{5}|C_4 \rangle \otimes |S_1^2 \rangle +\frac{\sqrt{3}}{5\sqrt{2}}|C_4 \rangle \otimes |S_2^2 \rangle +\frac{\sqrt{3}}{2\sqrt{10}}|C_2 \rangle \otimes |S_3^2 \rangle -\frac{1}{2\sqrt{10}}|C_4 \rangle \otimes |S_3^2 \rangle +\frac{\sqrt{3}}{4\sqrt{5}}|C_2 \rangle \otimes |S_4^2 \rangle \nonumber\\
  &&+\frac{1}{2\sqrt{5}}|C_4 \rangle \otimes |S_4^2 \rangle +\frac{1}{2\sqrt{10}}|C_5 \rangle \otimes |S_4^2 \rangle +\frac{\sqrt{3}}{4\sqrt{5}}|C_1 \rangle \otimes |S_5^2 \rangle +\frac{1}{2\sqrt{5}}|C_3 \rangle \otimes |S_5^2 \rangle \nonumber\\
  \begin{tabular}{|c|c|c|}
    \hline
    1 & 4 & 6 \\
    \hline
    2 & 5 \\
    \cline{1-2}
    3 \\
    \cline{1-1}
  \end{tabular}_{CS}
  &=&-\frac{4}{5}|C_5 \rangle \otimes |S_1^2 \rangle +\frac{\sqrt{3}}{5\sqrt{2}}|C_5 \rangle \otimes |S_2^2 \rangle +\frac{1}{\sqrt{10}}|C_5 \rangle \otimes |S_3^2 \rangle -\frac{\sqrt{3}}{2\sqrt{10}}|C_2 \rangle \otimes |S_4^2 \rangle +\frac{1}{2\sqrt{10}}|C_4 \rangle \otimes |S_4^2 \rangle \nonumber\\
  &&+\frac{\sqrt{3}}{2\sqrt{10}}|C_1 \rangle \otimes |S_5^2 \rangle -\frac{1}{2\sqrt{10}}|C_3 \rangle \otimes |S_5^2 \rangle \nonumber
\end{eqnarray}
\begin{eqnarray}
  \begin{tabular}{|c|c|}
    \hline
    1 & 2 \\
    \hline
    3 & 4 \\
    \hline
    5 \\
    \cline{1-1}
    6 \\
    \cline{1-1}
  \end{tabular}_{CS}
  &=&-\frac{3}{5}|C_1 \rangle \otimes |S_1^2 \rangle +\frac{\sqrt{6}}{5}|C_1 \rangle \otimes |S_2^2 \rangle -\frac{\sqrt{2}}{\sqrt{15}}|C_3 \rangle \otimes |S_3^2 \rangle +\frac{1}{\sqrt{15}}|C_3 \rangle \otimes |S_4^2 \rangle -\frac{1}{\sqrt{15}}|C_4 \rangle \otimes |S_5^2 \rangle \nonumber\\
  &&-\frac{\sqrt{2}}{\sqrt{15}}|C_5 \rangle \otimes |S_5^2 \rangle \nonumber\\
  \begin{tabular}{|c|c|}
    \hline
    1 & 3 \\
    \hline
    2 & 4 \\
    \hline
    5 \\
    \cline{1-1}
    6 \\
    \cline{1-1}
  \end{tabular}_{CS}
  &=&-\frac{3}{5}|C_2 \rangle \otimes |S_1^2 \rangle +\frac{\sqrt{6}}{5}|C_2 \rangle \otimes |S_2^2 \rangle -\frac{\sqrt{2}}{\sqrt{15}}|C_4 \rangle \otimes |S_3^2 \rangle -\frac{1}{\sqrt{15}}|C_4 \rangle \otimes |S_4^2 \rangle +\frac{\sqrt{2}}{\sqrt{15}}|C_5 \rangle \otimes |S_4^2 \rangle \nonumber\\
  &&-\frac{1}{\sqrt{15}}|C_3 \rangle \otimes |S_5^2 \rangle \nonumber\\
  \begin{tabular}{|c|c|}
    \hline
    1 & 2 \\
    \hline
    3 & 5 \\
    \hline
    4 \\
    \cline{1-1}
    6 \\
    \cline{1-1}
  \end{tabular}_{CS}
  &=&-\frac{3}{5}|C_3 \rangle \otimes |S_1^2 \rangle -\frac{2\sqrt{2}}{5\sqrt{3}}|C_3 \rangle \otimes |S_2^2 \rangle -\frac{\sqrt{2}}{\sqrt{15}}|C_1 \rangle \otimes |S_3^2 \rangle +\frac{\sqrt{2}}{3\sqrt{5}}|C_3 \rangle \otimes |S_3^2 \rangle +\frac{1}{\sqrt{15}}|C_1 \rangle \otimes |S_4^2 \rangle \nonumber\\
  &&+\frac{2}{3\sqrt{5}}|C_3 \rangle \otimes |S_4^2 \rangle -\frac{1}{\sqrt{15}}|C_2 \rangle \otimes |S_5^2 \rangle -\frac{2}{3\sqrt{5}}|C_4 \rangle \otimes |S_5^2 \rangle +\frac{\sqrt{2}}{3\sqrt{5}}|C_5 \rangle \otimes |S_5^2 \rangle \nonumber\\
  \begin{tabular}{|c|c|}
    \hline
    1 & 3 \\
    \hline
    2 & 5 \\
    \hline
    4 \\
    \cline{1-1}
    6 \\
    \cline{1-1}
  \end{tabular}_{CS}
  &=&-\frac{3}{5}|C_4 \rangle \otimes |S_1^2 \rangle -\frac{2\sqrt{2}}{5\sqrt{3}}|C_4 \rangle \otimes |S_2^2 \rangle -\frac{\sqrt{2}}{\sqrt{15}}|C_2 \rangle \otimes |S_3^2 \rangle +\frac{\sqrt{2}}{3\sqrt{5}}|C_4 \rangle \otimes |S_3^2 \rangle -\frac{1}{\sqrt{15}}|C_2 \rangle \otimes |S_4^2 \rangle \nonumber\\
  &&-\frac{2}{3\sqrt{5}}|C_4 \rangle \otimes |S_4^2 \rangle -\frac{\sqrt{2}}{3\sqrt{5}}|C_5 \rangle \otimes |S_4^2 \rangle -\frac{1}{\sqrt{15}}|C_1 \rangle \otimes |S_5^2 \rangle -\frac{2}{3\sqrt{5}}|C_3 \rangle \otimes |S_5^2 \rangle \nonumber\\
  \begin{tabular}{|c|c|}
    \hline
    1 & 4 \\
    \hline
    2 & 5 \\
    \hline
    3 \\
    \cline{1-1}
    6 \\
    \cline{1-1}
  \end{tabular}_{CS}
  &=&-\frac{3}{5}|C_5 \rangle \otimes |S_1^2 \rangle -\frac{2\sqrt{2}}{5\sqrt{3}}|C_5 \rangle \otimes |S_2^2 \rangle -\frac{2\sqrt{2}}{3\sqrt{5}}|C_5 \rangle \otimes |S_3^2 \rangle +\frac{\sqrt{2}}{\sqrt{15}}|C_2 \rangle \otimes |S_4^2 \rangle -\frac{\sqrt{2}}{3\sqrt{5}}|C_4 \rangle \otimes |S_4^2 \rangle \nonumber\\
  &&-\frac{\sqrt{2}}{\sqrt{15}}|C_1 \rangle \otimes |S_5^2 \rangle +\frac{\sqrt{2}}{3\sqrt{5}}|C_3 \rangle \otimes |S_5^2 \rangle \nonumber\\
  \begin{tabular}{|c|c|}
    \hline
    1 & 2 \\
    \hline
    3 & 6 \\
    \hline
    4 \\
    \cline{1-1}
    5 \\
    \cline{1-1}
  \end{tabular}_{CS}
  &=&-\frac{1}{\sqrt{3}}|C_3 \rangle \otimes |S_2^2 \rangle -\frac{2}{\sqrt{15}}|C_1 \rangle \otimes |S_3^2 \rangle -\frac{1}{3\sqrt{5}}|C_3 \rangle \otimes |S_3^2 \rangle +\frac{\sqrt{2}}{\sqrt{15}}|C_1 \rangle \otimes |S_4^2 \rangle -\frac{\sqrt{2}}{3\sqrt{5}}|C_3 \rangle \otimes |S_4^2 \rangle \nonumber\\
  &&-\frac{\sqrt{2}}{\sqrt{15}}|C_2 \rangle \otimes |S_5^2 \rangle +\frac{\sqrt{2}}{3\sqrt{5}}|C_4 \rangle \otimes |S_5^2 \rangle -\frac{1}{3\sqrt{5}}|C_5 \rangle \otimes |S_5^2 \rangle \nonumber\\
  \begin{tabular}{|c|c|}
    \hline
    1 & 3 \\
    \hline
    2 & 6 \\
    \hline
    4 \\
    \cline{1-1}
    5 \\
    \cline{1-1}
  \end{tabular}_{CS}
  &=&\frac{1}{\sqrt{3}}|C_4 \rangle \otimes |S_2^2 \rangle -\frac{2}{\sqrt{15}}|C_2 \rangle \otimes |S_3^2 \rangle -\frac{1}{3\sqrt{5}}|C_4 \rangle \otimes |S_3^2 \rangle -\frac{\sqrt{2}}{\sqrt{15}}|C_2 \rangle \otimes |S_4^2 \rangle +\frac{\sqrt{2}}{3\sqrt{5}}|C_4 \rangle \otimes |S_4^2 \rangle \nonumber\\
  &&+\frac{1}{3\sqrt{5}}|C_5 \rangle \otimes |S_4^2 \rangle -\frac{\sqrt{2}}{\sqrt{15}}|C_1 \rangle \otimes |S_5^2 \rangle +\frac{\sqrt{2}}{3\sqrt{5}}|C_3 \rangle \otimes |S_5^2 \rangle \nonumber\\
  \begin{tabular}{|c|c|}
    \hline
    1 & 4 \\
    \hline
    2 & 6 \\
    \hline
    3 \\
    \cline{1-1}
    5 \\
    \cline{1-1}
  \end{tabular}_{CS}
  &=&\frac{1}{\sqrt{3}}|C_5 \rangle \otimes |S_2^2 \rangle +\frac{2}{3\sqrt{5}}|C_5 \rangle \otimes |S_3^2 \rangle +\frac{2}{\sqrt{15}}|C_2 \rangle \otimes |S_4^2 \rangle +\frac{1}{3\sqrt{5}}|C_4 \rangle \otimes |S_4^2 \rangle -\frac{2}{\sqrt{15}}|C_1 \rangle \otimes |S_5^2 \rangle \nonumber\\
  &&-\frac{1}{3\sqrt{5}}|C_3 \rangle \otimes |S_5^2 \rangle \nonumber\\
  \begin{tabular}{|c|c|}
    \hline
    1 & 5 \\
    \hline
    2 & 6 \\
    \hline
    3 \\
    \cline{1-1}
    4 \\
    \cline{1-1}
  \end{tabular}_{CS}
  &=&-\frac{1}{\sqrt{3}}|C_5 \rangle \otimes |S_3^2 \rangle +\frac{1}{\sqrt{3}}|C_4 \rangle \otimes |S_4^2 \rangle -\frac{1}{\sqrt{3}}|C_3 \rangle \otimes |S_5^2 \rangle \nonumber
\end{eqnarray}
\subsection{S=3}
\begin{eqnarray}
  \begin{tabular}{|c|c|c|}
    \hline
    1 & 2 \\
    \hline
    3 & 4 \\
    \hline
    5 & 6 \\
    \hline
  \end{tabular}_{~CS}
  &=&|C_1 \rangle \otimes |S_1^3 \rangle ,~
  \begin{tabular}{|c|c|c|}
    \hline
    1 & 3 \\
    \hline
    2 & 4 \\
    \hline
    5 & 6 \\
    \hline
  \end{tabular}_{~CS}
  =|C_2 \rangle \otimes |S_1^3 \rangle ,~
  \begin{tabular}{|c|c|c|}
    \hline
    1 & 2 \\
    \hline
    3 & 5 \\
    \hline
    4 & 6 \\
    \hline
  \end{tabular}_{~CS}
  =|C_3 \rangle \otimes |S_1^3 \rangle ,~
  \begin{tabular}{|c|c|c|}
    \hline
    1 & 3 \\
    \hline
    2 & 5 \\
    \hline
    4 & 6 \\
    \hline
  \end{tabular}_{~CS}
  =|C_4 \rangle \otimes |S_1^3 \rangle \nonumber\\
  \begin{tabular}{|c|c|c|}
    \hline
    1 & 4 \\
    \hline
    2 & 5 \\
    \hline
    3 & 6 \\
    \hline
  \end{tabular}_{~CS}
  &=&|C_5 \rangle \otimes |S_1^3 \rangle \nonumber\\
\end{eqnarray}

\section{CS coupling of $q^5$}
\subsection{S=$\frac{1}{2}$}
\begin{eqnarray}
\begin{tabular}{|c|c|c|}
\hline
1 & 2 & 3 \\
\hline
4 & 5  \\
\cline{1-2}
\end{tabular}_{CS}
&=&\frac{1}{2}|C_1 \rangle \otimes |S_2^{\frac{1}{2}}\rangle + \frac{1}{2}|C_2 \rangle \otimes |S_3^{\frac{1}{2}}\rangle + \frac{1}{2}|C_3 \rangle \otimes |S_4^{\frac{1}{2}}\rangle + \frac{1}{2}|C_4 \rangle \otimes |S_5^{\frac{1}{2}}\rangle
\nonumber\\
\begin{tabular}{|c|c|c|}
\hline
1 & 2 & 4 \\
\hline
3 & 5  \\
\cline{1-2}
\end{tabular}_{CS}
&=&\frac{1}{2}|C_1 \rangle \otimes |S_1^{\frac{1}{2}}\rangle + \frac{1}{2\sqrt{2}}|C_1 \rangle \otimes |S_2^{\frac{1}{2}}\rangle -\frac{1}{2\sqrt{2}}|C_2 \rangle \otimes |S_3^{\frac{1}{2}}\rangle-\frac{1}{2\sqrt{2}}|C_3 \rangle \otimes |S_4^{\frac{1}{2}}\rangle+\frac{1}{2\sqrt{2}}|C_4 \rangle \otimes |S_5^{\frac{1}{2}}\rangle-\frac{1}{2}|C_5 \rangle \otimes |S_5^{\frac{1}{2}}\rangle
\nonumber\\
\begin{tabular}{|c|c|c|}
\hline
1 & 3 & 4 \\
\hline
2 & 5  \\
\cline{1-2}
\end{tabular}_{CS}
&=&\frac{1}{2}|C_2 \rangle \otimes |S_1^{\frac{1}{2}}\rangle-\frac{1}{2\sqrt{2}}|C_2 \rangle \otimes |S_2^{\frac{1}{2}}\rangle-\frac{1}{2\sqrt{2}}|C_1 \rangle \otimes |S_3^{\frac{1}{2}}\rangle+\frac{1}{2\sqrt{2}}|C_4 \rangle \otimes |S_4^{\frac{1}{2}}\rangle+\frac{1}{2}|C_5 \rangle \otimes |S_4^{\frac{1}{2}}\rangle+\frac{1}{2\sqrt{2}}|C_3 \rangle \otimes |S_5^{\frac{1}{2}}\rangle
\nonumber\\
\begin{tabular}{|c|c|c|}
\hline
1 & 2 & 5 \\
\hline
3 & 4  \\
\cline{1-2}
\end{tabular}_{CS}
&=&\frac{1}{2}|C_3 \rangle \otimes |S_1^{\frac{1}{2}}\rangle-\frac{1}{2\sqrt{2}}|C_3 \rangle \otimes |S_2^{\frac{1}{2}}\rangle+\frac{1}{2\sqrt{2}}|C_4 \rangle \otimes |S_3^{\frac{1}{2}}\rangle+\frac{1}{2}|C_5 \rangle \otimes |S_3^{\frac{1}{2}}\rangle-\frac{1}{2\sqrt{2}}|C_1 \rangle \otimes |S_4^{\frac{1}{2}}\rangle+\frac{1}{2\sqrt{2}}|C_2 \rangle \otimes |S_5^{\frac{1}{2}}\rangle
\nonumber\\
\begin{tabular}{|c|c|c|}
\hline
1 & 3 & 5 \\
\hline
2 & 4  \\
\cline{1-2}
\end{tabular}_{CS}
&=&\frac{1}{2}|C_4 \rangle \otimes |S_1^{\frac{1}{2}}\rangle+\frac{1}{2\sqrt{2}}|C_4 \rangle \otimes |S_2^{\frac{1}{2}}\rangle-\frac{1}{2}|C_5 \rangle \otimes |S_2^{\frac{1}{2}}\rangle+\frac{1}{2\sqrt{2}}|C_3 \rangle \otimes |S_3^{\frac{1}{2}}\rangle+\frac{1}{2\sqrt{2}}|C_2 \rangle \otimes |S_4^{\frac{1}{2}}\rangle+\frac{1}{2\sqrt{2}}|C_1 \rangle \otimes |S_5^{\frac{1}{2}}\rangle
\nonumber
\end{eqnarray}
\begin{eqnarray}
\begin{tabular}{|c|c|c|}
\hline
1 & 2 & 3 \\
\hline
4   \\
\cline{1-1}
5   \\
\cline{1-1}
\end{tabular}_{CS}
&=&\frac{\sqrt{3}}{2\sqrt{5}}|C_1 \rangle \otimes |S_2^{\frac{1}{2}}\rangle+\frac{1}{\sqrt{5}}|C_3 \rangle \otimes |S_2^{\frac{1}{2}}\rangle+\frac{\sqrt{3}}{2\sqrt{5}}|C_2 \rangle \otimes |S_3^{\frac{1}{2}}\rangle+\frac{1}{\sqrt{5}}|C_4 \rangle \otimes |S_3^{\frac{1}{2}}\rangle-\frac{\sqrt{3}}{2\sqrt{5}}|C_3 \rangle \otimes |S_4^{\frac{1}{2}}\rangle
\nonumber\\
&&-\frac{\sqrt{3}}{2\sqrt{5}}|C_4 \rangle \otimes |S_5^{\frac{1}{2}}\rangle
\nonumber\\
\begin{tabular}{|c|c|c|}
\hline
1 & 2 & 4 \\
\hline
3   \\
\cline{1-1}
5   \\
\cline{1-1}
\end{tabular}_{CS}
&=&\frac{\sqrt{3}}{2\sqrt{5}}|C_1 \rangle \otimes |S_1^{\frac{1}{2}}\rangle-\frac{1}{\sqrt{5}}|C_3 \rangle \otimes |S_1^{\frac{1}{2}}\rangle+\frac{\sqrt{3}}{2\sqrt{10}}|C_1 \rangle \otimes |S_2^{\frac{1}{2}}\rangle-\frac{\sqrt{3}}{2\sqrt{10}}|C_2 \rangle \otimes |S_3^{\frac{1}{2}}\rangle+\frac{1}{\sqrt{5}}|C_5 \rangle \otimes |S_3^{\frac{1}{2}}\rangle \nonumber\\
&&+\frac{\sqrt{3}}{2\sqrt{10}}|C_3 \rangle \otimes |S_4^{\frac{1}{2}}\rangle-\frac{\sqrt{3}}{2\sqrt{10}}|C_4 \rangle \otimes |S_5^{\frac{1}{2}}\rangle+\frac{\sqrt{3}}{2\sqrt{5}}|C_5 \rangle \otimes |S_5^{\frac{1}{2}}\rangle
\nonumber\\
\begin{tabular}{|c|c|c|}
\hline
1 & 3 & 4 \\
\hline
2   \\
\cline{1-1}
5   \\
\cline{1-1}
\end{tabular}_{CS}
&=&\frac{\sqrt{3}}{2\sqrt{5}}|C_2 \rangle \otimes |S_1^{\frac{1}{2}}\rangle-\frac{1}{\sqrt{5}}|C_4 \rangle \otimes |S_1^{\frac{1}{2}}\rangle-\frac{\sqrt{3}}{2\sqrt{10}}|C_2 \rangle \otimes |S_2^{\frac{1}{2}}\rangle-\frac{1}{\sqrt{5}}|C_5 \rangle \otimes |S_2^{\frac{1}{2}}\rangle-\frac{\sqrt{3}}{2\sqrt{10}}|C_1 \rangle \otimes |S_3^{\frac{1}{2}}\rangle \nonumber\\
&&-\frac{\sqrt{3}}{2\sqrt{10}}|C_4 \rangle \otimes |S_4^{\frac{1}{2}}\rangle-\frac{\sqrt{3}}{2\sqrt{5}}|C_5 \rangle \otimes |S_4^{\frac{1}{2}}\rangle-\frac{\sqrt{3}}{2\sqrt{10}}|C_3 \rangle \otimes |S_5^{\frac{1}{2}}\rangle
\nonumber\\
\begin{tabular}{|c|c|c|}
\hline
1 & 2 & 5 \\
\hline
3   \\
\cline{1-1}
4   \\
\cline{1-1}
\end{tabular}_{CS}
&=&-\frac{1}{2}|C_1 \rangle \otimes |S_1^{\frac{1}{2}}\rangle+\frac{1}{2\sqrt{2}}|C_1 \rangle \otimes |S_2^{\frac{1}{2}}\rangle-\frac{1}{2\sqrt{2}}|C_2 \rangle \otimes |S_3^{\frac{1}{2}}\rangle+\frac{1}{2\sqrt{2}}|C_3 \rangle \otimes |S_4^{\frac{1}{2}}\rangle-\frac{1}{2\sqrt{2}}|C_4 \rangle \otimes |S_5^{\frac{1}{2}}\rangle
\nonumber\\
&&-\frac{1}{2}|C_5 \rangle \otimes |S_5^{\frac{1}{2}}\rangle
\nonumber\\
\begin{tabular}{|c|c|c|}
\hline
1 & 3 & 5 \\
\hline
2   \\
\cline{1-1}
4   \\
\cline{1-1}
\end{tabular}_{CS}
&=&-\frac{1}{2}|C_2 \rangle \otimes |S_1^{\frac{1}{2}}\rangle-\frac{1}{2\sqrt{2}}|C_2 \rangle \otimes |S_2^{\frac{1}{2}}\rangle-\frac{1}{2\sqrt{2}}|C_1 \rangle \otimes |S_3^{\frac{1}{2}}\rangle-\frac{1}{2\sqrt{2}}|C_4 \rangle \otimes |S_4^{\frac{1}{2}}\rangle+\frac{1}{2}|C_5 \rangle \otimes |S_4^{\frac{1}{2}}\rangle
\nonumber\\
&&-\frac{1}{2\sqrt{2}}|C_3 \rangle \otimes |S_5^{\frac{1}{2}}\rangle
\nonumber\\
\begin{tabular}{|c|c|c|}
\hline
1 & 4 & 5 \\
\hline
2   \\
\cline{1-1}
3   \\
\cline{1-1}
\end{tabular}_{CS}
&=&\frac{1}{2}|C_2 \rangle \otimes |S_2^{\frac{1}{2}}\rangle-\frac{1}{2}|C_1 \rangle \otimes |S_3^{\frac{1}{2}}\rangle+\frac{1}{2}|C_4 \rangle \otimes |S_4^{\frac{1}{2}}\rangle-\frac{1}{2}|C_3 \rangle \otimes |S_5^{\frac{1}{2}}\rangle
\nonumber
\end{eqnarray}
\begin{eqnarray}
\begin{tabular}{|c|c|}
\hline
1 & 2  \\
\hline
3 & 4   \\
\cline{1-2}
5   \\
\cline{1-1}
\end{tabular}_{CS}
&=&-\frac{1}{2\sqrt{3}}|C_3 \rangle \otimes |S_1^{\frac{1}{2}}\rangle+\frac{1}{2\sqrt{6}}|C_3 \rangle \otimes |S_2^{\frac{1}{2}}\rangle-\frac{1}{2\sqrt{6}}|C_4 \rangle \otimes |S_3^{\frac{1}{2}}\rangle-\frac{1}{2\sqrt{3}}|C_5 \rangle \otimes |S_3^{\frac{1}{2}}\rangle-\frac{\sqrt{3}}{2\sqrt{2}}|C_1 \rangle \otimes |S_4^{\frac{1}{2}}\rangle
\nonumber\\
&&+\frac{\sqrt{3}}{2\sqrt{2}}|C_2 \rangle \otimes |S_5^{\frac{1}{2}}\rangle
\nonumber\\
\begin{tabular}{|c|c|}
\hline
1 & 3  \\
\hline
2 & 4   \\
\cline{1-2}
5   \\
\cline{1-1}
\end{tabular}_{CS}
&=&-\frac{1}{2\sqrt{3}}|C_4 \rangle \otimes |S_1^{\frac{1}{2}}\rangle-\frac{1}{2\sqrt{6}}|C_4 \rangle \otimes |S_2^{\frac{1}{2}}\rangle+\frac{1}{2\sqrt{3}}|C_5 \rangle \otimes |S_2^{\frac{1}{2}}\rangle-\frac{1}{2\sqrt{6}}|C_3 \rangle \otimes |S_3^{\frac{1}{2}}\rangle+\frac{\sqrt{3}}{2\sqrt{2}}|C_2 \rangle \otimes |S_4^{\frac{1}{2}}\rangle
\nonumber\\
&&+\frac{\sqrt{3}}{2\sqrt{2}}|C_1 \rangle \otimes |S_5^{\frac{1}{2}}\rangle
\nonumber\\
\begin{tabular}{|c|c|}
\hline
1 & 2  \\
\hline
3 & 5   \\
\cline{1-2}
4   \\
\cline{1-1}
\end{tabular}_{CS}
&=&-\frac{1}{2\sqrt{3}}|C_1 \rangle \otimes |S_1^{\frac{1}{2}}\rangle-\frac{1}{3}|C_3 \rangle \otimes |S_1^{\frac{1}{2}}\rangle+\frac{1}{2\sqrt{6}}|C_1 \rangle \otimes |S_2^{\frac{1}{2}}\rangle-\frac{\sqrt{2}}{3}|C_3 \rangle \otimes |S_2^{\frac{1}{2}}\rangle-\frac{1}{2\sqrt{6}}|C_2 \rangle \otimes |S_3^{\frac{1}{2}}\rangle
\nonumber\\
&&+\frac{\sqrt{2}}{3}|C_4 \rangle \otimes |S_3^{\frac{1}{2}}\rangle-\frac{1}{3}|C_5 \rangle \otimes |S_3^{\frac{1}{2}}\rangle-\frac{1}{2\sqrt{6}}|C_3 \rangle \otimes |S_4^{\frac{1}{2}}\rangle+\frac{1}{2\sqrt{6}}|C_4 \rangle \otimes |S_5^{\frac{1}{2}}\rangle+\frac{1}{2\sqrt{3}}|C_5 \rangle \otimes |S_5^{\frac{1}{2}}\rangle
\nonumber\\
\begin{tabular}{|c|c|}
\hline
1 & 3  \\
\hline
2 & 5   \\
\cline{1-2}
4   \\
\cline{1-1}
\end{tabular}_{CS}
&=&-\frac{1}{2\sqrt{3}}|C_2 \rangle \otimes |S_1^{\frac{1}{2}}\rangle-\frac{1}{3}|C_4 \rangle \otimes |S_1^{\frac{1}{2}}\rangle-\frac{1}{2\sqrt{6}}|C_2 \rangle \otimes |S_2^{\frac{1}{2}}\rangle+\frac{\sqrt{2}}{3}|C_4 \rangle \otimes |S_2^{\frac{1}{2}}\rangle+\frac{1}{3}|C_5 \rangle \otimes |S_2^{\frac{1}{2}}\rangle
\nonumber\\
&&-\frac{1}{2\sqrt{6}}|C_1 \rangle \otimes |S_3^{\frac{1}{2}}\rangle+\frac{\sqrt{2}}{3}|C_3 \rangle \otimes |S_3^{\frac{1}{2}}\rangle+\frac{1}{2\sqrt{6}}|C_4 \rangle \otimes |S_4^{\frac{1}{2}}\rangle-\frac{1}{2\sqrt{3}}|C_5 \rangle \otimes |S_4^{\frac{1}{2}}\rangle+\frac{1}{2\sqrt{6}}|C_3 \rangle \otimes |S_5^{\frac{1}{2}}\rangle
\nonumber\\
\begin{tabular}{|c|c|}
\hline
1 & 4  \\
\hline
2 & 5   \\
\cline{1-2}
3   \\
\cline{1-1}
\end{tabular}_{CS}
&=&\frac{2}{3}|C_5 \rangle \otimes |S_1^{\frac{1}{2}}\rangle+\frac{1}{2\sqrt{3}}|C_2 \rangle \otimes |S_2^{\frac{1}{2}}\rangle+\frac{1}{3}|C_4 \rangle \otimes |S_2^{\frac{1}{2}}\rangle-\frac{1}{2\sqrt{3}}|C_1 \rangle \otimes |S_3^{\frac{1}{2}}\rangle-\frac{1}{3}|C_3 \rangle \otimes |S_3^{\frac{1}{2}}\rangle
\nonumber\\
&&-\frac{1}{2\sqrt{3}}|C_4 \rangle \otimes |S_4^{\frac{1}{2}}\rangle+\frac{1}{2\sqrt{3}}|C_3 \rangle \otimes |S_5^{\frac{1}{2}}\rangle
\nonumber
\end{eqnarray}
\begin{eqnarray}
\begin{tabular}{|c|c|}
\hline
1 & 2  \\
\hline
3   \\
\cline{1-1}
4   \\
\cline{1-1}
5   \\
\cline{1-1}
\end{tabular}_{CS}
&=&-\frac{1}{\sqrt{6}}|C_1 \rangle \otimes |S_1^{\frac{1}{2}}\rangle+\frac{1}{3\sqrt{2}}|C_3 \rangle \otimes |S_1^{\frac{1}{2}}\rangle+\frac{1}{2\sqrt{3}}|C_1 \rangle \otimes |S_2^{\frac{1}{2}}\rangle+\frac{1}{3}|C_3 \rangle \otimes |S_2^{\frac{1}{2}}\rangle-\frac{1}{2\sqrt{3}}|C_2 \rangle \otimes |S_3^{\frac{1}{2}}\rangle
\nonumber\\
&&-\frac{1}{3}|C_4 \rangle \otimes |S_3^{\frac{1}{2}}\rangle+\frac{1}{3\sqrt{2}}|C_5 \rangle \otimes |S_3^{\frac{1}{2}}\rangle-\frac{1}{2\sqrt{3}}|C_3 \rangle \otimes |S_4^{\frac{1}{2}}\rangle+\frac{1}{2\sqrt{3}}|C_4 \rangle \otimes |S_5^{\frac{1}{2}}\rangle+\frac{1}{\sqrt{6}}|C_5 \rangle \otimes |S_5^{\frac{1}{2}}\rangle
\nonumber\\
\begin{tabular}{|c|c|}
\hline
1 & 3  \\
\hline
2   \\
\cline{1-1}
4   \\
\cline{1-1}
5   \\
\cline{1-1}
\end{tabular}_{CS}
&=&-\frac{1}{\sqrt{6}}|C_2 \rangle \otimes |S_1^{\frac{1}{2}}\rangle+\frac{1}{3\sqrt{2}}|C_4 \rangle \otimes |S_1^{\frac{1}{2}}\rangle-\frac{1}{2\sqrt{3}}|C_2 \rangle \otimes |S_2^{\frac{1}{2}}\rangle-\frac{1}{3}|C_4 \rangle \otimes |S_2^{\frac{1}{2}}\rangle-\frac{1}{3\sqrt{2}}|C_5 \rangle \otimes |S_2^{\frac{1}{2}}\rangle
\nonumber\\
&&-\frac{1}{2\sqrt{3}}|C_1 \rangle \otimes |S_3^{\frac{1}{2}}\rangle-\frac{1}{3}|C_3 \rangle \otimes |S_3^{\frac{1}{2}}\rangle+\frac{1}{2\sqrt{3}}|C_4 \rangle \otimes |S_4^{\frac{1}{2}}\rangle-\frac{1}{\sqrt{6}}|C_5 \rangle \otimes |S_4^{\frac{1}{2}}\rangle+\frac{1}{2\sqrt{3}}|C_3 \rangle \otimes |S_5^{\frac{1}{2}}\rangle
\nonumber\\
\begin{tabular}{|c|c|}
\hline
1 & 4  \\
\hline
2   \\
\cline{1-1}
3   \\
\cline{1-1}
5   \\
\cline{1-1}
\end{tabular}_{CS}
&=&-\frac{\sqrt{2}}{3}|C_5 \rangle \otimes |S_1^{\frac{1}{2}}\rangle+\frac{1}{\sqrt{6}}|C_2 \rangle \otimes |S_2^{\frac{1}{2}}\rangle-\frac{1}{3\sqrt{2}}|C_4 \rangle \otimes |S_2^{\frac{1}{2}}\rangle-\frac{1}{\sqrt{6}}|C_1 \rangle \otimes |S_3^{\frac{1}{2}}\rangle+\frac{1}{3\sqrt{2}}|C_3 \rangle \otimes |S_3^{\frac{1}{2}}\rangle
\nonumber\\
&&-\frac{1}{\sqrt{6}}|C_4 \rangle \otimes |S_4^{\frac{1}{2}}\rangle+\frac{1}{\sqrt{6}}|C_3 \rangle \otimes |S_5^{\frac{1}{2}}\rangle
\nonumber\\
\begin{tabular}{|c|c|}
\hline
1 & 5  \\
\hline
2   \\
\cline{1-1}
3   \\
\cline{1-1}
4   \\
\cline{1-1}
\end{tabular}_{CS}
&=&\frac{\sqrt{2}}{\sqrt{15}}|C_5 \rangle \otimes |S_1^{\frac{1}{2}}\rangle-\frac{\sqrt{2}}{\sqrt{15}}|C_4 \rangle \otimes |S_2^{\frac{1}{2}}\rangle+\frac{\sqrt{2}}{\sqrt{15}}|C_3 \rangle \otimes |S_3^{\frac{1}{2}}\rangle-\frac{\sqrt{3}}{\sqrt{10}}|C_2 \rangle \otimes |S_4^{\frac{1}{2}}\rangle+\frac{\sqrt{3}}{\sqrt{10}}|C_1 \rangle \otimes |S_5^{\frac{1}{2}}\rangle
\nonumber
\end{eqnarray}
\subsection{$S=\frac{3}{2}$}
\begin{eqnarray}
\begin{tabular}{|c|c|c|}
\hline
1 & 2 & 3  \\
\hline
4 & 5   \\
\cline{1-2}
\end{tabular}_{CS}
&=&-\frac{1}{2\sqrt{2}}|C_1 \rangle \otimes |S_3^{\frac{3}{2}}\rangle-\frac{\sqrt{3}}{2\sqrt{2}}|C_3 \rangle \otimes |S_3^{\frac{3}{2}}\rangle-\frac{1}{2\sqrt{2}}|C_2 \rangle \otimes |S_4^{\frac{3}{2}}\rangle-\frac{\sqrt{3}}{2\sqrt{2}}|C_4 \rangle \otimes |S_4^{\frac{3}{2}}\rangle
\nonumber\\
\begin{tabular}{|c|c|c|}
\hline
1 & 2 & 4  \\
\hline
3 & 5   \\
\cline{1-2}
\end{tabular}_{CS}
&=&-\frac{1}{2\sqrt{2}}|C_1 \rangle \otimes |S_2^{\frac{3}{2}}\rangle-\frac{1}{4}|C_1 \rangle \otimes |S_3^{\frac{3}{2}}\rangle+\frac{1}{4}|C_2 \rangle \otimes |S_4^{\frac{3}{2}}\rangle+\frac{\sqrt{3}}{2\sqrt{2}}|C_3 \rangle \otimes |S_2^{\frac{3}{2}}\rangle-\frac{\sqrt{3}}{2\sqrt{2}}|C_5 \rangle \otimes |S_4^{\frac{3}{2}}\rangle
\nonumber\\
\begin{tabular}{|c|c|c|}
\hline
1 & 3 & 4  \\
\hline
2 & 5   \\
\cline{1-2}
\end{tabular}_{CS}
&=&\frac{1}{4}|C_1 \rangle \otimes |S_4^{\frac{3}{2}}\rangle-\frac{1}{2\sqrt{2}}|C_2 \rangle \otimes |S_2^{\frac{3}{2}}\rangle+\frac{1}{4}|C_2 \rangle \otimes |S_3^{\frac{3}{2}}\rangle+\frac{\sqrt{3}}{2\sqrt{2}}|C_4 \rangle \otimes |S_2^{\frac{3}{2}}\rangle+\frac{\sqrt{3}}{2\sqrt{2}}|C_5 \rangle \otimes |S_3^{\frac{3}{2}}\rangle
\nonumber\\
\begin{tabular}{|c|c|c|}
\hline
1 & 2 & 5  \\
\hline
3 & 4   \\
\cline{1-2}
\end{tabular}_{CS}
&=&\frac{\sqrt{5}}{2\sqrt{2}}|C_1 \rangle \otimes |S_1^{\frac{3}{2}}\rangle+\frac{1}{2\sqrt{2}}|C_3 \rangle \otimes |S_2^{\frac{3}{2}}\rangle-\frac{1}{4}|C_3 \rangle \otimes |S_3^{\frac{3}{2}}\rangle+\frac{1}{4}|C_4 \rangle \otimes |S_4^{\frac{3}{2}}\rangle+\frac{1}{2\sqrt{2}}|C_5 \rangle \otimes |S_4^{\frac{3}{2}}\rangle
\nonumber\\
\begin{tabular}{|c|c|c|}
\hline
1 & 3 & 5  \\
\hline
2 & 4   \\
\cline{1-2}
\end{tabular}_{CS}
&=&\frac{\sqrt{5}}{2\sqrt{2}}|C_2 \rangle \otimes |S_1^{\frac{3}{2}}\rangle+\frac{1}{4}|C_3 \rangle \otimes |S_4^{\frac{3}{2}}\rangle+\frac{1}{2\sqrt{2}}|C_4 \rangle \otimes |S_2^{\frac{3}{2}}\rangle+\frac{1}{4}|C_4 \rangle \otimes |S_3^{\frac{3}{2}}\rangle-\frac{1}{2\sqrt{2}}|C_5 \rangle \otimes |S_3^{\frac{3}{2}}\rangle
\nonumber
\end{eqnarray}
\begin{eqnarray}
\begin{tabular}{|c|c|c|}
\hline
1 & 2 & 3  \\
\hline
4   \\
\cline{1-1}
5   \\
\cline{1-1}
\end{tabular}_{CS}
&=&\frac{\sqrt{3}}{2\sqrt{2}}|C_1 \rangle \otimes |S_3^{\frac{3}{2}}\rangle-\frac{1}{2\sqrt{2}}|C_3 \rangle \otimes |S_3^{\frac{3}{2}}\rangle+\frac{\sqrt{3}}{2\sqrt{2}}|C_2 \rangle \otimes |S_4^{\frac{3}{2}}\rangle-\frac{1}{2\sqrt{2}}|C_4 \rangle \otimes |S_4^{\frac{3}{2}}\rangle
\nonumber\\
\begin{tabular}{|c|c|c|}
\hline
1 & 2 & 4  \\
\hline
3   \\
\cline{1-1}
5   \\
\cline{1-1}
\end{tabular}_{CS}
&=&\frac{\sqrt{3}}{2\sqrt{2}}|C_1 \rangle \otimes |S_2^{\frac{3}{2}}\rangle+\frac{1}{2\sqrt{2}}|C_3 \rangle \otimes |S_2^{\frac{3}{2}}\rangle+\frac{\sqrt{3}}{4}|C_1 \rangle \otimes |S_3^{\frac{3}{2}}\rangle-\frac{\sqrt{3}}{4}|C_2 \rangle \otimes |S_4^{\frac{3}{2}}\rangle-\frac{1}{2\sqrt{2}}|C_5 \rangle \otimes |S_4^{\frac{3}{2}}\rangle
\nonumber\\
\begin{tabular}{|c|c|c|}
\hline
1 & 3 & 4  \\
\hline
2   \\
\cline{1-1}
5   \\
\cline{1-1}
\end{tabular}_{CS}
&=&\frac{\sqrt{3}}{2\sqrt{2}}|C_2 \rangle \otimes |S_2^{\frac{3}{2}}\rangle+\frac{1}{2\sqrt{2}}|C_4 \rangle \otimes |S_2^{\frac{3}{2}}\rangle-\frac{\sqrt{3}}{4}|C_2 \rangle \otimes |S_3^{\frac{3}{2}}\rangle+\frac{1}{2\sqrt{2}}|C_5 \rangle \otimes |S_3^{\frac{3}{2}}\rangle-\frac{\sqrt{3}}{4}|C_1 \rangle \otimes |S_4^{\frac{3}{2}}\rangle
\nonumber\\
\begin{tabular}{|c|c|c|}
\hline
1 & 2 & 5  \\
\hline
3   \\
\cline{1-1}
4   \\
\cline{1-1}
\end{tabular}_{CS}
&=&\frac{1}{\sqrt{2}}|C_3 \rangle \otimes |S_1^{\frac{3}{2}}\rangle+\frac{1}{2\sqrt{10}}|C_1 \rangle \otimes |S_2^{\frac{3}{2}}\rangle+\frac{\sqrt{3}}{2\sqrt{10}}|C_3 \rangle \otimes |S_2^{\frac{3}{2}}\rangle-\frac{1}{4\sqrt{5}}|C_1 \rangle \otimes |S_3^{\frac{3}{2}}\rangle
\nonumber\\
&&+\frac{\sqrt{3}}{2\sqrt{5}}|C_3 \rangle \otimes |S_3^{\frac{3}{2}}\rangle+\frac{1}{4\sqrt{5}}|C_2 \rangle \otimes |S_4^{\frac{3}{2}}\rangle-\frac{\sqrt{3}}{2\sqrt{5}}|C_4 \rangle \otimes |S_4^{\frac{3}{2}}\rangle+\frac{\sqrt{3}}{2\sqrt{10}}|C_5 \rangle \otimes |S_4^{\frac{3}{2}}\rangle
\nonumber\\
\begin{tabular}{|c|c|c|}
\hline
1 & 3 & 5  \\
\hline
2   \\
\cline{1-1}
4   \\
\cline{1-1}
\end{tabular}_{CS}
&=&\frac{1}{\sqrt{2}}|C_4 \rangle \otimes |S_1^{\frac{3}{2}}\rangle+\frac{1}{2\sqrt{10}}|C_2 \rangle \otimes |S_2^{\frac{3}{2}}\rangle+\frac{\sqrt{3}}{2\sqrt{10}}|C_4 \rangle \otimes |S_2^{\frac{3}{2}}\rangle+\frac{1}{4\sqrt{5}}|C_2 \rangle \otimes |S_3^{\frac{3}{2}}\rangle
\nonumber\\
&&-\frac{\sqrt{3}}{2\sqrt{5}}|C_4 \rangle \otimes |S_3^{\frac{3}{2}}\rangle-\frac{\sqrt{3}}{2\sqrt{10}}|C_5 \rangle \otimes |S_3^{\frac{3}{2}}\rangle+\frac{1}{4\sqrt{5}}|C_1 \rangle \otimes |S_4^{\frac{3}{2}}\rangle-\frac{\sqrt{3}}{2\sqrt{5}}|C_3 \rangle \otimes |S_4^{\frac{3}{2}}\rangle
\nonumber\\
\begin{tabular}{|c|c|c|}
\hline
1 & 4 & 5  \\
\hline
2   \\
\cline{1-1}
3   \\
\cline{1-1}
\end{tabular}_{CS}
&=&\frac{1}{\sqrt{2}}|C_5 \rangle \otimes |S_1^{\frac{3}{2}}\rangle-\frac{\sqrt{3}}{\sqrt{10}}|C_5 \rangle \otimes |S_2^{\frac{3}{2}}\rangle-\frac{1}{2\sqrt{10}}|C_2 \rangle \otimes |S_3^{\frac{3}{2}}\rangle-\frac{\sqrt{3}}{2\sqrt{10}}|C_4 \rangle \otimes |S_3^{\frac{3}{2}}\rangle
\nonumber\\
&&+\frac{1}{2\sqrt{10}}|C_1 \rangle \otimes |S_4^{\frac{3}{2}}\rangle+\frac{\sqrt{3}}{2\sqrt{10}}|C_3 \rangle \otimes |S_4^{\frac{3}{2}}\rangle
\nonumber
\end{eqnarray}
\begin{eqnarray}
\begin{tabular}{|c|c|}
\hline
1 & 2   \\
\hline
3 & 4   \\
\cline{1-2}
5   \\
\cline{1-1}
\end{tabular}_{CS}
&=&-\frac{\sqrt{3}}{2\sqrt{2}}|C_1 \rangle \otimes |S_1^{\frac{3}{2}}\rangle+\frac{\sqrt{5}}{2\sqrt{6}}|C_3 \rangle \otimes |S_2^{\frac{3}{2}}\rangle-\frac{\sqrt{5}}{4\sqrt{3}}|C_3 \rangle \otimes |S_3^{\frac{3}{2}}\rangle+\frac{\sqrt{5}}{4\sqrt{3}}|C_4 \rangle \otimes |S_4^{\frac{3}{2}}\rangle+\frac{\sqrt{5}}{2\sqrt{6}}|C_5 \rangle \otimes |S_4^{\frac{3}{2}}\rangle
\nonumber\\
\begin{tabular}{|c|c|}
\hline
1 & 3   \\
\hline
2 & 4   \\
\cline{1-2}
5   \\
\cline{1-1}
\end{tabular}_{CS}
&=&-\frac{\sqrt{3}}{2\sqrt{2}}|C_2 \rangle \otimes |S_1^{\frac{3}{2}}\rangle+\frac{\sqrt{5}}{2\sqrt{6}}|C_4 \rangle \otimes |S_2^{\frac{3}{2}}\rangle+\frac{\sqrt{5}}{4\sqrt{3}}|C_4 \rangle \otimes |S_3^{\frac{3}{2}}\rangle-\frac{\sqrt{5}}{2\sqrt{6}}|C_5 \rangle \otimes |S_3^{\frac{3}{2}}\rangle+\frac{\sqrt{5}}{4\sqrt{3}}|C_3 \rangle \otimes |S_4^{\frac{3}{2}}\rangle
\nonumber\\
\begin{tabular}{|c|c|}
\hline
1 & 2   \\
\hline
3 & 5   \\
\cline{1-2}
4   \\
\cline{1-1}
\end{tabular}_{CS}
&=&\frac{1}{\sqrt{6}}|C_3 \rangle \otimes |S_1^{\frac{3}{2}}\rangle+\frac{\sqrt{5}}{2\sqrt{6}}|C_1 \rangle \otimes |S_2^{\frac{3}{2}}\rangle-\frac{\sqrt{5}}{6\sqrt{2}}|C_3 \rangle \otimes |S_2^{\frac{3}{2}}\rangle-\frac{\sqrt{5}}{4\sqrt{3}}|C_1 \rangle \otimes |S_3^{\frac{3}{2}}\rangle
\nonumber\\
&&-\frac{\sqrt{5}}{6}|C_3 \rangle \otimes |S_3^{\frac{3}{2}}\rangle+\frac{\sqrt{5}}{4\sqrt{3}}|C_2 \rangle \otimes |S_4^{\frac{3}{2}}\rangle+\frac{\sqrt{5}}{6}|C_4 \rangle \otimes |S_4^{\frac{3}{2}}\rangle-\frac{\sqrt{5}}{6\sqrt{2}}|C_5 \rangle \otimes |S_4^{\frac{3}{2}}\rangle
\nonumber\\
\begin{tabular}{|c|c|}
\hline
1 & 3   \\
\hline
2 & 5   \\
\cline{1-2}
4   \\
\cline{1-1}
\end{tabular}_{CS}
&=&\frac{1}{\sqrt{6}}|C_4 \rangle \otimes |S_1^{\frac{3}{2}}\rangle+\frac{\sqrt{5}}{2\sqrt{6}}|C_2 \rangle \otimes |S_2^{\frac{3}{2}}\rangle-\frac{\sqrt{5}}{6\sqrt{2}}|C_4 \rangle \otimes |S_2^{\frac{3}{2}}\rangle+\frac{\sqrt{5}}{4\sqrt{3}}|C_2 \rangle \otimes |S_3^{\frac{3}{2}}\rangle
\nonumber\\
&&+\frac{\sqrt{5}}{6}|C_4 \rangle \otimes |S_3^{\frac{3}{2}}\rangle+\frac{\sqrt{5}}{6\sqrt{2}}|C_5 \rangle \otimes |S_3^{\frac{3}{2}}\rangle+\frac{\sqrt{5}}{4\sqrt{3}}|C_1 \rangle \otimes |S_4^{\frac{3}{2}}\rangle+\frac{\sqrt{5}}{6}|C_3 \rangle \otimes |S_4^{\frac{3}{2}}\rangle
\nonumber\\
\begin{tabular}{|c|c|}
\hline
1 & 4   \\
\hline
2 & 5   \\
\cline{1-2}
3   \\
\cline{1-1}
\end{tabular}_{CS}
&=&\frac{1}{\sqrt{6}}|C_5 \rangle \otimes |S_1^{\frac{3}{2}}\rangle+\frac{\sqrt{5}}{3\sqrt{2}}|C_5 \rangle \otimes |S_2^{\frac{3}{2}}\rangle-\frac{\sqrt{5}}{2\sqrt{6}}|C_2 \rangle \otimes |S_3^{\frac{3}{2}}\rangle+\frac{\sqrt{5}}{6\sqrt{2}}|C_4 \rangle \otimes |S_3^{\frac{3}{2}}\rangle
\nonumber\\
&&+\frac{\sqrt{5}}{2\sqrt{6}}|C_1 \rangle \otimes |S_4^{\frac{3}{2}}\rangle-\frac{\sqrt{5}}{6\sqrt{2}}|C_3 \rangle \otimes |S_4^{\frac{3}{2}}\rangle
\nonumber
\end{eqnarray}
\begin{eqnarray}
\begin{tabular}{|c|c|}
\hline
1 & 2   \\
\hline
3   \\
\cline{1-1}
4   \\
\cline{1-1}
5   \\
\cline{1-1}
\end{tabular}_{CS}
&=&-\frac{1}{\sqrt{3}}|C_3 \rangle \otimes |S_1^{\frac{3}{2}}\rangle+\frac{2}{\sqrt{15}}|C_1 \rangle \otimes |S_2^{\frac{3}{2}}\rangle+\frac{1}{3\sqrt{5}}|C_3 \rangle \otimes |S_2^{\frac{3}{2}}\rangle-\frac{\sqrt{2}}{\sqrt{15}}|C_1 \rangle \otimes |S_3^{\frac{3}{2}}\rangle
\nonumber\\
&&+\frac{\sqrt{2}}{3\sqrt{5}}|C_3 \rangle \otimes |S_3^{\frac{3}{2}}\rangle+\frac{\sqrt{2}}{\sqrt{15}}|C_2 \rangle \otimes |S_4^{\frac{3}{2}}\rangle-\frac{\sqrt{2}}{3\sqrt{5}}|C_4 \rangle \otimes |S_4^{\frac{3}{2}}\rangle+\frac{1}{3\sqrt{5}}|C_5 \rangle \otimes |S_4^{\frac{3}{2}}\rangle
\nonumber\\
\begin{tabular}{|c|c|}
\hline
1 & 3   \\
\hline
2   \\
\cline{1-1}
4   \\
\cline{1-1}
5   \\
\cline{1-1}
\end{tabular}_{CS}
&=&-\frac{1}{\sqrt{3}}|C_4 \rangle \otimes |S_1^{\frac{3}{2}}\rangle+\frac{2}{\sqrt{15}}|C_2 \rangle \otimes |S_2^{\frac{3}{2}}\rangle+\frac{1}{3\sqrt{5}}|C_4 \rangle \otimes |S_2^{\frac{3}{2}}\rangle+\frac{\sqrt{2}}{\sqrt{15}}|C_2 \rangle \otimes |S_3^{\frac{3}{2}}\rangle
\nonumber\\
&&-\frac{\sqrt{2}}{3\sqrt{5}}|C_4 \rangle \otimes |S_3^{\frac{3}{2}}\rangle-\frac{1}{3\sqrt{5}}|C_5 \rangle \otimes |S_3^{\frac{3}{2}}\rangle+\frac{\sqrt{2}}{\sqrt{15}}|C_1 \rangle \otimes |S_4^{\frac{3}{2}}\rangle-\frac{\sqrt{2}}{3\sqrt{5}}|C_3 \rangle \otimes |S_4^{\frac{3}{2}}\rangle
\nonumber\\
\begin{tabular}{|c|c|}
\hline
1 & 4   \\
\hline
2   \\
\cline{1-1}
3   \\
\cline{1-1}
5   \\
\cline{1-1}
\end{tabular}_{CS}
&=&-\frac{1}{\sqrt{3}}|C_5 \rangle \otimes |S_1^{\frac{3}{2}}\rangle-\frac{2}{3\sqrt{5}}|C_5 \rangle \otimes |S_2^{\frac{3}{2}}\rangle-\frac{2}{\sqrt{15}}|C_2 \rangle \otimes |S_3^{\frac{3}{2}}\rangle-\frac{1}{3\sqrt{5}}|C_4 \rangle \otimes |S_3^{\frac{3}{2}}\rangle
\nonumber\\
&&+\frac{2}{\sqrt{15}}|C_1 \rangle \otimes |S_4^{\frac{3}{2}}\rangle+\frac{1}{3\sqrt{5}}|C_3 \rangle \otimes |S_4^{\frac{3}{2}}\rangle
\nonumber\\
\begin{tabular}{|c|c|}
\hline
1 & 5   \\
\hline
2   \\
\cline{1-1}
3   \\
\cline{1-1}
4   \\
\cline{1-1}
\end{tabular}_{CS}
&=&\frac{1}{\sqrt{3}}|C_5 \rangle \otimes |S_2^{\frac{3}{2}}\rangle-\frac{1}{\sqrt{3}}|C_4 \rangle \otimes |S_3^{\frac{3}{2}}\rangle+\frac{1}{\sqrt{3}}|C_3 \rangle \otimes |S_4^{\frac{3}{2}}\rangle
\nonumber
\end{eqnarray}
\subsection{S=$\frac{5}{2}$}
\begin{eqnarray}
\begin{tabular}{|c|c|}
\hline
1 & 2   \\
\hline
3 & 4   \\
\cline{1-2}
5   \\
\cline{1-1}
\end{tabular}_{CS}
=|C_1 \rangle \otimes |S_1^{\frac{5}{2}}\rangle.
\quad
\begin{tabular}{|c|c|}
\hline
1 & 3   \\
\hline
2 & 4   \\
\cline{1-2}
5   \\
\cline{1-1}
\end{tabular}_{CS}
=|C_2 \rangle \otimes |S_1^{\frac{5}{2}}\rangle.
\quad
\begin{tabular}{|c|c|}
\hline
1 & 2   \\
\hline
3 & 5   \\
\cline{1-2}
4   \\
\cline{1-1}
\end{tabular}_{CS}
=|C_4 \rangle \otimes |S_1^{\frac{5}{2}}\rangle.
\quad
\begin{tabular}{|c|c|}
\hline
1 & 3   \\
\hline
2 & 5   \\
\cline{1-2}
4   \\
\cline{1-1}
\end{tabular}_{CS}
=|C_4 \rangle \otimes |S_1^{\frac{5}{2}}\rangle.
\quad
\begin{tabular}{|c|c|}
\hline
1 & 4   \\
\hline
2 & 5   \\
\cline{1-2}
3   \\
\cline{1-1}
\end{tabular}_{CS}
=|C_5 \rangle \otimes |S_1^{\frac{5}{2}}\rangle.
\nonumber\\
\end{eqnarray}

\section{Spin basis transformation : $q^5 \rightarrow q^5Q$}
\begin{itemize}
\item S=0 : $|S^{\frac{1}{2}}_1\rangle \rightarrow |S^0_1\rangle$, \quad $|S^{\frac{1}{2}}_2\rangle \rightarrow |S^0_2\rangle$, \quad $|S^{\frac{1}{2}}_3\rangle \rightarrow |S^0_3\rangle$, \quad $|S^{\frac{1}{2}}_4\rangle \rightarrow |S^0_4\rangle$, \quad $|S^{\frac{1}{2}}_5\rangle \rightarrow |S^0_5\rangle$
\item S=1 : $|S^{\frac{1}{2}}_1\rangle \rightarrow |S^1_5\rangle$, \quad $|S^{\frac{1}{2}}_2\rangle \rightarrow |S^1_6\rangle$, \quad $|S^{\frac{1}{2}}_3\rangle \rightarrow |S^1_7\rangle$, \quad $|S^{\frac{1}{2}}_4\rangle \rightarrow |S^1_8\rangle$, \quad $|S^{\frac{1}{2}}_5\rangle \rightarrow |S^1_9\rangle$, \\ $|S^{\frac{3}{2}}_1\rangle \rightarrow |S^1_1\rangle$, \quad $|S^{\frac{3}{2}}_2\rangle \rightarrow |S^1_2\rangle$, \quad$|S^{\frac{3}{2}}_3\rangle         \rightarrow |S^1_3\rangle$, \quad$|S^{\frac{3}{2}}_4\rangle \rightarrow |S^1_4\rangle$
\item S=2 : $|S^{\frac{3}{2}}_1\rangle \rightarrow |S^2_2\rangle$, \quad $|S^{\frac{3}{2}}_2\rangle \rightarrow |S^2_3\rangle$, \quad $|S^{\frac{3}{2}}_3\rangle \rightarrow |S^2_4\rangle$, \quad $|S^{\frac{3}{2}}_4\rangle \rightarrow |S^2_5\rangle$, \quad $|S^{\frac{5}{2}}_1\rangle \rightarrow |S^2_1\rangle$
\item S=3 : $|S^{\frac{5}{2}}_1\rangle \rightarrow |S^3_1\rangle$
\end{itemize}

\section{Flavor, color and spin state of $q^5Q$}
\begin{eqnarray}
\psi_{1,S=0}&=&-\frac{\sqrt{3}}{4\sqrt{2}}|C_2\rangle \otimes |F_1\rangle \otimes |S_1^0\rangle +\frac{1}{4\sqrt{2}}|C_4\rangle \otimes |F_1\rangle \otimes |S_1^0\rangle+\frac{\sqrt{3}}{4\sqrt{2}}|C_1\rangle \otimes |F_2\rangle \otimes |S_1^0\rangle -\frac{1}{4\sqrt{2}}|C_3\rangle \otimes |F_2\rangle \otimes |S_1^0\rangle \nonumber\\
&&+\frac{\sqrt{3}}{8}|C_2\rangle \otimes |F_1\rangle \otimes |S_2^0\rangle +\frac{1}{8}|C_4\rangle \otimes |F_1\rangle \otimes |S_2^0\rangle -\frac{1}{4\sqrt{2}}|C_5\rangle \otimes |F_1\rangle \otimes |S_2^0\rangle +\frac{\sqrt{3}}{8}|C_1\rangle \otimes |F_2\rangle \otimes |S_2^0\rangle \nonumber\\
&&+\frac{1}{8}|C_3\rangle \otimes |F_2\rangle \otimes |S_2^0\rangle +\frac{\sqrt{3}}{8}|C_1\rangle \otimes |F_1\rangle \otimes |S_3^0\rangle +\frac{1}{8}|C_3\rangle \otimes |F_1\rangle \otimes |S_3^0\rangle -\frac{\sqrt{3}}{8}|C_2\rangle \otimes |F_2\rangle \otimes |S_3^0\rangle \nonumber\\
&&-\frac{1}{8}|C_4\rangle \otimes |F_2\rangle \otimes |S_3^0\rangle -\frac{1}{4\sqrt{2}}|C_5\rangle \otimes |F_2\rangle \otimes |S_3^0\rangle +\frac{1}{8}|C_2\rangle \otimes |F_1\rangle \otimes |S_4^0\rangle -\frac{\sqrt{3}}{8}|C_4\rangle \otimes |F_1\rangle \otimes |S_4^0\rangle \nonumber\\
&&-\frac{\sqrt{3}}{4\sqrt{2}}|C_5\rangle \otimes |F_1\rangle \otimes |S_4^0\rangle +\frac{1}{8}|C_1\rangle \otimes |F_2\rangle \otimes |S_4^0\rangle -\frac{\sqrt{3}}{8}|C_3\rangle \otimes |F_2\rangle \otimes |S_4^0\rangle +\frac{1}{8}|C_1\rangle \otimes |F_1\rangle \otimes |S_5^0\rangle \nonumber\\
&&-\frac{\sqrt{3}}{8}|C_3\rangle \otimes |F_1\rangle \otimes |S_5^0\rangle -\frac{1}{8}|C_2\rangle \otimes |F_2\rangle \otimes |S_5^0\rangle +\frac{\sqrt{3}}{8}|C_4\rangle \otimes |F_2\rangle \otimes |S_5^0\rangle -\frac{\sqrt{3}}{4\sqrt{2}}|C_5\rangle \otimes |F_2\rangle \otimes |S_5^0\rangle \nonumber
\end{eqnarray}
\begin{eqnarray}
\psi_{2,S=0}&=&-\frac{\sqrt{3}}{4}|C_2\rangle \otimes |F_1\rangle \otimes |S_1^0\rangle +\frac{1}{4}|C_4\rangle \otimes |F_1\rangle \otimes |S_1^0\rangle+\frac{\sqrt{3}}{4}|C_1\rangle \otimes |F_2\rangle \otimes |S_1^0\rangle -\frac{1}{4}|C_3\rangle \otimes |F_2\rangle \otimes |S_1^0\rangle \nonumber\\
&&+\frac{1}{4}|C_5\rangle \otimes |F_1\rangle \otimes |S_2^0\rangle +\frac{1}{4}|C_5\rangle \otimes |F_2\rangle \otimes |S_3^0\rangle +\frac{\sqrt{3}}{4}|C_5\rangle \otimes |F_1\rangle \otimes |S_4^0\rangle +\frac{\sqrt{3}}{4}|C_5\rangle \otimes |F_2\rangle \otimes |S_5^0\rangle \nonumber
\end{eqnarray}
\begin{eqnarray}
\psi_{3,S=0}&=&-\frac{1}{4\sqrt{6}}|C_2\rangle \otimes |F_1\rangle \otimes |S_1^0\rangle +\frac{1}{12\sqrt{2}}|C_4\rangle \otimes |F_1\rangle \otimes |S_1^0\rangle+\frac{1}{4\sqrt{6}}|C_1\rangle \otimes |F_2\rangle \otimes |S_1^0\rangle -\frac{1}{12\sqrt{2}}|C_3\rangle \otimes |F_2\rangle \otimes |S_1^0\rangle \nonumber\\
&&-\frac{1}{8\sqrt{3}}|C_2\rangle \otimes |F_1\rangle \otimes |S_2^0\rangle +\frac{7}{24}|C_4\rangle \otimes |F_1\rangle \otimes |S_2^0\rangle -\frac{1}{12\sqrt{2}}|C_5\rangle \otimes |F_1\rangle \otimes |S_2^0\rangle -\frac{1}{8\sqrt{3}}|C_1\rangle \otimes |F_2\rangle \otimes |S_2^0\rangle \nonumber\\
&&+\frac{7}{24}|C_3\rangle \otimes |F_2\rangle \otimes |S_2^0\rangle -\frac{1}{8\sqrt{3}}|C_1\rangle \otimes |F_1\rangle \otimes |S_3^0\rangle +\frac{7}{24}|C_3\rangle \otimes |F_1\rangle \otimes |S_3^0\rangle +\frac{1}{8\sqrt{3}}|C_2\rangle \otimes |F_2\rangle \otimes |S_3^0\rangle \nonumber\\
&&-\frac{7}{24}|C_4\rangle \otimes |F_2\rangle \otimes |S_3^0\rangle -\frac{1}{12\sqrt{2}}|C_5\rangle \otimes |F_2\rangle \otimes |S_3^0\rangle -\frac{3}{8}|C_2\rangle \otimes |F_1\rangle \otimes |S_4^0\rangle +\frac{1}{8\sqrt{3}}|C_4\rangle \otimes |F_1\rangle \otimes |S_4^0\rangle \nonumber\\
&&-\frac{1}{4\sqrt{6}}|C_5\rangle \otimes |F_1\rangle \otimes |S_4^0\rangle -\frac{3}{8}|C_1\rangle \otimes |F_2\rangle \otimes |S_4^0\rangle +\frac{1}{8\sqrt{3}}|C_3\rangle \otimes |F_2\rangle \otimes |S_4^0\rangle -\frac{3}{8}|C_1\rangle \otimes |F_1\rangle \otimes |S_5^0\rangle \nonumber\\
&&+\frac{1}{8\sqrt{3}}|C_3\rangle \otimes |F_1\rangle \otimes |S_5^0\rangle +\frac{3}{8}|C_2\rangle \otimes |F_2\rangle \otimes |S_5^0\rangle -\frac{1}{8\sqrt{3}}|C_4\rangle \otimes |F_2\rangle \otimes |S_5^0\rangle -\frac{1}{4\sqrt{6}}|C_5\rangle \otimes |F_2\rangle \otimes |S_5^0\rangle \nonumber
\end{eqnarray}
\begin{eqnarray}
\psi_{4,S=0}&=&-\frac{1}{2\sqrt{3}}|C_2\rangle \otimes |F_1\rangle \otimes |S_1^0\rangle +\frac{1}{6}|C_4\rangle \otimes |F_1\rangle \otimes |S_1^0\rangle+\frac{1}{2\sqrt{3}}|C_1\rangle \otimes |F_2\rangle \otimes |S_1^0\rangle -\frac{1}{6}|C_3\rangle \otimes |F_2\rangle \otimes |S_1^0\rangle \nonumber\\
&&-\frac{1}{2\sqrt{6}}|C_2\rangle \otimes |F_1\rangle \otimes |S_2^0\rangle -\frac{1}{3\sqrt{2}}|C_4\rangle \otimes |F_1\rangle \otimes |S_2^0\rangle -\frac{1}{6}|C_5\rangle \otimes |F_1\rangle \otimes |S_2^0\rangle -\frac{1}{2\sqrt{6}}|C_1\rangle \otimes |F_2\rangle \otimes |S_2^0\rangle \nonumber\\
&&-\frac{1}{3\sqrt{2}}|C_3\rangle \otimes |F_2\rangle \otimes |S_2^0\rangle -\frac{1}{2\sqrt{6}}|C_1\rangle \otimes |F_1\rangle \otimes |S_3^0\rangle -\frac{1}{3\sqrt{2}}|C_3\rangle \otimes |F_1\rangle \otimes |S_3^0\rangle +\frac{1}{2\sqrt{6}}|C_2\rangle \otimes |F_2\rangle \otimes |S_3^0\rangle \nonumber\\
&&+\frac{1}{3\sqrt{2}}|C_4\rangle \otimes |F_2\rangle \otimes |S_3^0\rangle -\frac{1}{6}|C_5\rangle \otimes |F_2\rangle \otimes |S_3^0\rangle +\frac{1}{2\sqrt{6}}|C_4\rangle \otimes |F_1\rangle \otimes |S_4^0\rangle -\frac{1}{2\sqrt{3}}|C_5\rangle \otimes |F_1\rangle \otimes |S_4^0\rangle \nonumber\\
&&+\frac{1}{2\sqrt{6}}|C_3\rangle \otimes |F_2\rangle \otimes |S_4^0\rangle +\frac{1}{2\sqrt{6}}|C_3\rangle \otimes |F_1\rangle \otimes |S_5^0\rangle -\frac{1}{2\sqrt{6}}|C_4\rangle \otimes |F_2\rangle \otimes |S_5^0\rangle -\frac{1}{2\sqrt{3}}|C_5\rangle \otimes |F_2\rangle \otimes |S_5^0\rangle \nonumber
\end{eqnarray}
\begin{eqnarray}
\psi_{1,S=1}&=&-\frac{\sqrt{3}}{4\sqrt{2}}|C_2\rangle \otimes |F_1\rangle \otimes |S_5^1\rangle +\frac{1}{4\sqrt{2}}|C_4\rangle \otimes |F_1\rangle \otimes |S_5^1\rangle+\frac{\sqrt{3}}{4\sqrt{2}}|C_1\rangle \otimes |F_2\rangle \otimes |S_5^1\rangle -\frac{1}{4\sqrt{2}}|C_3\rangle \otimes |F_2\rangle \otimes |S_5^1\rangle \nonumber\\
&&+\frac{\sqrt{3}}{8}|C_2\rangle \otimes |F_1\rangle \otimes |S_6^1\rangle +\frac{1}{8}|C_4\rangle \otimes |F_1\rangle \otimes |S_6^1\rangle -\frac{1}{4\sqrt{2}}|C_5\rangle \otimes |F_1\rangle \otimes |S_6^1\rangle +\frac{\sqrt{3}}{8}|C_1\rangle \otimes |F_2\rangle \otimes |S_6^1\rangle \nonumber\\
&&+\frac{1}{8}|C_3\rangle \otimes |F_2\rangle \otimes |S_6^1\rangle +\frac{\sqrt{3}}{8}|C_1\rangle \otimes |F_1\rangle \otimes |S_7^1\rangle +\frac{1}{8}|C_3\rangle \otimes |F_1\rangle \otimes |S_7^1\rangle -\frac{\sqrt{3}}{8}|C_2\rangle \otimes |F_2\rangle \otimes |S_7^1\rangle \nonumber\\
&&-\frac{1}{8}|C_4\rangle \otimes |F_2\rangle \otimes |S_7^1\rangle -\frac{1}{4\sqrt{2}}|C_5\rangle \otimes |F_2\rangle \otimes |S_7^1\rangle +\frac{1}{8}|C_2\rangle \otimes |F_1\rangle \otimes |S_8^1\rangle -\frac{\sqrt{3}}{8}|C_4\rangle \otimes |F_1\rangle \otimes |S_8^1\rangle \nonumber\\
&&-\frac{\sqrt{3}}{4\sqrt{2}}|C_5\rangle \otimes |F_1\rangle \otimes |S_8^1\rangle +\frac{1}{8}|C_1\rangle \otimes |F_2\rangle \otimes |S_8^1\rangle -\frac{\sqrt{3}}{8}|C_3\rangle \otimes |F_2\rangle \otimes |S_8^1\rangle +\frac{1}{8}|C_1\rangle \otimes |F_1\rangle \otimes |S_9^1\rangle \nonumber\\
&&-\frac{\sqrt{3}}{8}|C_3\rangle \otimes |F_1\rangle \otimes |S_9^1\rangle -\frac{1}{8}|C_2\rangle \otimes |F_2\rangle \otimes |S_9^1\rangle +\frac{\sqrt{3}}{8}|C_4\rangle \otimes |F_2\rangle \otimes |S_9^1\rangle -\frac{\sqrt{3}}{4\sqrt{2}}|C_5\rangle \otimes |F_2\rangle \otimes |S_9^1\rangle \nonumber
\end{eqnarray}
\begin{eqnarray}
\psi_{2,S=1}&=&-\frac{\sqrt{3}}{4}|C_2\rangle \otimes |F_1\rangle \otimes |S_5^1\rangle +\frac{1}{4}|C_4\rangle \otimes |F_1\rangle \otimes |S_5^1\rangle+\frac{\sqrt{3}}{4}|C_1\rangle \otimes |F_2\rangle \otimes |S_5^1\rangle -\frac{1}{4}|C_3\rangle \otimes |F_2\rangle \otimes |S_5^1\rangle \nonumber\\
&&+\frac{1}{4}|C_5\rangle \otimes |F_1\rangle \otimes |S_6^1\rangle +\frac{1}{4}|C_5\rangle \otimes |F_2\rangle \otimes |S_7^1\rangle +\frac{\sqrt{3}}{4}|C_5\rangle \otimes |F_1\rangle \otimes |S_8^1\rangle +\frac{\sqrt{3}}{4}|C_5\rangle \otimes |F_2\rangle \otimes |S_9^1\rangle \nonumber
\end{eqnarray}
\begin{eqnarray}
\psi_{3,S=1}&=&-\frac{1}{4\sqrt{6}}|C_2\rangle \otimes |F_1\rangle \otimes |S_5^1\rangle +\frac{1}{12\sqrt{2}}|C_4\rangle \otimes |F_1\rangle \otimes |S_5^1\rangle+\frac{1}{4\sqrt{6}}|C_1\rangle \otimes |F_2\rangle \otimes |S_5^1\rangle -\frac{1}{12\sqrt{2}}|C_3\rangle \otimes |F_2\rangle \otimes |S_5^1\rangle \nonumber\\
&&-\frac{1}{8\sqrt{3}}|C_2\rangle \otimes |F_1\rangle \otimes |S_6^1\rangle +\frac{7}{24}|C_4\rangle \otimes |F_1\rangle \otimes |S_6^1\rangle -\frac{1}{12\sqrt{2}}|C_5\rangle \otimes |F_1\rangle \otimes |S_6^1\rangle -\frac{1}{8\sqrt{3}}|C_1\rangle \otimes |F_2\rangle \otimes |S_6^1\rangle \nonumber\\
&&+\frac{7}{24}|C_3\rangle \otimes |F_2\rangle \otimes |S_6^1\rangle -\frac{1}{8\sqrt{3}}|C_1\rangle \otimes |F_1\rangle \otimes |S_7^1\rangle +\frac{7}{24}|C_3\rangle \otimes |F_1\rangle \otimes |S_7^1\rangle +\frac{1}{8\sqrt{3}}|C_2\rangle \otimes |F_2\rangle \otimes |S_7^1\rangle \nonumber\\
&&-\frac{7}{24}|C_4\rangle \otimes |F_2\rangle \otimes |S_7^1\rangle -\frac{1}{12\sqrt{2}}|C_5\rangle \otimes |F_2\rangle \otimes |S_7^1\rangle -\frac{3}{8}|C_2\rangle \otimes |F_1\rangle \otimes |S_8^1\rangle +\frac{1}{8\sqrt{3}}|C_4\rangle \otimes |F_1\rangle \otimes |S_8^1\rangle \nonumber\\
&&-\frac{1}{4\sqrt{6}}|C_5\rangle \otimes |F_1\rangle \otimes |S_8^1\rangle -\frac{3}{8}|C_1\rangle \otimes |F_2\rangle \otimes |S_8^1\rangle +\frac{1}{8\sqrt{3}}|C_3\rangle \otimes |F_2\rangle \otimes |S_8^1\rangle -\frac{3}{8}|C_1\rangle \otimes |F_1\rangle \otimes |S_9^1\rangle \nonumber\\
&&+\frac{1}{8\sqrt{3}}|C_3\rangle \otimes |F_1\rangle \otimes |S_9^1\rangle +\frac{3}{8}|C_2\rangle \otimes |F_2\rangle \otimes |S_9^1\rangle -\frac{1}{8\sqrt{3}}|C_4\rangle \otimes |F_2\rangle \otimes |S_9^1\rangle -\frac{1}{4\sqrt{6}}|C_5\rangle \otimes |F_2\rangle \otimes |S_9^1\rangle \nonumber
\end{eqnarray}
\begin{eqnarray}
\psi_{4,S=1}&=&-\frac{1}{2\sqrt{3}}|C_2\rangle \otimes |F_1\rangle \otimes |S_5^1\rangle +\frac{1}{6}|C_4\rangle \otimes |F_1\rangle \otimes |S_5^1\rangle+\frac{1}{2\sqrt{3}}|C_1\rangle \otimes |F_2\rangle \otimes |S_5^1\rangle -\frac{1}{6}|C_3\rangle \otimes |F_2\rangle \otimes |S_5^1\rangle \nonumber\\
&&-\frac{1}{2\sqrt{6}}|C_2\rangle \otimes |F_1\rangle \otimes |S_6^1\rangle -\frac{1}{3\sqrt{2}}|C_4\rangle \otimes |F_1\rangle \otimes |S_6^1\rangle -\frac{1}{6}|C_5\rangle \otimes |F_1\rangle \otimes |S_6^1\rangle -\frac{1}{2\sqrt{6}}|C_1\rangle \otimes |F_2\rangle \otimes |S_6^1\rangle \nonumber\\
&&-\frac{1}{3\sqrt{2}}|C_3\rangle \otimes |F_2\rangle \otimes |S_6^1\rangle -\frac{1}{2\sqrt{6}}|C_1\rangle \otimes |F_1\rangle \otimes |S_7^1\rangle -\frac{1}{3\sqrt{2}}|C_3\rangle \otimes |F_1\rangle \otimes |S_7^1\rangle +\frac{1}{2\sqrt{6}}|C_2\rangle \otimes |F_2\rangle \otimes |S_7^1\rangle \nonumber\\
&&+\frac{1}{3\sqrt{2}}|C_4\rangle \otimes |F_2\rangle \otimes |S_7^1\rangle -\frac{1}{6}|C_5\rangle \otimes |F_2\rangle \otimes |S_7^1\rangle +\frac{1}{2\sqrt{6}}|C_4\rangle \otimes |F_1\rangle \otimes |S_8^1\rangle -\frac{1}{2\sqrt{3}}|C_5\rangle \otimes |F_1\rangle \otimes |S_8^1\rangle \nonumber\\
&&+\frac{1}{2\sqrt{6}}|C_3\rangle \otimes |F_2\rangle \otimes |S_8^1\rangle +\frac{1}{2\sqrt{6}}|C_3\rangle \otimes |F_1\rangle \otimes |S_9^1\rangle -\frac{1}{2\sqrt{6}}|C_4\rangle \otimes |F_2\rangle \otimes |S_9^1\rangle -\frac{1}{2\sqrt{3}}|C_5\rangle \otimes |F_2\rangle \otimes |S_9^1\rangle \nonumber
\end{eqnarray}
\begin{eqnarray}
\psi_{5,S=1}&=&\frac{\sqrt{5}}{8}|C_2\rangle \otimes |F_1\rangle \otimes |S_1^1\rangle -\frac{\sqrt{5}}{8}|C_1\rangle \otimes |F_2\rangle \otimes |S_1^1\rangle+\frac{\sqrt{3}}{8}|C_2\rangle \otimes |F_1\rangle \otimes |S_2^1\rangle -\frac{1}{4}|C_4\rangle \otimes |F_1\rangle \otimes |S_2^1\rangle \nonumber\\
&&-\frac{\sqrt{3}}{8}|C_1\rangle \otimes |F_2\rangle \otimes |S_2^1\rangle +\frac{1}{4}|C_3\rangle \otimes |F_2\rangle \otimes |S_2^1\rangle -\frac{\sqrt{3}}{8\sqrt{2}}|C_2\rangle \otimes |F_1\rangle \otimes |S_3^1\rangle +\frac{1}{8\sqrt{2}}|C_4\rangle \otimes |F_1\rangle \otimes |S_3^1\rangle \nonumber\\
&&-\frac{1}{2}|C_5\rangle \otimes |F_1\rangle \otimes |S_3^1\rangle -\frac{\sqrt{3}}{8\sqrt{2}}|C_1\rangle \otimes |F_2\rangle \otimes |S_3^1\rangle +\frac{1}{8\sqrt{2}}|C_3\rangle \otimes |F_2\rangle \otimes |S_3^1\rangle -\frac{\sqrt{3}}{8\sqrt{2}}|C_1\rangle \otimes |F_1\rangle \otimes |S_4^1\rangle \nonumber\\
&&+\frac{1}{8\sqrt{2}}|C_3\rangle \otimes |F_1\rangle \otimes |S_4^1\rangle +\frac{\sqrt{3}}{8\sqrt{2}}|C_2\rangle \otimes |F_2\rangle \otimes |S_4^1\rangle -\frac{1}{8\sqrt{2}}|C_4\rangle \otimes |F_2\rangle \otimes |S_4^1\rangle -\frac{1}{2}|C_5\rangle \otimes |F_2\rangle \otimes |S_4^1\rangle \nonumber
\end{eqnarray}
\begin{eqnarray}
\psi_{6,S=1}&=&\frac{\sqrt{3}}{4\sqrt{2}}|C_4\rangle \otimes |F_1\rangle \otimes |S_1^1\rangle -\frac{\sqrt{3}}{4\sqrt{2}}|C_3\rangle \otimes |F_2\rangle \otimes |S_1^1\rangle -\frac{\sqrt{3}}{2\sqrt{10}}|C_2\rangle \otimes |F_1\rangle \otimes |S_2^1\rangle -\frac{1}{4\sqrt{10}}|C_4\rangle \otimes |F_1\rangle \otimes |S_2^1\rangle \nonumber\\
&&+\frac{\sqrt{3}}{2\sqrt{10}}|C_1\rangle \otimes |F_2\rangle \otimes |S_2^1\rangle +\frac{1}{4\sqrt{10}}|C_3\rangle \otimes |F_2\rangle \otimes |S_2^1\rangle +\frac{3\sqrt{3}}{8\sqrt{5}}|C_2\rangle \otimes |F_1\rangle \otimes |S_3^1\rangle -\frac{3}{8\sqrt{5}}|C_4\rangle \otimes |F_1\rangle \otimes |S_3^1\rangle \nonumber\\
&&-\frac{1}{\sqrt{10}}|C_5\rangle \otimes |F_1\rangle \otimes |S_3^1\rangle +\frac{3\sqrt{3}}{8\sqrt{5}}|C_1\rangle \otimes |F_2\rangle \otimes |S_3^1\rangle -\frac{3}{8\sqrt{5}}|C_3\rangle \otimes |F_2\rangle \otimes |S_3^1\rangle +\frac{3\sqrt{3}}{8\sqrt{5}}|C_1\rangle \otimes |F_1\rangle \otimes |S_4^1\rangle \nonumber\\
&&-\frac{3}{8\sqrt{5}}|C_3\rangle \otimes |F_1\rangle \otimes |S_4^1\rangle -\frac{3\sqrt{3}}{8\sqrt{5}}|C_2\rangle \otimes |F_2\rangle \otimes |S_4^1\rangle +\frac{3}{8\sqrt{5}}|C_4\rangle \otimes |F_2\rangle \otimes |S_4^1\rangle -\frac{1}{\sqrt{10}}|C_5\rangle \otimes |F_2\rangle \otimes |S_4^1\rangle \nonumber
\end{eqnarray}
\begin{eqnarray}
\psi_{7,S=1}&=&\frac{3}{8}|C_2\rangle \otimes |F_1\rangle \otimes |S_1^1\rangle +\frac{1}{4\sqrt{3}}|C_4\rangle \otimes |F_1\rangle \otimes |S_1^1\rangle -\frac{3}{8}|C_1\rangle \otimes |F_2\rangle \otimes |S_1^1\rangle -\frac{1}{4\sqrt{3}}|C_3\rangle \otimes |F_2\rangle \otimes |S_1^1\rangle \nonumber\\
&&+\frac{\sqrt{5}}{8\sqrt{3}}|C_2\rangle \otimes |F_1\rangle \otimes |S_2^1\rangle -\frac{\sqrt{5}}{6}|C_4\rangle \otimes |F_1\rangle \otimes |S_2^1\rangle -\frac{\sqrt{5}}{8\sqrt{3}}|C_1\rangle \otimes |F_2\rangle \otimes |S_2^1\rangle +\frac{\sqrt{5}}{6}|C_3\rangle \otimes |F_2\rangle \otimes |S_2^1\rangle \nonumber\\
&&+\frac{\sqrt{5}}{8\sqrt{6}}|C_2\rangle \otimes |F_1\rangle \otimes |S_3^1\rangle -\frac{\sqrt{5}}{24\sqrt{2}}|C_4\rangle \otimes |F_1\rangle \otimes |S_3^1\rangle +\frac{\sqrt{5}}{6}|C_5\rangle \otimes |F_1\rangle \otimes |S_3^1\rangle +\frac{\sqrt{5}}{8\sqrt{6}}|C_1\rangle \otimes |F_2\rangle \otimes |S_3^1\rangle \nonumber\\
&&-\frac{\sqrt{5}}{24\sqrt{2}}|C_3\rangle \otimes |F_2\rangle \otimes |S_3^1\rangle +\frac{\sqrt{5}}{8\sqrt{6}}|C_1\rangle \otimes |F_1\rangle \otimes |S_4^1\rangle -\frac{\sqrt{5}}{24\sqrt{2}}|C_3\rangle \otimes |F_1\rangle \otimes |S_4^1\rangle -\frac{\sqrt{5}}{8\sqrt{6}}|C_2\rangle \otimes |F_2\rangle \otimes |S_4^1\rangle \nonumber\\
&&+\frac{\sqrt{5}}{24\sqrt{2}}|C_4\rangle \otimes |F_2\rangle \otimes |S_4^1\rangle+\frac{\sqrt{5}}{6}|C_5\rangle \otimes |F_2\rangle \otimes |S_4^1\rangle
\end{eqnarray}
\begin{eqnarray}
\psi_{8,S=1}&=&-\frac{1}{\sqrt{6}}|C_4\rangle \otimes |F_1\rangle \otimes |S_1^1\rangle +\frac{1}{\sqrt{6}}|C_3\rangle \otimes |F_2\rangle \otimes |S_1^1\rangle +\frac{\sqrt{2}}{\sqrt{15}}|C_2\rangle \otimes |F_1\rangle \otimes |S_2^1\rangle +\frac{1}{3\sqrt{10}}|C_4\rangle \otimes |F_1\rangle \otimes |S_2^1\rangle \nonumber\\
&&-\frac{\sqrt{2}}{\sqrt{15}}|C_1\rangle \otimes |F_2\rangle \otimes |S_2^1\rangle -\frac{1}{3\sqrt{10}}|C_3\rangle \otimes |F_2\rangle \otimes |S_2^1\rangle +\frac{1}{\sqrt{15}}|C_2\rangle \otimes |F_1\rangle \otimes |S_3^1\rangle -\frac{1}{3\sqrt{5}}|C_4\rangle \otimes |F_1\rangle \otimes |S_3^1\rangle \nonumber\\
&&-\frac{1}{3\sqrt{10}}|C_5\rangle \otimes |F_1\rangle \otimes |S_3^1\rangle +\frac{1}{\sqrt{15}}|C_1\rangle \otimes |F_2\rangle \otimes |S_3^1\rangle -\frac{1}{3\sqrt{5}}|C_3\rangle \otimes |F_2\rangle \otimes |S_3^1\rangle +\frac{1}{\sqrt{15}}|C_1\rangle \otimes |F_1\rangle \otimes |S_4^1\rangle \nonumber\\
&&-\frac{1}{3\sqrt{5}}|C_3\rangle \otimes |F_1\rangle \otimes |S_4^1\rangle -\frac{1}{\sqrt{15}}|C_2\rangle \otimes |F_2\rangle \otimes |S_4^1\rangle +\frac{1}{3\sqrt{5}}|C_4\rangle \otimes |F_2\rangle \otimes |S_4^1\rangle -\frac{1}{3\sqrt{10}}|C_5\rangle \otimes |F_2\rangle \otimes |S_4^1\rangle \nonumber
\end{eqnarray}
\begin{eqnarray}
\psi_{1,S=2}&=&\frac{\sqrt{5}}{8}|C_2\rangle \otimes |F_1\rangle \otimes |S_2^2\rangle -\frac{\sqrt{5}}{8}|C_1\rangle \otimes |F_2\rangle \otimes |S_2^2\rangle+\frac{\sqrt{3}}{8}|C_2\rangle \otimes |F_1\rangle \otimes |S_3^2\rangle -\frac{1}{4}|C_4\rangle \otimes |F_1\rangle \otimes |S_3^2\rangle \nonumber\\
&&-\frac{\sqrt{3}}{8}|C_1\rangle \otimes |F_2\rangle \otimes |S_3^2\rangle +\frac{1}{4}|C_3\rangle \otimes |F_2\rangle \otimes |S_3^2\rangle -\frac{\sqrt{3}}{8\sqrt{2}}|C_2\rangle \otimes |F_1\rangle \otimes |S_4^2\rangle +\frac{1}{8\sqrt{2}}|C_4\rangle \otimes |F_1\rangle \otimes |S_4^2\rangle \nonumber\\
&&-\frac{1}{2}|C_5\rangle \otimes |F_1\rangle \otimes |S_4^2\rangle -\frac{\sqrt{3}}{8\sqrt{2}}|C_1\rangle \otimes |F_2\rangle \otimes |S_4^2\rangle +\frac{1}{8\sqrt{2}}|C_3\rangle \otimes |F_2\rangle \otimes |S_4^2\rangle -\frac{\sqrt{3}}{8\sqrt{2}}|C_1\rangle \otimes |F_1\rangle \otimes |S_5^2\rangle \nonumber\\
&&+\frac{1}{8\sqrt{2}}|C_3\rangle \otimes |F_1\rangle \otimes |S_5^2\rangle +\frac{\sqrt{3}}{8\sqrt{2}}|C_2\rangle \otimes |F_2\rangle \otimes |S_5^2\rangle -\frac{1}{8\sqrt{2}}|C_4\rangle \otimes |F_2\rangle \otimes |S_5^2\rangle -\frac{1}{2}|C_5\rangle \otimes |F_2\rangle \otimes |S_5^2\rangle \nonumber
\end{eqnarray}
\begin{eqnarray}
\psi_{2,S=2}&=&\frac{\sqrt{3}}{4\sqrt{2}}|C_4\rangle \otimes |F_1\rangle \otimes |S_2^2\rangle -\frac{\sqrt{3}}{4\sqrt{2}}|C_3\rangle \otimes |F_2\rangle \otimes |S_2^2\rangle -\frac{\sqrt{3}}{2\sqrt{10}}|C_2\rangle \otimes |F_1\rangle \otimes |S_3^2\rangle -\frac{1}{4\sqrt{10}}|C_4\rangle \otimes |F_1\rangle \otimes |S_3^2\rangle \nonumber\\
&&+\frac{\sqrt{3}}{2\sqrt{10}}|C_1\rangle \otimes |F_2\rangle \otimes |S_3^2\rangle +\frac{1}{4\sqrt{10}}|C_3\rangle \otimes |F_2\rangle \otimes |S_3^2\rangle +\frac{3\sqrt{3}}{8\sqrt{5}}|C_2\rangle \otimes |F_1\rangle \otimes |S_4^2\rangle -\frac{3}{8\sqrt{5}}|C_4\rangle \otimes |F_1\rangle \otimes |S_4^2\rangle \nonumber\\
&&-\frac{1}{\sqrt{10}}|C_5\rangle \otimes |F_1\rangle \otimes |S_4^2\rangle +\frac{3\sqrt{3}}{8\sqrt{5}}|C_1\rangle \otimes |F_2\rangle \otimes |S_4^2\rangle -\frac{3}{8\sqrt{5}}|C_3\rangle \otimes |F_2\rangle \otimes |S_4^2\rangle +\frac{3\sqrt{3}}{8\sqrt{5}}|C_1\rangle \otimes |F_1\rangle \otimes |S_5^2\rangle \nonumber\\
&&-\frac{3}{8\sqrt{5}}|C_3\rangle \otimes |F_1\rangle \otimes |S_5^2\rangle -\frac{3\sqrt{3}}{8\sqrt{5}}|C_2\rangle \otimes |F_2\rangle \otimes |S_5^2\rangle +\frac{3}{8\sqrt{5}}|C_4\rangle \otimes |F_2\rangle \otimes |S_5^2\rangle -\frac{1}{\sqrt{10}}|C_5\rangle \otimes |F_2\rangle \otimes |S_5^2\rangle \nonumber
\end{eqnarray}
\begin{eqnarray}
\psi_{3,S=2}&=&\frac{3}{8}|C_2\rangle \otimes |F_1\rangle \otimes |S_2^2\rangle +\frac{1}{4\sqrt{3}}|C_4\rangle \otimes |F_1\rangle \otimes |S_2^2\rangle -\frac{3}{8}|C_1\rangle \otimes |F_2\rangle \otimes |S_2^2\rangle -\frac{1}{4\sqrt{3}}|C_3\rangle \otimes |F_2\rangle \otimes |S_2^2\rangle \nonumber\\
&&+\frac{\sqrt{5}}{8\sqrt{3}}|C_2\rangle \otimes |F_1\rangle \otimes |S_3^2\rangle -\frac{\sqrt{5}}{6}|C_4\rangle \otimes |F_1\rangle \otimes |S_3^2\rangle -\frac{\sqrt{5}}{8\sqrt{3}}|C_1\rangle \otimes |F_2\rangle \otimes |S_3^2\rangle +\frac{\sqrt{5}}{6}|C_3\rangle \otimes |F_2\rangle \otimes |S_3^2\rangle \nonumber\\
&&+\frac{\sqrt{5}}{8\sqrt{6}}|C_2\rangle \otimes |F_1\rangle \otimes |S_4^2\rangle -\frac{\sqrt{5}}{24\sqrt{2}}|C_4\rangle \otimes |F_1\rangle \otimes |S_4^2\rangle +\frac{\sqrt{5}}{6}|C_5\rangle \otimes |F_1\rangle \otimes |S_4^2\rangle +\frac{\sqrt{5}}{8\sqrt{6}}|C_1\rangle \otimes |F_2\rangle \otimes |S_4^2\rangle \nonumber\\
&&-\frac{\sqrt{5}}{24\sqrt{2}}|C_3\rangle \otimes |F_2\rangle \otimes |S_4^2\rangle +\frac{\sqrt{5}}{8\sqrt{6}}|C_1\rangle \otimes |F_1\rangle \otimes |S_5^2\rangle -\frac{\sqrt{5}}{24\sqrt{2}}|C_3\rangle \otimes |F_1\rangle \otimes |S_5^2\rangle -\frac{\sqrt{5}}{8\sqrt{6}}|C_2\rangle \otimes |F_2\rangle \otimes |S_5^2\rangle \nonumber\\
&&+\frac{\sqrt{5}}{24\sqrt{2}}|C_4\rangle \otimes |F_2\rangle \otimes |S_5^2\rangle+\frac{\sqrt{5}}{6}|C_5\rangle \otimes |F_2\rangle \otimes |S_5^2\rangle
\end{eqnarray}
\begin{eqnarray}
\psi_{4,S=2}&=&-\frac{1}{\sqrt{6}}|C_4\rangle \otimes |F_1\rangle \otimes |S_2^2\rangle +\frac{1}{\sqrt{6}}|C_3\rangle \otimes |F_2\rangle \otimes |S_2^2\rangle +\frac{\sqrt{2}}{\sqrt{15}}|C_2\rangle \otimes |F_1\rangle \otimes |S_3^2\rangle +\frac{1}{3\sqrt{10}}|C_4\rangle \otimes |F_1\rangle \otimes |S_3^2\rangle \nonumber\\
&&-\frac{\sqrt{2}}{\sqrt{15}}|C_1\rangle \otimes |F_2\rangle \otimes |S_3^2\rangle -\frac{1}{3\sqrt{10}}|C_3\rangle \otimes |F_2\rangle \otimes |S_3^2\rangle +\frac{1}{\sqrt{15}}|C_2\rangle \otimes |F_1\rangle \otimes |S_4^2\rangle -\frac{1}{3\sqrt{5}}|C_4\rangle \otimes |F_1\rangle \otimes |S_4^2\rangle \nonumber\\
&&-\frac{1}{3\sqrt{10}}|C_5\rangle \otimes |F_1\rangle \otimes |S_4^2\rangle +\frac{1}{\sqrt{15}}|C_1\rangle \otimes |F_2\rangle \otimes |S_4^2\rangle -\frac{1}{3\sqrt{5}}|C_3\rangle \otimes |F_2\rangle \otimes |S_4^2\rangle +\frac{1}{\sqrt{15}}|C_1\rangle \otimes |F_1\rangle \otimes |S_5^2\rangle \nonumber\\
&&-\frac{1}{3\sqrt{5}}|C_3\rangle \otimes |F_1\rangle \otimes |S_5^2\rangle -\frac{1}{\sqrt{15}}|C_2\rangle \otimes |F_2\rangle \otimes |S_5^2\rangle +\frac{1}{3\sqrt{5}}|C_4\rangle \otimes |F_2\rangle \otimes |S_5^2\rangle -\frac{1}{3\sqrt{10}}|C_5\rangle \otimes |F_2\rangle \otimes |S_5^2\rangle \nonumber
\end{eqnarray}
\begin{eqnarray}
\psi_{5,S=2}&=&-\frac{\sqrt{3}}{2\sqrt{2}}|C_2\rangle \otimes |F_1\rangle \otimes |S_1^2\rangle +\frac{1}{2\sqrt{2}}|C_4\rangle \otimes |F_1\rangle \otimes |S_1^2\rangle +\frac{\sqrt{3}}{2\sqrt{2}}|C_1\rangle \otimes |F_2\rangle \otimes |S_1^2\rangle -\frac{1}{2\sqrt{2}}|C_3\rangle \otimes |F_2\rangle \otimes |S_1^2\rangle \nonumber
\end{eqnarray}
\begin{eqnarray}
\psi_{1,S=3}&=&-\frac{\sqrt{3}}{2\sqrt{2}}|C_2\rangle \otimes |F_1\rangle \otimes |S_1^3\rangle +\frac{1}{2\sqrt{2}}|C_4\rangle \otimes |F_1\rangle \otimes |S_1^3\rangle +\frac{\sqrt{3}}{2\sqrt{2}}|C_1\rangle \otimes |F_2\rangle \otimes |S_1^3\rangle -\frac{1}{2\sqrt{2}}|C_3\rangle \otimes |F_2\rangle \otimes |S_1^3\rangle \nonumber
\end{eqnarray}

\section{Matrix elements of $\lambda_i \lambda_j$ and $\lambda_i \lambda_j \sigma_i \cdot \sigma_j$}
We represent the matrix elements of $\lambda_i \lambda_j=\langle\psi|\lambda_i \lambda_j|\psi \rangle$ and $\lambda_i \lambda_j \sigma_i \cdot \sigma_j=\langle \psi|\lambda_i \lambda_j \sigma_i \cdot \sigma_j|\psi \rangle$ only for S=0,1,2,3.
\subsection{S=0}

\begin{eqnarray}
&&\lambda_1 \lambda_2 =
\left(
\begin{array}{cccc}
 -\frac{7}{6} & \frac{1}{\sqrt{2}} & -\frac{1}{6} & -\frac{\sqrt{2}}{3} \\
 \frac{1}{\sqrt{2}} & -\frac{5}{3} & \frac{1}{3 \sqrt{2}} & \frac{2}{3} \\
 -\frac{1}{6} & \frac{1}{3 \sqrt{2}} & -\frac{13}{18} & -\frac{\sqrt{2}}{9} \\
 -\frac{\sqrt{2}}{3} & \frac{2}{3} & -\frac{\sqrt{2}}{9} & -\frac{10}{9}
\end{array}
\right)
,~
\lambda_1 \lambda_4=
\left(
\begin{array}{cccc}
 -\frac{11}{12} & -\frac{1}{2 \sqrt{2}} & -\frac{1}{12} & \frac{\sqrt{2}}{3} \\
 -\frac{1}{2 \sqrt{2}} & -\frac{1}{2} & -\frac{1}{6 \sqrt{2}} & -\frac{1}{3} \\
 -\frac{1}{12} & -\frac{1}{6 \sqrt{2}} & -\frac{41}{36} & \frac{2 \sqrt{2}}{9} \\
 \frac{\sqrt{2}}{3} & -\frac{1}{3} & \frac{2 \sqrt{2}}{9} & -\frac{10}{9}
\end{array}
\right)
,~
\lambda_4 \lambda_5 =
\left(
\begin{array}{cccc}
 -\frac{5}{3} & 0 & 1 & -\sqrt{2} \\
 0 & -\frac{8}{3} & 0 & 0 \\
 1 & 0 & -\frac{5}{3} & -\sqrt{2} \\
 -\sqrt{2} & 0 & -\sqrt{2} & -\frac{2}{3}
\end{array}
\right)
,\nonumber\\
&&\lambda_1 \lambda_6 =
\left(
\begin{array}{cccc}
 -\frac{7}{6} & -\frac{1}{\sqrt{2}} & \frac{1}{2} & 0 \\
 -\frac{1}{\sqrt{2}} & -1 & -\frac{1}{3 \sqrt{2}} & -\frac{2}{3} \\
 \frac{1}{2} & -\frac{1}{3 \sqrt{2}} & -\frac{29}{18} & -\frac{2
   \sqrt{2}}{9} \\
 0 & -\frac{2}{3} & -\frac{2 \sqrt{2}}{9} & -\frac{8}{9}
\end{array}
\right)
,~
\lambda_5 \lambda_6 =
\left(
\begin{array}{cccc}
 -\frac{11}{12} & \frac{3}{2 \sqrt{2}} & -\frac{3}{4} & 0 \\
 \frac{3}{2 \sqrt{2}} & -\frac{7}{6} & \frac{1}{2 \sqrt{2}} & 1 \\
 -\frac{3}{4} & \frac{1}{2 \sqrt{2}} & -\frac{1}{4} &
   \frac{\sqrt{2}}{3} \\
 0 & 1 & \frac{\sqrt{2}}{3} & -\frac{4}{3}
\end{array}
\right).
\end{eqnarray}
\begin{eqnarray}
&&\lambda_1 \lambda_2 \sigma_1 \sigma_2 = \left(
\begin{array}{cccc}
 \frac{5}{6} & -\frac{5}{3 \sqrt{2}} & \frac{1}{18} & \frac{\sqrt{2}}{9} \\
 -\frac{5}{3 \sqrt{2}} & 1 & -\frac{5}{9 \sqrt{2}} & -\frac{10}{9} \\
 \frac{1}{18} & -\frac{5}{9 \sqrt{2}} & \frac{37}{54} & \frac{\sqrt{2}}{27} \\
 \frac{\sqrt{2}}{9} & -\frac{10}{9} & \frac{\sqrt{2}}{27} & \frac{22}{27}
\end{array}
\right)
,~
\lambda_1 \lambda_4 \sigma_1 \sigma_4 = \left(
\begin{array}{cccc}
 \frac{11}{4} & \frac{5}{6 \sqrt{2}} & \frac{1}{36} &
   -\frac{\sqrt{2}}{9} \\
 \frac{5}{6 \sqrt{2}} & \frac{23}{18} & \frac{5}{18 \sqrt{2}} &
   \frac{5}{9} \\
 \frac{1}{36} & \frac{5}{18 \sqrt{2}} & \frac{17}{108} & -\frac{2
   \sqrt{2}}{27} \\
 -\frac{\sqrt{2}}{9} & \frac{5}{9} & -\frac{2 \sqrt{2}}{27} &
   -\frac{50}{27}
\end{array}
\right), \nonumber\\
&&\lambda_4 \lambda_5 \sigma_4 \sigma_5 = \left(
\begin{array}{cccc}
 -3 & 0 & -\frac{1}{3} & \frac{\sqrt{2}}{3} \\
 0 & -\frac{8}{3} & 0 & 0 \\
 -\frac{1}{3} & 0 & -3 & \frac{\sqrt{2}}{3} \\
 \frac{\sqrt{2}}{3} & 0 & \frac{\sqrt{2}}{3} & -\frac{10}{3}
\end{array}
\right)
,~
\lambda_1 \lambda_6 \sigma_1 \sigma_6 = \left(
\begin{array}{cccc}
 \frac{5}{6} & \frac{5}{3 \sqrt{2}} & -\frac{1}{6} & 0 \\
 \frac{5}{3 \sqrt{2}} & \frac{31}{9} & \frac{5}{9 \sqrt{2}} &
   \frac{10}{9} \\
 -\frac{1}{6} & \frac{5}{9 \sqrt{2}} & -\frac{43}{54} & -\frac{22
   \sqrt{2}}{27} \\
 0 & \frac{10}{9} & -\frac{22 \sqrt{2}}{27} & -\frac{4}{27}
\end{array}
\right), \nonumber\\
&&\lambda_5 \lambda_6 \sigma_5 \sigma_6 = \left(
\begin{array}{cccc}
 \frac{11}{4} & -\frac{5}{2 \sqrt{2}} & \frac{1}{4} & 0 \\
 -\frac{5}{2 \sqrt{2}} & \frac{17}{6} & -\frac{5}{6 \sqrt{2}} &
   -\frac{5}{3} \\
 \frac{1}{4} & -\frac{5}{6 \sqrt{2}} & -\frac{101}{36} & \frac{11
   \sqrt{2}}{9} \\
 0 & -\frac{5}{3} & \frac{11 \sqrt{2}}{9} & \frac{20}{9}
\end{array}
\right).
\end{eqnarray}
\subsection{S=1}
\begin{eqnarray}
&&\lambda_1 \lambda_2 =
\left(
\begin{array}{cccccccc}
 -\frac{7}{6} & \frac{1}{\sqrt{2}} & -\frac{1}{6} & -\frac{\sqrt{2}}{3} & 0 & 0 & 0 & 0 \\
 \frac{1}{\sqrt{2}} & -\frac{5}{3} & \frac{1}{3 \sqrt{2}} & \frac{2}{3} & 0 & 0 & 0 & 0 \\
 -\frac{1}{6} & \frac{1}{3 \sqrt{2}} & -\frac{13}{18} & -\frac{\sqrt{2}}{9} & 0 & 0 & 0 & 0 \\
 -\frac{\sqrt{2}}{3} & \frac{2}{3} & -\frac{\sqrt{2}}{9} & -\frac{10}{9} & 0 & 0 & 0 & 0 \\
 0 & 0 & 0 & 0 & -\frac{5}{3} & -\sqrt{\frac{2}{5}} & \frac{\sqrt{5}}{3} & -\frac{\sqrt{\frac{2}{5}}}{3} \\
 0 & 0 & 0 & 0 & -\sqrt{\frac{2}{5}} & -\frac{16}{15} & \frac{\sqrt{2}}{3} & -\frac{2}{15} \\
 0 & 0 & 0 & 0 & \frac{\sqrt{5}}{3} & \frac{\sqrt{2}}{3} & -\frac{11}{9} & \frac{\sqrt{2}}{9} \\
 0 & 0 & 0 & 0 & -\frac{\sqrt{\frac{2}{5}}}{3} & -\frac{2}{15} & \frac{\sqrt{2}}{9} & -\frac{32}{45}
\end{array}
\right), \nonumber\\
&&\lambda_1 \lambda_4=
\left(
\begin{array}{cccccccc}
 -\frac{11}{12} & -\frac{1}{2 \sqrt{2}} & -\frac{1}{12} & \frac{\sqrt{2}}{3} & 0 & 0 & 0 & 0 \\
 -\frac{1}{2 \sqrt{2}} & -\frac{1}{2} & -\frac{1}{6 \sqrt{2}} & -\frac{1}{3} & 0 & 0 & 0 & 0 \\
 -\frac{1}{12} & -\frac{1}{6 \sqrt{2}} & -\frac{41}{36} & \frac{2 \sqrt{2}}{9} & 0 & 0 & 0 & 0 \\
 \frac{\sqrt{2}}{3} & -\frac{1}{3} & \frac{2 \sqrt{2}}{9} & -\frac{10}{9} & 0 & 0 & 0 & 0 \\
 0 & 0 & 0 & 0 & -\frac{13}{24} & \frac{3}{4 \sqrt{10}} & -\frac{5 \sqrt{5}}{24} & \frac{\sqrt{\frac{2}{5}}}{3} \\
 0 & 0 & 0 & 0 & \frac{3}{4 \sqrt{10}} & -\frac{19}{20} & -\frac{7}{12 \sqrt{2}} & \frac{4}{15} \\
 0 & 0 & 0 & 0 & -\frac{5 \sqrt{5}}{24} & -\frac{7}{12 \sqrt{2}} & -\frac{67}{72} & \frac{\sqrt{2}}{9} \\
 0 & 0 & 0 & 0 & \frac{\sqrt{\frac{2}{5}}}{3} & \frac{4}{15} & \frac{\sqrt{2}}{9} & -\frac{56}{45}
\end{array}
\right), \nonumber\\
&&\lambda_4 \lambda_5 =
\left(
\begin{array}{cccccccc}
 -\frac{5}{3} & 0 & 1 & -\sqrt{2} & 0 & 0 & 0 & 0 \\
 0 & -\frac{8}{3} & 0 & 0 & 0 & 0 & 0 & 0 \\
 1 & 0 & -\frac{5}{3} & -\sqrt{2} & 0 & 0 & 0 & 0 \\
 -\sqrt{2} & 0 & -\sqrt{2} & -\frac{2}{3} & 0 & 0 & 0 & 0 \\
 0 & 0 & 0 & 0 & -\frac{29}{12} & \frac{3}{2 \sqrt{10}} & \frac{\sqrt{5}}{4} & -\sqrt{\frac{2}{5}} \\
 0 & 0 & 0 & 0 & \frac{3}{2 \sqrt{10}} & -\frac{53}{30} & \frac{3}{2 \sqrt{2}} & -\frac{6}{5} \\
 0 & 0 & 0 & 0 & \frac{\sqrt{5}}{4} & \frac{3}{2 \sqrt{2}} & -\frac{17}{12} & -\sqrt{2} \\
 0 & 0 & 0 & 0 & -\sqrt{\frac{2}{5}} & -\frac{6}{5} & -\sqrt{2} & -\frac{16}{15}
\end{array}
\right), \nonumber\\
&&\lambda_1 \lambda_6 =
\left(
\begin{array}{cccccccc}
 -\frac{7}{6} & -\frac{1}{\sqrt{2}} & \frac{1}{2} & 0 & 0 & 0 & 0 & 0 \\
 -\frac{1}{\sqrt{2}} & -1 & -\frac{1}{3 \sqrt{2}} & -\frac{2}{3} & 0 & 0 & 0 & 0 \\
 \frac{1}{2} & -\frac{1}{3 \sqrt{2}} & -\frac{29}{18} & -\frac{2 \sqrt{2}}{9} & 0 & 0 & 0 & 0 \\
 0 & -\frac{2}{3} & -\frac{2 \sqrt{2}}{9} & -\frac{8}{9} & 0 & 0 & 0 & 0 \\
 0 & 0 & 0 & 0 & -\frac{11}{12} & \frac{\sqrt{\frac{5}{2}}}{2} & -\frac{\sqrt{5}}{4} & 0 \\
 0 & 0 & 0 & 0 & \frac{\sqrt{\frac{5}{2}}}{2} & -\frac{13}{10} & -\frac{1}{6 \sqrt{2}} & -\frac{4}{15} \\
 0 & 0 & 0 & 0 & -\frac{\sqrt{5}}{4} & -\frac{1}{6 \sqrt{2}} & -\frac{37}{36} & -\frac{4 \sqrt{2}}{9} \\
 0 & 0 & 0 & 0 & 0 & -\frac{4}{15} & -\frac{4 \sqrt{2}}{9} & -\frac{64}{45}
\end{array}
\right), \nonumber\\
&&\lambda_5 \lambda_6 =
\left(
\begin{array}{cccccccc}
 -\frac{11}{12} & \frac{3}{2 \sqrt{2}} & -\frac{3}{4} & 0 & 0 & 0 & 0 & 0 \\
 \frac{3}{2 \sqrt{2}} & -\frac{7}{6} & \frac{1}{2 \sqrt{2}} & 1 & 0 & 0 & 0 & 0 \\
 -\frac{3}{4} & \frac{1}{2 \sqrt{2}} & -\frac{1}{4} & \frac{\sqrt{2}}{3} & 0 & 0 & 0 & 0 \\
 0 & 1 & \frac{\sqrt{2}}{3} & -\frac{4}{3} & 0 & 0 & 0 & 0 \\
 0 & 0 & 0 & 0 & -\frac{31}{24} & -\frac{3 \sqrt{\frac{5}{2}}}{4} & \frac{3 \sqrt{5}}{8} & 0 \\
 0 & 0 & 0 & 0 & -\frac{3 \sqrt{\frac{5}{2}}}{4} & -\frac{43}{60} & \frac{1}{4 \sqrt{2}} & \frac{2}{5} \\
 0 & 0 & 0 & 0 & \frac{3 \sqrt{5}}{8} & \frac{1}{4 \sqrt{2}} & -\frac{9}{8} & \frac{2 \sqrt{2}}{3} \\
 0 & 0 & 0 & 0 & 0 & \frac{2}{5} & \frac{2 \sqrt{2}}{3} & -\frac{8}{15}
\end{array}
\right).
\end{eqnarray}
\begin{eqnarray}
&&\lambda_1 \lambda_2 \sigma_1 \sigma_2=
\left(
\begin{array}{cccccccc}
 \frac{5}{6} & -\frac{5}{3 \sqrt{2}} & \frac{1}{18} & \frac{\sqrt{2}}{9} & 0 & 0 & 0 & 0 \\
 -\frac{5}{3 \sqrt{2}} & 1 & -\frac{5}{9 \sqrt{2}} & -\frac{10}{9} & 0 & 0 & 0 & 0 \\
 \frac{1}{18} & -\frac{5}{9 \sqrt{2}} & \frac{37}{54} & \frac{\sqrt{2}}{27} & 0 & 0 & 0 & 0 \\
 \frac{\sqrt{2}}{9} & -\frac{10}{9} & \frac{\sqrt{2}}{27} & \frac{22}{27} & 0 & 0 & 0 & 0 \\
 0 & 0 & 0 & 0 & \frac{7}{6} & \frac{7}{3 \sqrt{10}} & -\frac{11 \sqrt{5}}{18} & \frac{\sqrt{\frac{2}{5}}}{9} \\
 0 & 0 & 0 & 0 & \frac{7}{3 \sqrt{10}} & \frac{3}{5} & -\frac{7}{9 \sqrt{2}} & \frac{34}{45} \\
 0 & 0 & 0 & 0 & -\frac{11 \sqrt{5}}{18} & -\frac{7}{9 \sqrt{2}} & \frac{19}{54} & -\frac{\sqrt{2}}{27} \\
 0 & 0 & 0 & 0 & \frac{\sqrt{\frac{2}{5}}}{9} & \frac{34}{45} & -\frac{\sqrt{2}}{27} & -\frac{16}{135}
\end{array}
\right), \nonumber\\
&&\lambda_1 \lambda_4 \sigma_1 \sigma_4=
\left(
\begin{array}{cccccccc}
 \frac{11}{4} & \frac{5}{6 \sqrt{2}} & \frac{1}{36} & -\frac{\sqrt{2}}{9} & 0 & 0 & 0 & 0 \\
 \frac{5}{6 \sqrt{2}} & \frac{23}{18} & \frac{5}{18 \sqrt{2}} & \frac{5}{9} & 0 & 0 & 0 & 0 \\
 \frac{1}{36} & \frac{5}{18 \sqrt{2}} & \frac{17}{108} & -\frac{2 \sqrt{2}}{27} & 0 & 0 & 0 & 0
   \\
 -\frac{\sqrt{2}}{9} & \frac{5}{9} & -\frac{2 \sqrt{2}}{27} & -\frac{50}{27} & 0 & 0 & 0 & 0 \\
 0 & 0 & 0 & 0 & \frac{15}{8} & -\frac{13}{12 \sqrt{10}} & \frac{23 \sqrt{5}}{72} &
   -\frac{\sqrt{\frac{2}{5}}}{9} \\
 0 & 0 & 0 & 0 & -\frac{13}{12 \sqrt{10}} & \frac{31}{36} & \frac{17}{36 \sqrt{2}} &
   -\frac{4}{9} \\
 0 & 0 & 0 & 0 & \frac{23 \sqrt{5}}{72} & \frac{17}{36 \sqrt{2}} & -\frac{71}{216} &
   -\frac{\sqrt{2}}{27} \\
 0 & 0 & 0 & 0 & -\frac{\sqrt{\frac{2}{5}}}{9} & -\frac{4}{9} & -\frac{\sqrt{2}}{27} &
   -\frac{56}{27}
\end{array}
\right), \nonumber\\
&&\lambda_4 \lambda_5 \sigma_4 \sigma_5=
\left(
\begin{array}{cccccccc}
 -3 & 0 & -\frac{1}{3} & \frac{\sqrt{2}}{3} & 0 & 0 & 0 & 0 \\
 0 & -\frac{8}{3} & 0 & 0 & 0 & 0 & 0 & 0 \\
 -\frac{1}{3} & 0 & -3 & \frac{\sqrt{2}}{3} & 0 & 0 & 0 & 0 \\
 \frac{\sqrt{2}}{3} & 0 & \frac{\sqrt{2}}{3} & -\frac{10}{3} & 0 & 0 & 0 & 0 \\
 0 & 0 & 0 & 0 & -\frac{11}{4} & -\frac{1}{2 \sqrt{10}} & -\frac{\sqrt{5}}{12} &
   \frac{\sqrt{\frac{2}{5}}}{3} \\
 0 & 0 & 0 & 0 & -\frac{1}{2 \sqrt{10}} & -\frac{89}{30} & -\frac{1}{2 \sqrt{2}} & \frac{2}{5}
   \\
 0 & 0 & 0 & 0 & -\frac{\sqrt{5}}{12} & -\frac{1}{2 \sqrt{2}} & -\frac{37}{12} &
   \frac{\sqrt{2}}{3} \\
 0 & 0 & 0 & 0 & \frac{\sqrt{\frac{2}{5}}}{3} & \frac{2}{5} & \frac{\sqrt{2}}{3} &
   -\frac{16}{5}
\end{array}
\right), \nonumber\\
&&\lambda_1 \lambda_6 \sigma_1 \sigma_6=
\left(
\begin{array}{cccccccc}
 -\frac{5}{18} & -\frac{5}{9 \sqrt{2}} & \frac{1}{18} & 0 & -\frac{16}{9} & -\frac{2
   \sqrt{10}}{9} & -\frac{4 \sqrt{5}}{9} & 0 \\
 -\frac{5}{9 \sqrt{2}} & -\frac{31}{27} & -\frac{5}{27 \sqrt{2}} & -\frac{10}{27} & -\frac{2
   \sqrt{2}}{3} & \frac{56}{27 \sqrt{5}} & \frac{2 \sqrt{10}}{27} & \frac{32}{27 \sqrt{5}} \\
 \frac{1}{18} & -\frac{5}{27 \sqrt{2}} & \frac{43}{162} & \frac{22 \sqrt{2}}{81} & \frac{4}{9}
   & -\frac{22 \sqrt{\frac{2}{5}}}{27} & -\frac{88 \sqrt{5}}{81} & -\frac{32
   \sqrt{\frac{2}{5}}}{81} \\
 0 & -\frac{10}{27} & \frac{22 \sqrt{2}}{81} & \frac{4}{81} & 0 & \frac{56}{27 \sqrt{5}} &
   -\frac{8 \sqrt{10}}{81} & -\frac{464}{81 \sqrt{5}} \\
 -\frac{16}{9} & -\frac{2 \sqrt{2}}{3} & \frac{4}{9} & 0 & \frac{95}{36} & -\frac{5
   \sqrt{\frac{5}{2}}}{6} & \frac{7 \sqrt{5}}{36} & 0 \\
 -\frac{2 \sqrt{10}}{9} & \frac{56}{27 \sqrt{5}} & -\frac{22 \sqrt{\frac{2}{5}}}{27} &
   \frac{56}{27 \sqrt{5}} & -\frac{5 \sqrt{\frac{5}{2}}}{6} & \frac{5}{54} & \frac{47}{54
   \sqrt{2}} & -\frac{20}{27} \\
 -\frac{4 \sqrt{5}}{9} & \frac{2 \sqrt{10}}{27} & -\frac{88 \sqrt{5}}{81} & -\frac{8
   \sqrt{10}}{81} & \frac{7 \sqrt{5}}{36} & \frac{47}{54 \sqrt{2}} & \frac{307}{324} & \frac{28
   \sqrt{2}}{81} \\
 0 & \frac{32}{27 \sqrt{5}} & -\frac{32 \sqrt{\frac{2}{5}}}{81} & -\frac{464}{81 \sqrt{5}} & 0
   & -\frac{20}{27} & \frac{28 \sqrt{2}}{81} & \frac{80}{81}
\end{array}
\right), \nonumber\\
&&\lambda_5 \lambda_6 \sigma_5 \sigma_6=
\left(
\begin{array}{cccccccc}
 -\frac{11}{12} & \frac{5}{6 \sqrt{2}} & -\frac{1}{12} & 0 & -\frac{4}{3} & \frac{\sqrt{10}}{3}
   & \frac{2 \sqrt{5}}{3} & 0 \\
 \frac{5}{6 \sqrt{2}} & -\frac{17}{18} & \frac{5}{18 \sqrt{2}} & \frac{5}{9} & \sqrt{2} &
   \frac{32}{9 \sqrt{5}} & -\frac{\sqrt{10}}{9} & -\frac{16}{9 \sqrt{5}} \\
 -\frac{1}{12} & \frac{5}{18 \sqrt{2}} & \frac{101}{108} & -\frac{11 \sqrt{2}}{27} &
   -\frac{2}{3} & \frac{11 \sqrt{\frac{2}{5}}}{9} & \frac{8 \sqrt{5}}{27} & \frac{16
   \sqrt{\frac{2}{5}}}{27} \\
 0 & \frac{5}{9} & -\frac{11 \sqrt{2}}{27} & -\frac{20}{27} & 0 & -\frac{28}{9 \sqrt{5}} &
   \frac{4 \sqrt{10}}{27} & -\frac{128}{27 \sqrt{5}} \\
 -\frac{4}{3} & \sqrt{2} & -\frac{2}{3} & 0 & \frac{65}{24} & \frac{5 \sqrt{\frac{5}{2}}}{4} &
   -\frac{7 \sqrt{5}}{24} & 0 \\
 \frac{\sqrt{10}}{3} & \frac{32}{9 \sqrt{5}} & \frac{11 \sqrt{\frac{2}{5}}}{9} & -\frac{28}{9
   \sqrt{5}} & \frac{5 \sqrt{\frac{5}{2}}}{4} & -\frac{5}{36} & -\frac{47}{36 \sqrt{2}} &
   \frac{10}{9} \\
 \frac{2 \sqrt{5}}{3} & -\frac{\sqrt{10}}{9} & \frac{8 \sqrt{5}}{27} & \frac{4 \sqrt{10}}{27} &
   -\frac{7 \sqrt{5}}{24} & -\frac{47}{36 \sqrt{2}} & \frac{557}{216} & -\frac{14 \sqrt{2}}{27}
   \\
 0 & -\frac{16}{9 \sqrt{5}} & \frac{16 \sqrt{\frac{2}{5}}}{27} & -\frac{128}{27 \sqrt{5}} & 0 &
   \frac{10}{9} & -\frac{14 \sqrt{2}}{27} & -\frac{40}{27}
\end{array}
\right).
\end{eqnarray}
\subsection{S=2}
\begin{eqnarray}
&&\lambda_1 \lambda_2 =
\left(
\begin{array}{ccccc}
 -\frac{5}{3} & -\sqrt{\frac{2}{5}} & \frac{\sqrt{5}}{3} & -\frac{\sqrt{\frac{2}{5}}}{3} & 0 \\
 -\sqrt{\frac{2}{5}} & -\frac{16}{15} & \frac{\sqrt{2}}{3} & -\frac{2}{15} & 0 \\
 \frac{\sqrt{5}}{3} & \frac{\sqrt{2}}{3} & -\frac{11}{9} & \frac{\sqrt{2}}{9} & 0 \\
 -\frac{\sqrt{\frac{2}{5}}}{3} & -\frac{2}{15} & \frac{\sqrt{2}}{9} & -\frac{32}{45} & 0 \\
 0 & 0 & 0 & 0 & -\frac{2}{3}
\end{array}
\right)
, \quad
\lambda_1 \lambda_4=
\left(
\begin{array}{ccccc}
 -\frac{13}{24} & \frac{3}{4 \sqrt{10}} & -\frac{5 \sqrt{5}}{24} & \frac{\sqrt{\frac{2}{5}}}{3} & 0
   \\
 \frac{3}{4 \sqrt{10}} & -\frac{19}{20} & -\frac{7}{12 \sqrt{2}} & \frac{4}{15} & 0 \\
 -\frac{5 \sqrt{5}}{24} & -\frac{7}{12 \sqrt{2}} & -\frac{67}{72} & \frac{\sqrt{2}}{9} & 0 \\
 \frac{\sqrt{\frac{2}{5}}}{3} & \frac{4}{15} & \frac{\sqrt{2}}{9} & -\frac{56}{45} & 0 \\
 0 & 0 & 0 & 0 & -1
\end{array}
\right)
, \nonumber\\
&&\lambda_4 \lambda_5 =
\left(
\begin{array}{ccccc}
 -\frac{29}{12} & \frac{3}{2 \sqrt{10}} & \frac{\sqrt{5}}{4} & -\sqrt{\frac{2}{5}} & 0 \\
 \frac{3}{2 \sqrt{10}} & -\frac{53}{30} & \frac{3}{2 \sqrt{2}} & -\frac{6}{5} & 0 \\
 \frac{\sqrt{5}}{4} & \frac{3}{2 \sqrt{2}} & -\frac{17}{12} & -\sqrt{2} & 0 \\
 -\sqrt{\frac{2}{5}} & -\frac{6}{5} & -\sqrt{2} & -\frac{16}{15} & 0 \\
 0 & 0 & 0 & 0 & -\frac{8}{3}
\end{array}
\right)
,\quad
\lambda_1 \lambda_6 =
\left(
\begin{array}{ccccc}
 -\frac{11}{12} & \frac{\sqrt{\frac{5}{2}}}{2} & -\frac{\sqrt{5}}{4} & 0 & 0 \\
 \frac{\sqrt{\frac{5}{2}}}{2} & -\frac{13}{10} & -\frac{1}{6 \sqrt{2}} & -\frac{4}{15} & 0 \\
 -\frac{\sqrt{5}}{4} & -\frac{1}{6 \sqrt{2}} & -\frac{37}{36} & -\frac{4 \sqrt{2}}{9} & 0 \\
 0 & -\frac{4}{15} & -\frac{4 \sqrt{2}}{9} & -\frac{64}{45} & 0 \\
 0 & 0 & 0 & 0 & -2
\end{array}
\right)
, \nonumber\\
&&\lambda_5 \lambda_6 =
\left(
\begin{array}{ccccc}
 -\frac{31}{24} & -\frac{3 \sqrt{\frac{5}{2}}}{4} & \frac{3 \sqrt{5}}{8} & 0 & 0 \\
 -\frac{3 \sqrt{\frac{5}{2}}}{4} & -\frac{43}{60} & \frac{1}{4 \sqrt{2}} & \frac{2}{5} & 0 \\
 \frac{3 \sqrt{5}}{8} & \frac{1}{4 \sqrt{2}} & -\frac{9}{8} & \frac{2 \sqrt{2}}{3} & 0 \\
 0 & \frac{2}{5} & \frac{2 \sqrt{2}}{3} & -\frac{8}{15} & 0 \\
 0 & 0 & 0 & 0 & \frac{1}{3}
\end{array}
\right).
\end{eqnarray}
\begin{eqnarray}
&&\lambda_1 \lambda_2 \sigma_1 \sigma_2 = \left(
\begin{array}{ccccc}
 \frac{7}{6} & \frac{7}{3 \sqrt{10}} & -\frac{11 \sqrt{5}}{18} & \frac{\sqrt{\frac{2}{5}}}{9} & 0
   \\
 \frac{7}{3 \sqrt{10}} & \frac{3}{5} & -\frac{7}{9 \sqrt{2}} & \frac{34}{45} & 0 \\
 -\frac{11 \sqrt{5}}{18} & -\frac{7}{9 \sqrt{2}} & \frac{19}{54} & -\frac{\sqrt{2}}{27} & 0 \\
 \frac{\sqrt{\frac{2}{5}}}{9} & \frac{34}{45} & -\frac{\sqrt{2}}{27} & -\frac{16}{135} & 0 \\
 0 & 0 & 0 & 0 & -\frac{2}{3}
\end{array}
\right)
,~
\lambda_1 \lambda_4 \sigma_1 \sigma_4 = \left(
\begin{array}{ccccc}
 \frac{15}{8} & -\frac{13}{12 \sqrt{10}} & \frac{23 \sqrt{5}}{72} & -\frac{\sqrt{\frac{2}{5}}}{9} & 0 \\
 -\frac{13}{12 \sqrt{10}} & \frac{31}{36} & \frac{17}{36 \sqrt{2}} & -\frac{4}{9} & 0 \\
 \frac{23 \sqrt{5}}{72} & \frac{17}{36 \sqrt{2}} & -\frac{71}{216} & -\frac{\sqrt{2}}{27} & 0 \\
 -\frac{\sqrt{\frac{2}{5}}}{9} & -\frac{4}{9} & -\frac{\sqrt{2}}{27} & -\frac{56}{27} & 0 \\
 0 & 0 & 0 & 0 & -1
\end{array}
\right), \nonumber\\
&&\lambda_4 \lambda_5 \sigma_4 \sigma_5 = \left(
\begin{array}{ccccc}
 -\frac{11}{4} & -\frac{1}{2 \sqrt{10}} & -\frac{\sqrt{5}}{12} & \frac{\sqrt{\frac{2}{5}}}{3} & 0 \\
 -\frac{1}{2 \sqrt{10}} & -\frac{89}{30} & -\frac{1}{2 \sqrt{2}} & \frac{2}{5} & 0 \\
 -\frac{\sqrt{5}}{12} & -\frac{1}{2 \sqrt{2}} & -\frac{37}{12} & \frac{\sqrt{2}}{3} & 0 \\
 \frac{\sqrt{\frac{2}{5}}}{3} & \frac{2}{5} & \frac{\sqrt{2}}{3} & -\frac{16}{5} & 0 \\
 0 & 0 & 0 & 0 & -\frac{8}{3}
\end{array}
\right)
,~
\lambda_1 \lambda_6 \sigma_1 \sigma_6 = \left(
\begin{array}{ccccc}
 -\frac{19}{12} & \frac{\sqrt{\frac{5}{2}}}{2} & -\frac{7}{12 \sqrt{5}} & 0 & \frac{4}{3 \sqrt{5}} \\
 \frac{\sqrt{\frac{5}{2}}}{2} & -\frac{1}{18} & -\frac{47}{90 \sqrt{2}} & \frac{4}{9} & \frac{8
   \sqrt{2}}{15} \\
 -\frac{7}{12 \sqrt{5}} & -\frac{47}{90 \sqrt{2}} & -\frac{307}{540} & -\frac{28 \sqrt{2}}{135} &
   -\frac{92}{45} \\
 0 & \frac{4}{9} & -\frac{28 \sqrt{2}}{135} & -\frac{16}{27} & -\frac{32 \sqrt{2}}{45} \\
 \frac{4}{3 \sqrt{5}} & \frac{8 \sqrt{2}}{15} & -\frac{92}{45} & -\frac{32 \sqrt{2}}{45} & \frac{14}{5}
\end{array}
\right), \nonumber\\
&&\lambda_5 \lambda_6 \sigma_5 \sigma_6 = \left(
\begin{array}{ccccc}
 -\frac{13}{8} & -\frac{3 \sqrt{\frac{5}{2}}}{4} & \frac{7}{8 \sqrt{5}} & 0 & -\frac{2}{\sqrt{5}} \\
 -\frac{3 \sqrt{\frac{5}{2}}}{4} & \frac{1}{12} & \frac{47}{60 \sqrt{2}} & -\frac{2}{3} & -\frac{4
   \sqrt{2}}{5} \\
 \frac{7}{8 \sqrt{5}} & \frac{47}{60 \sqrt{2}} & -\frac{557}{360} & \frac{14 \sqrt{2}}{45} & -\frac{26}{15}
   \\
 0 & -\frac{2}{3} & \frac{14 \sqrt{2}}{45} & \frac{8}{9} & \frac{16 \sqrt{2}}{15} \\
 -\frac{2}{\sqrt{5}} & -\frac{4 \sqrt{2}}{5} & -\frac{26}{15} & \frac{16 \sqrt{2}}{15} & -\frac{7}{15}
\end{array}
\right).
\end{eqnarray}
\subsection{S=3}
\begin{eqnarray}
  \lambda_1 \lambda_2 = -\frac{2}{3},~ \lambda_1 \lambda_4 = -1,~ \lambda_4 \lambda_5 = -\frac{8}{3},~ \lambda_1 \lambda_6 = -2,~ \lambda_5 \lambda_6 = \frac{1}{3}.
\end{eqnarray}
\begin{eqnarray}
  \lambda_1 \lambda_2 \sigma_1 \sigma_2 = -\frac{2}{3},~ \lambda_1 \lambda_4 \sigma_1 \sigma_4 = -1,~ \lambda_4 \lambda_5 \sigma_4 \sigma_5 = -\frac{8}{3},~ \lambda_1 \lambda_6 \sigma_1 \sigma_6 = -2,~ \lambda_5 \lambda_6 \sigma_5 \sigma_6 = \frac{1}{3}.
\end{eqnarray}
For each spin, $\sum_{i<j}^6 \lambda_i \lambda_j=-16 I_n$ where $I_n$ is the $n \times n$ identity matrix.
\end{widetext}

\end{document}